\documentclass[10pt,preprint2]{emulateapj}
\usepackage{natbib}

\shorttitle{VINE - performance}
\shortauthors{Nelson et al.}

\begin{document}

\newcommand{\vineI}{Vine1}
\newcommand{\vineII}{Vine2}

\newcommand{\cpc}{Comp. Phys. Comm.}
\newcommand{\cpr}{Comp. Phys. Rep.}
\newcommand{\siamcomp}{SIAM J. Scient. Comp.}
\newcommand{\jcomp}{J. Comp. Phys.}
\newcommand{\m}{\mathbf}
\newcommand{\f}{\frac}
\newcommand{\beq}{\begin{equation}}
\newcommand{\eeq}{\end{equation}}
\newcommand{\beqa}{\begin{eqnarray}}
\newcommand{\eeqa}{\end{eqnarray}}

\title{VINE -- A numerical code for simulating astrophysical systems
using particles II: Implementation and performance characteristics}

\author{Andrew F. Nelson\altaffilmark{1,2}, M. Wetzstein\altaffilmark{3,4},
  and T. Naab\altaffilmark{4,5} }
\email{andy.nelson@lanl.gov}

\altaffiltext{1}{Los Alamos National Laboratory, HPC-5, MS B272,
                    Los Alamos, NM, 87545, USA }
\altaffiltext{2}{UKAFF Fellow}
\altaffiltext{3}{Department of Astrophysical Sciences, Princeton University,
                     Princeton, NJ 08544, USA}

\altaffiltext{4}{Universit\"ats-Sternwarte, Scheinerstr. 1,
                     81679 M\"unchen, Germany}
\altaffiltext{5}{Institute of Astronomy, Maddingley Road, Cambridge,
                      United Kingdom}

\begin{abstract}
We continue our presentation of VINE. In this paper, we begin with a
description of relevant architectural properties of the serial and
shared memory parallel computers on which VINE is intended to run, and
describe their influences on the design of the code itself. We
continue with a detailed description of a number of optimizations made
to the layout of the particle data in memory and to our implementation
of a binary tree used to access that data for use in gravitational
force calculations and searches for SPH neighbor particles. We
describe the modifications to the code necessary to obtain forces
efficiently from special purpose `GRAPE' hardware, the interfaces
required to allow transparent substitution of those forces in the code
instead of those obtained from the tree, and the modifications
necessary to use both tree and GRAPE together as a fused GRAPE/tree
combination. We conclude with an extensive series of performance
tests, which demonstrate that the code can be run efficiently and
without modification in serial on small workstations or in parallel
using the OpenMP compiler directives on large scale, shared memory
parallel machines. We analyze the effects of the code optimizations
and estimate that they improve its overall performance by more than an
order of magnitude over that obtained by many other tree codes. Scaled
parallel performance of the gravity and SPH calculations, together the
most costly components of most simulations, is nearly linear up to at
least 120 processors on moderate sized test problems using the
Origin~3000 architecture, and to the maximum machine sizes available
to us on several other architectures. At similar accuracy, performance
of VINE, used in GRAPE-tree mode, is approximately a factor two slower
than that of VINE, used in host-only mode. Further optimizations of
the GRAPE/host communications could improve the speed by as much as a
factor of three, but have not yet been implemented in VINE. Finally,
we find that although parallel performance on small problems may reach
a plateau beyond which more processors bring no additional speedup,
performance never decreases, a factor important for running large
simulations on many processors with individual time steps, where only
a small fraction of the total particles require updates at any given
moment.

\end{abstract}



\section{Introduction}\label{sec:intro}

In a companion paper \citep*[][hereafter \vineI]{vineI}, we describe
the physics implemented in our numerical code, VINE, the integrators
used to evolve systems forward in time, the implementation of
hydrodynamics using SPH (Smoothed Particle Hydrodynamics), the
gravitational force solver and of the implementation of periodic
boundaries. We also describe several test problems on which we
demonstrate the code and its performance relative to the Gadget~2 code
of \citet{springel_g2}. In this paper, we describe our implementation
of the techniques used to calculate gravitational forces and determine
neighbors needed for SPH calculations. We also discuss in detail the
optimizations made to maximize the code's performance.

In order to maximize scientific productivity, a numerical code must be
able to perform simulations with the lowest possible computational
expense while still maintaining an accurate realization of the
evolution. The exact definition of an `accurate realization of the
evolution' will in general be problem dependent: for one problem, one
technique might yield entirely acceptable results, while for another
it produces nonsense. VINE has been designed to be both very flexible
and very modular, in order to allow the choices of specific algorithms
to be made by its users, rather than by its writers. If one algorithm
or setting turns out to be inappropriate, another may easily be
selected or added to the code.

Of the two main components in the particles' integration schemes,
update and derivative calculation, the latter is by far the most
computationally expensive, but the former may have greater influence
on the total run time. For example, stability properties may require
smaller time steps be made with one integrator rather than another, or
one integrator may require more derivative calculations per time step
than another. In \vineI, we discussed the two integrators implemented
in VINE, each of which require different constraints to obtain a given
computational accuracy at a given expense. We also described our
implementation of an individual time step scheme, by which each
particle can be evolved forward in time at its own rate, greatly
reducing the total computational expense when systems with large
dynamical ranges are modeled.

In this paper, we focus our attention on techniques that can make
derivative calculations more efficient. Our purpose will be twofold.
First, we describe our implementation of a binary tree based scheme
for efficiently determining gravitational forces on particles and
lists of neighbors particles for use in SPH calculations and the
alternative options we have implemented for using special purpose
hardware for those calculations if it is available. Secondly, because
efficient calculations require both efficient algorithms and efficient
implementation of those algorithms on computational hardware, we
describe in some detail the the low level optimizations we have
implemented to improve the performance of the code itself on modern,
microprocessor based computers. It is our hope that in describing the
optimizations in detail, other users of VINE will come to understand
the principles important for obtaining good performance, so that they
are able to incorporate features of their own into VINE with
comparatively little effort, and with equally optimized performance.
Alternatively, because we recognize that some may have considerable
investments in their own codes, our descriptions may be of use in
making incremental modifications to those codes that also improve
performance.

We begin in section \ref{sec:implement} with a discussion of the
properties of the computers on which we expect that VINE will be used.
In section \ref{sec:tree} we describe the construction of the binary
tree used to organize particles so that calculations may be performed
efficiently on them. In section \ref{sec:treewalk}, we describe
efficient methods for accessing the data in the tree, and in section
\ref{sec:gravity} we describe optimizations of the calculations
themselves. Next, in section \ref{sec:overview} we finish the
description of VINE with an overview of the code itself and the
hardware requirements needed to run a simulation of a given size. The
overall performance of each of the four major portions of the code
that interact with the tree is described in section \ref{sec:perf},
including a discussion of parallel efficiency and tunable parameters
to increase performance. Finally, in section \ref{sec_summary} we
summarize the features of VINE, suggest further improvements that may
be made to it and give a web site where the code may be obtained
electronically.

\section{Practical issues relevant for obtaining good
performance}\label{sec:implement}

Because efficient computation requires knowledge of both the
calculations being performed and the machines on which they are
performed, we describe here some of the more important properties of
the computers on which we expect this code to be run, and the impact
that their strengths and limitations have for the design of the code.

We have designed VINE to run efficiently on microprocessors commonly
used in workstations and shared memory parallel computers. If
available, and at the user's option, VINE can utilize special purpose
`GRAPE' hardware to calculate mutual gravitational forces of particles
on each other. In this section, we review some of the most significant
architectural features of systems employing such hardware, describing
their strengths and limitations. We will conclude by pointing out
constraints that motivate some of the optimizations in VINE.

\subsection{Running simulations on one processor}\label{sec:impl-single}

Without question, the most important hardware constraint we encounter
in optimizing VINE for fast performance is that imposed by memory
latency. As a class, all microprocessors share the constraint that
loading or storing a value to or from memory is much more expensive
than, for example, adding them together. Depending on the processor, a
single calculation may take one or at most a few clock cycles, and
several calculations may be processed at the same time on the same
processor. On the other hand, loading and storing may take as many as
hundreds or thousands of clock cycles. The universal solution has been
to incorporate relatively small caches of memory into the processor
itself, from which values can be accessed very quickly, assuming that
they are already present in the cache.

Typically, modern processors include two or more levels of cache,
labeled `L1', `L2' etc, which are divided into a number of `lines' and
`sets'. The L1 cache is the smallest and fastest and higher cache
levels are slower and larger. A cache line consists of $\sim 32-128$
bytes of memory and is the smallest increment of information that can
be loaded into or stored from that cache at a time. Even if a program
requires only a single value from a given range of memory, say a
single integer (4 bytes) or a single double precision (8 bytes) real
value, that load also brings several additional, possibly unneeded
values into the cache.  Any given address in main memory can be loaded
into any one of exactly $n_s$ lines, where $n_s$ is the number of sets
(the set `associativity'). Typically, $n_s\sim2-4$.

For larger scale memory requirements, processors access a main memory,
for which access times are much longer. In practice, access to
the entire memory at one time does not occur because a translation
must be made between the virtual address by which the program refers
to some value and the physical address at which that value is actually
stored. The processor holds a finite number of such translations to
`pages' of main memory which, depending on the system, may be as small
as 4~kB or as large as 32~MB and on some, may even be selected at run
time by the user for a given job. The processor stores the address
conversions for a number of virtual/physical pages in a special cache
called the `Translation Lookaside Buffer' (TLB). Accessing a value
resident on a page already mapped by the TLB, but not currently
resident in cache may take as many as a few hundred clock cycles, due
to the slower speed of main memory relative to the processor.
Accessing a value from a page not resident in the TLB requires in
addition that a new address translation be calculated, replacing
replace one of those currently in residence in the TLB. While some
processors have special circuitry to assist in the calculation so that
it adds only a small additional delay, others require an intervention
by the operating system, resulting in additional delays of a few
hundred cycles.

\subsection{Running simulations with GRAPE hardware}\label{sec:impl-grape}

The most costly component of simulating any system including self
gravity is the calculation of the mutual gravitational forces of
particles on each other. While costly, it is also simple to describe
and implement, and therefore lends itself to a specialized solution.
Rather than perform the calculations on a general purpose processor
used for all the calculations, special-purpose processors called GRAPE
(for \textbf{GRA}vity Pi\textbf{PE}) have been
developed \citep[see e.g.][]{sugimoto90,makino_taiji98} to perform the
gravity calculation instead. 

Although GRAPE hardware is specifically designed to accelerate
computation of gravitational interactions, it cannot perform any of
the other calculations required for the simulation to proceed.  It
must therefore be attached to a general purpose computer (the `host')
and data must be sent to it for calculation, then results returned to
the host for use in the rest of the calculations. In some
circumstances, the data transfer time (i.e. the `memory bandwidth')
can in fact be comparable to or greater than the calculation time, due
both to the actual transfer speed and to latencies required to begin
or complete each transfer. An important example of such inefficiency
will be when only a few particles require force calculations, such as
will occur when individual time steps (see \vineI) are used in a
simulation, or when the interaction lists are relatively short, such
as will occur when the hardware is used in combination with a tree. 

Another limitation of GRAPE systems is that they implement a reduced
precision internal numerical representation, which constrains the
dynamic range of the input data (particle masses and positions) and
increases the errors in forces returned. GRAPE-3 \citep{okumura93} for
example, produces errors of order of $2\%$, while GRAPE-5
\citep{kawai2000} produces errors of $0.3\%$. On the other hand, even
numbered revisions, such as GRAPE-2 \citep{ito91}, GRAPE-4
\citep{mtes97} and GRAPE-6 \citep{makino2003}, implement much higher
internal precision numerical representations, and do not produce such
large errors.

Because the errors in the reduced precision variants are uncorrelated,
they do not impose problems for the time evolution of collisionless
simulations \citep{mie90}, and may be used without difficulty in
models of such systems. On the other hand, the high precision GRAPE
variants may be used to model both collisionless and collisional
systems such as stellar clusters, where individual particles in the
simulation represent individual stars in the cluster.

So far, a number of simulations using GRAPE processors in
combination with the tree and individual time steps have been
performed using VINE \citep{naab2003,naab2006,bell2006}. 

\subsection{Running simulations on many processors in
parallel}\label{sec:impl-para}

Even with the fastest processors available, the time required to
perform simulations of systems at high resolution may exceed the time
that the scientist running the simulation can remain patient.
Fortunately, the required work can be shared among more than one
processor so that time required becomes much smaller, and parallel
computing architectures are commercially available at costs that make
them affordable for use in scientific research. 

In the abstract, work required to complete a given task must be
divisible into sub-units that are independent of all other such units
in order to run correctly in parallel, and those sub-units must be
parceled out to be run at the same time on different processors. Data
required to complete the task must not be changed while any work
remain incomplete and the relevant results of previous tasks must be
communicated to different processors, so that a single coherent
picture of the entire simulation exists at all times. In order to run
efficiently, the work units must be of similar size, so that they can
be divided evenly between all processors and so that little time is
wasted while some processors sit idle and others complete
disproportionate shares. Secondly, communication must be fast, so that
relatively little time is spent synchronizing the results of different
processors. Finally, the amount of work that cannot be broken into
independent parts must be small.

For shared memory systems, partitioning naturally occurs at a very low
level in the code, often at the level of individual loops themselves.
An important factor limiting performance in this case is that the work
per loop iteration can be extremely small, so that the overhead
required to distribute work among processors becomes significant. In
other cases, parallelizing loops at higher levels of organizational
structure may be possible, but will not have identical quantities of
work per iteration, leading to load imbalance unless some intelligence
can be applied to parceling out each iteration.

Additionally, in large scale shared memory systems that employ so
called Non-Uniform Memory Architectures, or `NUMAs', access times to
main memory are not the same for all locations, but instead depend on
where a particular value is stored relative to the processor that
attempts to load it. For simple parallelized loops, such as those used
by the integrator to update particle data at each time step, a
significant and otherwise hidden source of load imbalance can
originate from this source. The same number of loop iterations,
containing identical mathematical operations, may require different
amounts of time to complete when run on different processors because
data required to complete one processor's iterations are
coincidentally found closer to it than are the data required to
complete some other processor's iterations. 

Finally, in recent hardware architectures, two or more processor
`cores' are present on the same chip, and some components on that chip
(e.g. cache, memory controllers, or simply the pins by which data are
transfered on and off the chip itself) are effectively shared between
several logical processors. If the execution profile of code running
on all cores of a chip leads to competition for access to a given
resource, then the contention that results will degrade the overall
performance of the code. In this context, limits on the overall memory
bandwidth of data transfered to the processor may become of particular
concern, in addition to the memory latency issues discussed above.
Processors may `starve' for work, while waiting for data to be loaded
from memory. To date, we have not observed any impacts on the
performance in VINE due to such issues however. 

\subsection{Implications for designing efficient
numerical software}\label{sec:impl-design}

VINE has been developed for use on single and multi-processor shared
memory architectures using the OpenMP \citep{openmp99} suite of
compiler directives to share work among the processors. In this
section, we will therefore focus on issues relevant for obtaining good
serial and parallel performance on shared memory architectures. We
note, however, that many of the issues presented here are common to
both shared and distributed memory paradigms, and so many
recommendations about optimization strategies will apply to
distributed memory parallelism as well.

Both on small scales and large, memory access latencies dominate the
list of performance constraints for large simulations.  The two
important points to note are that first, accessing cached data is very
fast compared to accessing main memory and second, that the caches are
typically much smaller than the total memory required for a typical
simulation. This means that less immediately needed quantities may be
overwritten by loads of other quantities if other calculations require
them. Information may need to be loaded into cache from main memory
many times if it is required for many calculations. We are therefore
well advised to reuse previously cached data to the fullest extent
possible before discarding it for other more immediately useful data.

For example, we will often require a calculation of the distance
between two particles. In three dimensions, such a calculation
requires that six quantities be loaded into the processor for
calculation, consisting of the three spatial coordinates of each
particle. We might consider storing each component of position in
separate arrays, one for each direction. However, this will inevitably
lead in practice to six costly loads from main memory for a single
distance calculation. On the other hand, with the discussion in
section \ref{sec:impl-single} in mind, we know that a single load
operation will load at least four double precision values from main
memory into the L1 cache. If we instead store particle positions in
adjacent memory locations, then all three coordinates can be loaded
into the L1 cache with a single load from main memory, and the second
and third loads to the processor are nearly cost free because they are
already resident in L1. In this context, a two dimensional array with
one dimension corresponding to the $x$, $y$ or $z$ component of the
position, and the other to the specific particle, fits our
requirements exactly if we also take care to arrange that the `fast'
index (which, because VINE is written in Fortran, will be the first)
defines the position components, and the slow index defines the
particle. 

On a more general level, we will frequently require that the same
values be accessed many times. For example, spatially adjacent
particles will have nearly identical lists of neighbors for SPH
calculations and nearly node interaction lists for gravity
calculations. All calculations involving those lists must load the
same quantities for these particles many times. As
\citet{warren_salmon95} point out, if we can arrange that their data
can remain in cache after their first load from main memory, the
calculations will proceed much faster. Two optimizations are possible
here, implemented not only here, but in other codes as well 
\citep{warren_salmon95,springel_g2,stadel01}. First, we note
that if we can arrange that neighbor searches for one particle, be
immediately followed by neighbor searches for a nearby particle so
that the data for their nearly identical neighbor lists (and
prospective neighbors) are already in cache, then calculations using
those data will proceed much faster due to decreased load times.
Second, if physically nearby particles are also located nearby in the
system's memory, then the correlated spatial positions and memory
locations ensure that data are ordinarily found on the same set of
physical pages in memory, providing better cache reuse and fewer TLB
misses. In section \ref{sec:tree-post}, we describe how sorting the
particles and tree nodes accomplishes both optimizations.

On still larger scales, tree traversals will span huge ranges of
physical memory, but access data in that memory only sparsely and
perform only a small number of calculations per access. The actual
calculation of gravitational forces will span similarly large ranges
of memory, with only slightly more calculations per access. For
example, a typical calculation of the gravitational force on a
particle in a simulation comparable to one of our test simulations in
section \ref{sec:test-probs} below, will require the summation of
multipole moment contributions from several hundred to several
thousand nodes and atoms, and require several times this many nodes be
examined for acceptability. Although the quantity of data for these
nodes is comparatively small, it is spread thinly over a volume of
memory that may be many gigabytes in size. As a consequence,
consecutive calculations will lose all benefits from cache or TLB
reuse even if done for particles with identical interaction lists. The
size of an interaction list means that interaction data near the
beginning of a list will be evicted from cache by data near the end of
the list, requiring that it be reloaded when the second particle's
force is calculated. The same effect will occur in the TLB, with
address translations for later nodes in the list evicting those that
came earlier. 

Two straightforward optimizations to the code itself and a third
hardware optimization both increase the raw speed of the calculations
and reduce the impact of sparse memory access patterns substantially.
First, sections \ref{sec:grav-trav} and \ref{sec:find-nay} describe a
method for performing a single tree traversal for groups of particles
at the same time, reducing the overall number of redundant accesses to
tree data. Second, section \ref{sec:treegrav-opts} describes a method
for loading small segments of the full interaction list data into a
small temporary array tuned to the size of the L1 cache, which is then
used for calculations on many particles occur before being discarded
in favor of later nodes in the interaction list. Finally, we utilize
the option of using large hardware pages on systems where they
available and can be used efficiently. 

Sections \ref{sec:perform-grav}, \ref{sec:perform-sph-reo} and
\ref{sec:perform-build} describe the benefits realized by each of the
optimizations discussed in this section on the gravity, SPH and tree
build calculations which are, respectively, the three most costly
operations in particle simulations. Each of the memory management
optimizations discussed above will also act to avoid memory latencies
inherent in NUMA systems as well, by concentrating memory accesses in
a small, moving footprint that is accessed repeatedly. Therefore,
optimizing the code for high performance in serial mode will have the
additional benefit that parallel performance may also benefit. 

\section{Building a tree: organizing particles for efficient
access}\label{sec:tree}

As we discussed in \vineI, lists of neighbor particles and approximate
gravitational force computations can be obtained with the help of a
tree data structure to organize particles. Using a tree, groups of
particles that may potentially interact with another may be qualified
or disqualified with a single calculation, using a node in the tree as
an approximate substitute for some large number of exact interactions.

A variety of techniques for building and implementing tree structures
have been discussed in the literature. One common technique
\citep{barnes_hut86} (hereafter a `BH tree') builds the tree in a `top
down' fashion by artificially tessellating space into successively
smaller cubes. The first cube contains the entire system of particles
and if any cube contains more than one particle, it is split into
eight smaller cubes of equal volume (hence the name oct-tree for this
type of tree structure). The procedure is repeated until the cubes on
the lowest level (the leaves of the tree) contain either exactly one
particle or no particle at all. A similar technique originating in
\citet{Bentley79} is used in the PKDgrav code \citep{stadel01} and its
later incarnation, Gasoline \citep{wadsley2004}, which also includes
hydrodynamics with SPH. This tree algorithm builds a balanced, so
called `$k-d$' tree (referring to $k$ data per node in $d$ dimensions),
by recursively bisecting space. The tree is constructed from top down,
by choosing a direction (typically the direction for which particles
are most extended in space) along which to bisect the population of
particles, so that half of the particles lie on one branch of the tree
and half on the other. A third technique
\citep{press86,jernigan_porter89,bbcp90}, occasionally referred to as a
`Press tree' in honor of one of its inventors, builds a tree from the
`bottom up' by associating mutual nearest neighbor particles or nodes
with each other as successively higher nodes in the tree until only
one node is left. Since exactly two particles or nodes are grouped
together, the latter two types of trees are commonly grouped together
and called binary trees. 

In this section, we first briefly discuss some literature debating the
merits of each of the above tree algorithms and some reasons for
choosing the Press tree for use in VINE, followed by a description of
the method used to construct the tree used in VINE. Because the
details of the tree structure itself will have a major effect on the
total run time of the rest of the code, we will outline our tree
construction and post processing in some detail, although many aspects
of it may be found in the references above.

\subsection{What kind of tree is best?}\label{sec:tree-type}

Use of tree based force calculation algorithms for $N$-body
simulations became common beginning in the 1980s (see citations
above), and debates over which sort of tree algorithm is better or
worse for which sort of calculation have continued since that time. We
attempt here to summarize some of this debate and the choices that
influenced our decision to continue using the Press tree, present in
versions of the code on which VINE is based \citep{bbcp90}.

The basic metric for the quality of a given tree algorithm can be
summarized essentially as a statement of the science goals of the tree
code's user: more, and more realistic, simulations of interesting
physical systems are better. Therefore, when used as a tool to
accelerate calculations of interparticle forces, the best tree
algorithm is the one which produces more accurate (or at least
`accurate enough' for a given problem) forces more quickly than any
other, so that more and larger simulations can be completed in a given
time. Unfortunately, this metric has not always been used in studies
of tree code quality, sometimes leading to conclusions that directly
contradict each other.

For example, a study by \citet{anderson99} compares the performance of
BH trees with $k-d$ trees using largely analytic arguments. He
concludes that a spatially balanced\footnote{In Anderson's
nomenclature, density balancing is defined by the criterion described
in the last section, that nodes at each level contain equal numbers of
particles. In contrast, nodes in a spatially balanced $k-d$ tree may
have unequal particle counts, and instead divide space equally at each
level, as is done in oct-trees.} $k-d$ tree will provide performance
superior to that of a BH tree. In contrast, \citet{waltz02} compare a
BH tree to a $k-d$ tree in direct code to code tests but make the
opposite conclusion, that BH trees ordinarily provide superior
performance to $k-d$ trees. 

We can begin to resolve the contradictions by examining the metrics
used to define better performance in each case. For example,
\citet{anderson99} uses the average length of an interaction list as a
proxy metric for the actual calculation rates, neglecting the cost of
tree traversals entirely. He finds that $k-d$ tree traversals result
in shorter lists, meaning that $k-d$ trees are preferable. On the
other hand, \citet{waltz02} use the average number of nodes touched
during a traversal as their primary metric. They find that several
times more nodes are opened in binary trees than in oct-trees, meaning
that oct-trees are preferable. In fact, in terms of the interaction
list lengths, the results of the two studies actually confirm each
other--\citet{waltz02} also find that lists some 20-30\% shorter
result from $k-d$ trees, but attach little weight to this fact in
their conclusions.

Interpreting these conclusions in any general case depends strongly on
how well the proxy metrics actually correlate with the total cost of a
calculation. In other words: are the metrics relevant only for a given
study, using a given code? Given the conclusions, we may reasonably
infer that \citet{waltz02} found that traversals were the dominant
calculation cost, while \citet{anderson99} found the opposite, which
lead directly to their opposing conclusions. However, neither the
\citet{anderson99} nor the \citet{waltz02} studies discuss the
relative costs of traversal and evaluation in any detail and further,
it is not clear that their results are generalizable. For example, few
codes in common use implement the comparatively costly `priority
queue' node opening strategy of \citet{salmon_warren94}, as
\citet{waltz02} do. Also, \citet{waltz02} make their comparisons to
density balanced $k-d$ trees, while \citet{anderson99} points out that
using this variant may lead both to poor error properties and to lower
calculation rates. 

A much earlier study by \citet{makino90b} appears to be more generally
applicable, and compares the performance of a BH tree to a Press tree.
His results are largely consistent with the later studies in that he
too finds many more nodes need to be opened for the case of the binary
tree, but that shorter interaction lists are required to provide the
same force accuracy. As a result, the per interaction cost is some
50\% larger for the binary tree than for the Press tree, but since
fewer interactions are required, he finds near parity in performance
at the same accuracy. He concludes that secondary factors, such as the
tree construction cost, become important in decisions to use one or
another. Here again, although he notes that Press trees are an order
of magnitude more costly to construct, they can be reused while an
oct-tree must be rebuild at each step, so that the cost is amortized.
Makino concludes that because the costs of the traversals and force
calculations using either method are comparable, and that lesser
factors such as the tree construction time are also comparable, the
algorithmic efficiency on different architectures may be important in
influencing the decision to use one or the other.

We have chosen to implement a nearest neighbor binary tree based on
the algorithm described in \citet{press86} and \citet{bbcp90} in VINE.
Since a consistent conclusion of the studies above is that traversal
costs are higher in such trees, we also implement the grouped tree
traversal strategy of \citet{barnes90}, which acts to reduce the
costs to more acceptable levels. At settings typical for production
use, we will find that those costs are indeed low, at some 10-20\% of
the total. A nearest neighbor tree also has the advantage that
particles tend to be grouped in the tree in the same manner that they
are grouped in space, rather than being grouped into artificially
bounded cubes in which particles may be well separated in space but
still found in the same cube, or quite close but found in different
cubes. As an example, consider a centrally condensed, spherical
particle distribution contained at the center of a `master' cube. At
the very first subdivision of that cube into sub-cubes, particles near
the center may be separated into different tree branches, while the
nearest neighbor tree instead groups them together.  It is therefore a
first step towards our goal of performing neighbor searches for
adjacent particles at similar times during the full calculation in
order to gain the advantages of cache reuse.

\subsection{Tree construction}\label{sec:tree-build}

Nearest neighbor tree construction depends upon the ability to
efficiently determine the nearest neighbors of all particles or nodes
for which no nearest neighbor has yet been found and to associate such
pairs into higher order nodes. Both because the exposition in
\citet{bbcp90} was quite brief and to describe our modifications to
it, we will describe the process we use in detail here. 

First, we note that in order to ensure modularity and maximize the
efficiency of accessing the tree data, we implement completely
independent data structures for the tree and for the driver of the
integration. The first action of the tree build routine is therefore
to receive a list of positions, masses and smoothing lengths (set to a
fixed value for $N$-body particles) from the driver and store them in
dedicated memory local to the tree module, after which the actual
construction begins. 

\citet{bbcp90} employ a temporary hash grid overlaid on the particle
positions to associate each particle with a specific grid zone. Then
for each zone, they define a linked list of particles occupying that
zone to determine suitably small lists of candidate particles to
examine for nearest neighbor status. In spirit, this method is the
same as the so called `friends of friends' method \citep{hockney81},
but employs a slightly more sophisticated method for defining the hash
grid.

We create the hash grid by first sorting the particles' positions in
each spatial dimension, $d$, and then dividing the sorted lists into
$(N_p/n_h)^{1/d}$ equal length sections. The average of the
coordinates of the two particles on either side an adjoining section
then defines the position of the boundary for each zone in each
direction. Thus, the grid is unequally spaced in the positions of each
grid boundary, but equally spaced in number of particles per grid
coordinate. The quantity $n_h$ defines the expected average number of
particles per zone. Typically we find that an average of $n_h\sim3-5$
is most computationally efficient, but the best hashing factor also
depends on the level of particle clustering found in a given
simulation.

After creating the grid, we assign a one dimensional coordinate key to
each particle, which defines its position in the grid as 
\begin{equation}\label{eq:tree-key}
K_p  = i + (j-1) n_x + (k-1) n_x n_y
\end{equation}
where $i, j$ and $k$ are the ordered triple defining a particle's
position, and $n_x$ and $n_y$ are the number of hash grid zones in the
$x$ and $y$ coordinate dimensions of the grid. When 2D simulations are
performed, $k=1$ so that the last term is always zero. With a key in
hand for each particle, a single loop through the particle list is
sufficient to define a set of linked lists containing the particles
resident in each hash grid zone. Lists for each zone are characterized
by a `head-of-list' node and a zero terminated list of `next'
pointers, defining the next node or particle in the list. Traversing
the list for a given grid zone is then equivalent to accessing a list
of candidate particles within a small, well defined region, as we
require. The appropriate list for any given particle is available
directly from examining its coordinate key, which identifies the
appropriate head of list node.

In order to determine nearest neighbor status from the lists of
candidates, we first determine distances to all candidates in a
particle's own zone and the distance to the nearest zone boundary. If
any boundary is closer than the nearest neighbor particle so far
discovered, then the search region is expanded in the direction of the
nearest zone boundary. The search continues until the nearest neighbor
is closer than any zone boundary. Once found, we store the nearest
neighbor's identity and continue until we have determined the nearest
neighbors of all unassociated particles and nodes. 

After all of the nearest neighbors have been found, we check each node
for {\it mutual} nearest neighbor status, defined by the condition
that a node's nearest neighbor has that node as its own nearest
neighbor. If the condition fails, the node remains on the unassociated
node list until the next round. If it succeeds, we create a new
unassociated parent node from the pair and remove them from the list
of unassociated nodes. The sibling pointer for each of its two
children are updated to point to each other, and a daughter pointer is
defined to point to one of the children, defining the `left' child.
Each pointer is an integer array whose value contains the array index
of the daughter or sibling node (daughter pointers for particles are
set to zero). We calculate the mass of the new node and its position
is defined as the center of mass of the two children.

The entire procedure is repeated with the remaining unassociated nodes
until only one (the `root' node) is left. When complete, the tree
consists of exactly $2N_p-1$ nodes, of which $N_p$ correspond to the
original particles (`atoms').

\subsection{Optimizations of the tree build}\label{sec:build-opts}

So far the tree construction algorithm is identical to that outlined
in \citet{bbcp90}. Here, we introduce several optimizations of the
original method, which greatly accelerate the construction over that
required in the original method. 

Using the recipe above, unassociated nodes are examined and nearest
neighbors found in an arbitrary order, depending on their location in
the list given to the build. Instead, we recognize that the candidate
nearest neighbor examinations will be most efficient if nodes that are
spatially adjacent are examined consecutively, so that their data are
already located in the cache. Therefore, rather than looping over the
list of nodes, we loop over blocked sub-regions of our hash grid,
examining all nodes in each region before going on to the next. This
ordering ensures a high probability that recently examined nodes will
still be found in the cache when they are re-examined as a prospective
nearest neighbor for a nearby node. In addition, newly created nodes
will be placed in nearby memory locations, increasing the benefits of
the correlations between memory and spatial locations as the build
proceeds to later iterations. 

At each iteration of the tree construction, a new grid is created,
requiring that the positions of the unassociated nodes be sorted in
each direction. Instead of performing sorts at each level, we have
found that inserting the new nodes into the original hash grid is a
vastly less expensive alternative because the position of the new node
is already known and its grid zone can be immediately determined. To
retain efficiency as the number of nodes per zone decreases, we revise
the hash grid and the associated particle keys by cyclically
decreasing the number of zones in each of the coordinate directions by
a factor of two.

After the first iteration, the linked lists of unassociated nodes for
each grid zone will contain both newly created nodes and nodes for
which no associate could be found in the previous iteration. In
addition to the head of list pointer for each zone, we also define a
head of list pointer for the old nodes on the list, so that we can
either traverse the full list of new and old nodes for the zone, or
only the new nodes. Distinguishing between old and newly created nodes
is useful because no `old' node will be closer to a given node than
the nearest neighbor that has already been determined for it. We can
therefore reduce succeeding searches for nearest neighbors to the list
of newly created nodes as long as the previously identified nearest
neighbor also remains unassociated. This optimization is especially
beneficial and when the particle distribution is very inhomogeneous,
because in such cases relatively fewer new nodes are created per level
and many redundant examinations can be avoided.

\subsection{Parallelizing the tree build}\label{sec:build-parallel}

Building the tree in parallel requires that the work on each level of
the construction be divided into three distinct steps, each of which
must be completed before the next can begin. First, a nearest neighbor
must be assigned for each unassociated node and, second, mutual
nearest neighbor status must be determined for pairs of unassociated
nodes and a new node created from each pair and, third, newly created
nodes must be placed in the hash grid. 

Distributing iterations of the loop over sub-regions among processors
is sufficient to parallelize the work required in both of the first
two steps. A synchronization point is required between the two steps
however to ensure that all nearest neighbors determinations have been
made before any mutual nearest neighbor determinations are made. The
distinction is necessary to eliminate a potential race condition that
would otherwise exist in the association. Without the synchronization,
two nodes residing in different sub-regions (and handled by different
processors) that are in fact mutual nearest neighbors could be passed
over because one of those nodes has simply not yet been assigned any
nearest neighbor at all.

As nodes are associated and created, they must be placed in a specific
memory location. An additional synchronization is required in the form
of an atomic update to a counter that defines the location of the last
node already stored in memory. Each processor increments the counter
only one time per level by an amount equal to the number of nodes
found in its part of the search. Contention for access to the counter
is therefore minimal.

The computational cost of setting up the initial hash grid is
dominated almost entirely by the cost of sorting the particle
positions in each coordinate direction. We have implemented the
parallel quick sort described by \citet{SH97}, in which particle
positions are divided into $N_{proc}$ segments, each of which is
sorted on one processor, then merged with a series of
$\log_4(N_{proc})$ merge cycles to obtain the final, globally sorted
list. Actual placement of particles in the hash grid has not been
parallelized because in the current version of the code, we found
performance loss rather than gain. Revising the hash grid is
parallelized by partitioning the zones along one coordinate dimension
among the processors, then re-associating nodes in deleted zones with
the newly revised grid. 

\subsection{Tree post processing: laying the groundwork for efficient
access}\label{sec:tree-post}

After the build is complete, we perform a number of post processing
operations on the resulting tree in order to prepare it for efficient
access and use. In considering what post processing will be useful, we
recall our requirement that tree traversals for spatially nearby
particles should occur consecutively. A list of nodes and particles
sorted according to the order they are encountered in the tree
traversal will satisfy this requirement but presumes a tree traversal
strategy, which we will discuss in detail in section
\ref{sec:treewalk} below. For current purposes, it is sufficient to
note that our traversal effectively converts the tree structure into
an ordered list of nodes and particles.

Minimally, this conversion requires that the daughter and sibling
pointers for each node be re-associated into linked lists, which then
define the order nodes are encountered in a traversal. In practice,
only the right daughter nodes require re-association if we generalize
the definition of `sibling' for right daughter nodes to point instead
to its parent's, grandparent's (or great-grandparent's etc) sibling.
For consistency, we define a fictitious sibling of the root node,
which also serves as the sibling of all nodes on the extreme right
branch of the tree.

We extend the conversion process further by actually reordering the
placement of the data in the computer's memory as well, to correspond
to the order that data are encountered in a traversal. Then the data
for particles or nodes that are nearby in physical space are stored in
locations that are nearby in the computer's memory and repeated
accesses become more efficient. Because the particle data are copied
into completely distinct memory locations for use in the tree, there
are actually two distinct sets of data that can be reordered
independently of each other. First, the data defining the node
information in the tree may be reordered, so that calculations
accessing the tree are accelerated. Second, the data defining the
particles themselves may be reordered, so that calculations that also
(or only) access the particle data may be accelerated.

Reordering the tree data requires that we perform a tree traversal
(section \ref{sec:traversal} below) that opens all nodes in succession
to establish what the optimal ordering will actually be, and that we
then relocate each node in turn to correspond to that ordering. During
the traversal, we create a list of nodes linking the consecutive
ordering of the tree traversal with the arbitrary ordering of the tree
as it was built. Each entry in the list then corresponds to a specific
tree node and is thereafter identified as such. Pointers to daughter
and sibling nodes are similarly identified and the associations of the
original ordering are updated to reflect the re-identification.

In practice, we find that both associating the original ordering to
the revised ordering is needed, and vice versa, so we create two
reordering lists corresponding to each case. We save these lists so
that when node data are revised (section \ref{sec:tree-update}), they
can be used to distribute the particle data into optimal ordering for
rapid traversals, at the cost of a single copy of the particle data
into the tree data structures. Apart from the copy-in process during
the tree revision (which must be performed whether or not the tree
data are reordered), no references to the auxiliary list are required
and there is no cost to the reordering. 

Reordering the particle data also requires one full traversal, in
order to obtain a list of all particles only, ordered by their
appearance in the tree. Once obtained, the reordering requires only a
scratch array into which to copy the particle data, then to restore it
in the new ordering using the list. Sections \ref{sec:perform-sph-reo}
and \ref{sec:perform-build} describe the benefits of reordering the
particle data. The frequency of reordering required to retain these
speedups will of course be very problem dependent, however we have
found for some typical problems that a frequency of once every few
tens to hundreds of time steps is sufficient, so the cost of this
reordering is insignificant.

Finally, we create an ordered list of `clump' nodes, which are defined
by the condition that each clump node contains $N_{\rm cl}$ or fewer
particles, but whose parent contains more than $N_{\rm cl}$ particles
or is spatially separated from its sibling by more than a critical
distance (specified at runtime). These clumps are used in the tree
traversals for the gravity and neighbor search calculations as
described in sections \ref{sec:grav-trav} and \ref{sec:find-nay}
below. Various traversals used by the code require that clumps be
detected during the traversal so that they may be set aside for
special handling. In order to allow detection both simply and
inexpensively, we further redefine the daughter and sibling pointers
to clump nodes to be negative valued integers. Their absolute values
are then the true array index of the clump node. Comparison of the
pointer to its absolute value then determines a node's status as a
clump or ordinary node.

Taken together, all of the post processing steps described in this
section require less than 5\% of the total time required for the tree
construction itself, and so are a small computational expense. We
shall show that even this small cost pays enormous dividends when the
tree is accessed, making the added coding complexity also worthwhile.

\subsection{Data contained in the tree and required by
calculations}\label{sec:tree-data}

After construction, a number of other data remain to be calculated for
each tree node, including multipole moments and convergence radii
needed to determine their acceptability for use in various
calculations. We use the composition formulae originally defined in
\citet{bbcp90} to determine the mass, position and quadrupole moments
of each node, starting from the particle data only. For a node, $n$,
with daughters ${\rm d1}$ and ${\rm d2}$, these formulae are
respectively:
\begin{equation}\label{eq:treenode-mass}
M_{\rm n} = M_{\rm d1} + M_{\rm d2},
\end{equation}
\begin{equation}\label{eq:treenode-pos}
\mbox{\boldmath$ X_{\rm n} \,$} = 
                {{M_{\rm d1} \mbox{\boldmath$ X_{\rm d1} \,$} +
                  M_{\rm d2} \mbox{\boldmath$ X_{\rm d2} \,$}}\over
                 {M_{\rm d1} + M_{\rm d2}}}
\end{equation}
and
\begin{eqnarray}\label{eq:treenode-quad}
\mbox{\boldmath$ Q_{\rm n} \,$} & =  & \mbox{\boldmath$ Q_{\rm d1} \,$} +
                                 \mbox{\boldmath$ Q_{\rm d2} \,$} +
              {{M_{\rm d1}M_{\rm d2}}\over{M_{\rm d1}+M_{\rm d2}}}\times
                               \nonumber  \\
                          &    &
                 (\mbox{\boldmath$X_{\rm d1} - X_{\rm d2} \,$})\otimes
                 (\mbox{\boldmath$X_{\rm d1} - X_{\rm d2} \,$}).
\end{eqnarray}
In addition, we require both the actual size of each node and its
multipole convergence radius. The size of a node is conservatively
specified by the condition 
\begin{eqnarray}\label{eq:treenode-size}
h_{\rm n} & = & \max( {{M_{\rm d1}}\over{M_{\rm n}}}
                 | \mbox{\boldmath$ X_{\rm d2} \,$} -
                   \mbox{\boldmath$ X_{\rm d1} \,$} | + h_{\rm d2},
                   \nonumber \\
          &   &
                  {{M_{\rm d2}}\over{M_{\rm n}}}
                 | \mbox{\boldmath$ X_{\rm d2} \,$} -
                   \mbox{\boldmath$ X_{\rm d1} \,$} | + h_{\rm d1}  ).
\end{eqnarray}
The size of individual particles is defined to be either their
smoothing length or the gravitational softening parameter $\epsilon$,
in the case of $N$-body particles. Except for individual particles,
the node sizes computed using equation \ref{eq:treenode-size} is
always larger than strictly necessary, but we have found that the
computational cost of this conservative definition is not particularly
large. An important exception is that we have found it advantageous to
specify the node size exactly for clumps and all of their descendant
nodes for use in the SPH neighbor searches, because many fewer nodes
must be opened and examined. Therefore, for this subset of nodes, we
recalculate the size of node, $n$, as the maximum distance of any
particle, $i$, contained in the node, from its center of mass: 
\begin{equation}\label{eq:treenode-size-exact}
h_{\rm n} = \max_i ( 
                 | \mbox{\boldmath$ X_{\rm n} \,$} -
                   \mbox{\boldmath$ X_i       \,$}  | + h_i   ).
\end{equation}

As described in \vineI \ (section 4.2), VINE implements three runtime
selectable options, referred to as `Multipole Acceptance Criteria' or
MACs, for determining the acceptability of a given node for use in the
gravitational force calculation on a particle. We refer to them as the
`geometric', the `SW absolute error' and the `Gadget' MACs,
respectively. Each is based on a different implementation of the
convergence radius of a multipole expansion of the force from that
node. One datum for each node is required to implement each of the
three criteria and, depending on which criterion was chosen by the
user at runtime, VINE selects which of the three data to calculate and
store. In each case, only the portion of the criterion that remains
invariant for all node examinations is stored rather than the full MAC
definition, in order to minimize computational cost during tree
traversals. A user defined accuracy parameter, $\theta$, is also
required to complete the specification of the criterion, but takes on
different interpretations for each of the different MACs, as defined
below.

When the geometric MAC is selected, VINE stores the quantity
\begin{equation}\label{eq:geometric-mac}
R_{crit} = {{h_j}\over{\theta}}
\end{equation}
for each node $j$, where $\theta$ is interpreted as a user chosen,
dimensionless value between zero and one, parameterizing the minimum
acceptable distance at which a node may be used in the gravity
calculation. When the absolute error criterion of
\citet{salmon_warren94} (the 'SW' MAC) is selected, VINE stores the
quantity:
\begin{equation}\label{eq:SW-mac}
R_{crit} = \sqrt{ {{h_j^2}\over{4}} + 
              \sqrt{ {{3{\rm Tr}\, {\mbox{\boldmath$ Q_j$}}}\over{\theta}}}}
\end{equation}
where in this case $\theta$ is a value defining the maximum absolute
error in the acceleration that a single node may contribute to the sum
and ${\mbox{\boldmath$ Q_j$}}$ is the quadrupole moment tensor for
node $j$. Finally, when the Gadget MAC of \citet{springel2001} is
selected, VINE stores the quantity:
\begin{equation}\label{eq:gadget-mac}
\Phi_{crit} = {{M_j h_j^4}\over{\theta }},
\end{equation}
where the gravitational constant is set to $G=1$ and we have replaced
the variable $R_{crit}$ with $\Phi_{crit}$ in this definition to make
clear the fact that the saved portion of the criterion does not have
units of length. For the Gadget MAC, $\theta$ is interpreted as the
maximum magnitude of the relative error in the force allowable to any
single acceptable node.

To these, we add an additional criterion that we will find useful,
describing the condition that two nodes are in physical contact, or in
other words, are `neighbors': 
\begin{equation}\label{eq:neigh-mac}
r_{ij}^2 < (h_i + h_j)^2,
\end{equation}
where the nodes' physical extents are given by $h_i$ and $h_j$,
through either equation \ref{eq:treenode-size} or
\ref{eq:treenode-size-exact}. Thus, whether a node actually represents
only one or very many particles, the condition for tree nodes to be
neighbors is equivalent to the SPH condition that defines neighbor
status for individual particles (see \vineI, \ section 3.2).

\subsection{Updating vs. rebuilding the
tree}\label{sec:tree-update}

Even though the tree is a nearest neighbor tree, none of the
calculations using it require that the nearest neighbor property be
satisfied. Therefore we may reuse it in future time steps as long as
it remains efficient to do so, revising only the quantities defined in
section \ref{sec:tree-data} and amortizing the cost of its
construction over many time steps. Section \ref{sec:freq-rebuild}
illustrates the conditions that affect the frequency of rebuilds,
relative to revisions.

As for the build, when data in the tree are updated, particle data are
received from the calling routine and stored (in the sorted tree
ordering) in dedicated, locally defined arrays. Node data are created
from the node composition formulae above, with quadrupole moments for
particles set to zero. Given these formulae, a necessary condition for
creating or updating information for a given node is that all its
descendant nodes have been already been updated, and leads naturally
from the bottom to top (leaf nodes to root node) update suggested by
\citet{hern90}. We have not found this method to be efficient however,
either for updates on a single processor or in parallel, because the
work per level is small and the data are widely scattered in memory.

Instead we take notice of the fact that the node reordering done to
improve performance of the tree traversals can also be used for the
tree updates, when used in reverse. In other words, while a tree
traversal that starts from the root and opens every node in turn is
guaranteed to examine every parent node before its children, a tree
traversal that starts from the tree's termination node and proceeds
towards the root is guaranteed to examine every child node before its
parent. Structured in this way, a tree update requires a single, long
loop over all nodes. Updates for spatially nearby nodes are performed
naturally at the same time, allowing substantial speedup due to
improved cache reuse. Updates in parallel are done by splitting the
tree into approximately equal sized branches, to be handled separately
by different processors. A small number of isolated nodes (typically
$\sim10$) near the root of the tree which cannot easily be associated
with any sub branch, are updated in a second, post processing step.

\section{Traversing the Tree}\label{sec:treewalk}

The ultimate goal of any simulation is to model a physical system, so
all calculations not leading directly to that end are in some sense
wasted. In this context, and even though use of a tree will be
enormously beneficial compared to its alternatives, both the processes
of building a tree and extracting data from it are wasted: neither is
directly related to evolving any particles forward in time or to
calculating derivatives for any quantity or particle. Therefore we are
well advised to reduce, as far as possible, the time spent in such
activities.

Given a tree structure, what is the most efficient way to determine
which nodes in the tree are useful as is and which must be refined
further? A variety of tree traversal strategies have been discussed in
the literature
\citep{barnes90,makino90a,hern90,bbcp90,warren_salmon95,dubinski96}
which are efficient for use on various kinds of hardware, from vector
based machines to single CPU microprocessor based machines, to shared
or distributed memory parallel architectures. VINE employs elements
from several of these methods to obtain various information from the
tree. In this section, we first examine a flexible prototype traversal
and then describe adaptations of it, used to obtain a variety of
information from the tree required for the gravity calculation and the
neighbor determination for SPH.

Throughout this discussion, we will refer to a `sink' particle or node
to be a point for which data are required from the tree or interaction
is to be calculated. A `source' particle or node is an entry in the
tree that is tested for or used to determine a contribution to that
interaction. In principle, the tree could be traversed and
interactions calculated for any point in space, whether or not that
point is actually a node in the tree, however we have implemented only
traversals for nodes and particles that are tree members. 

\subsection{A prototype tree traversal}\label{sec:traversal}

Traversing a tree requires first, one or more tests that determine the
acceptability of a node, second, a prescription for determining which
node will be the next to be examined and, finally, a termination
condition for the traversal. Without reference to gathering any
particular information from the tree, the results of a node
examination fall into three categories. A node may either pass or fail
its acceptability criterion, or we may defer examination until later
based on some other, special property of the node. VINE tests a tree
node first for its special properties, then for its acceptability,
using the stackless `follow the left wall' rule discussed in
\citet{makino90a}, known more generally as a `depth first' tree
traversal. Because it effectively turns the tree traversal which opens
all nodes into a space filling curve, in practical effect, this
traversal amounts to a variant of the Peano-Hilbert or Morton ordering
employed by grid based tree codes
\citep{springel2001,warren_salmon95}, applicable instead to a nearest
neighbor tree. 

\begin{figure*}
\epsscale{1.0}
\plotone{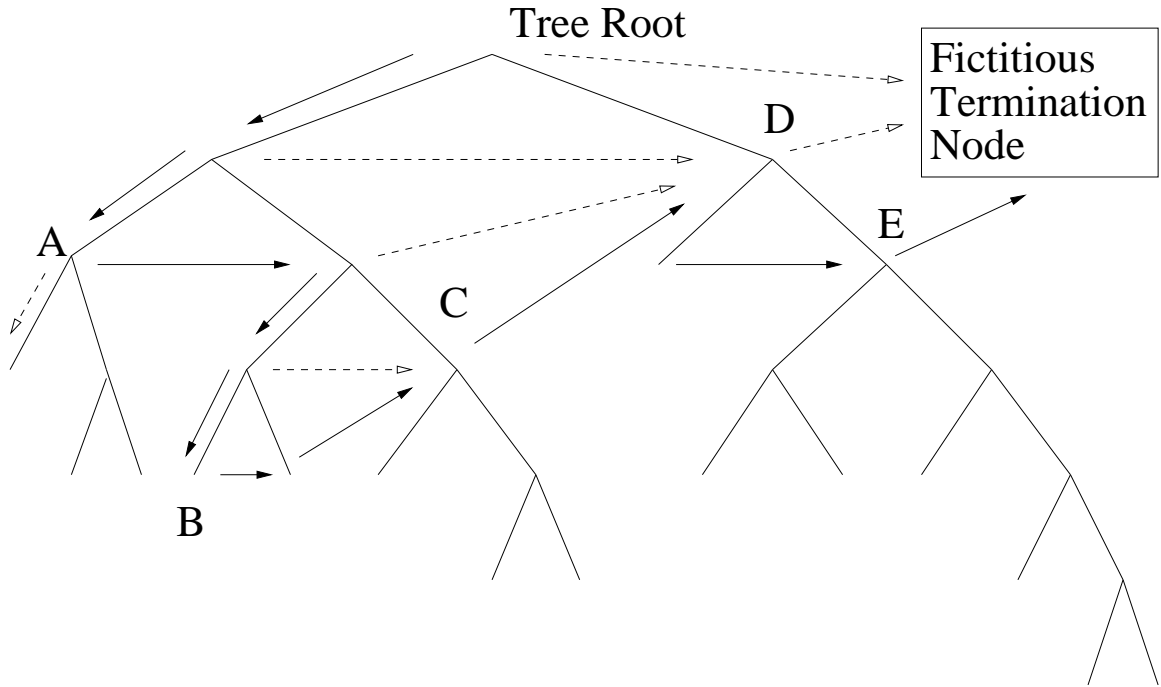}
\caption{\label{fig:tree-trav}
Graphic description of the follow the left wall tree traversal used in
VINE. The direction taken after each node examination is shown with a
solid arrow, while the discarded direction is shown with a dashed
arrow.}
\end{figure*}

In a follow the left wall traversal the tree nodes are converted into
an ordered list, for which the path of a traversal that opens every
node in turn will always take the leftmost branch of the tree that has
not already been examined. Accepting a node and advancing to its
sibling is therefore equivalent to placing a reference to the node on
an interaction list, then dropping some number ($>1$) of entries down
the list, or exactly one if the node is an atom. The implementation in
VINE defines two linked lists which contain respectively, the sibling
of the node and its left daughter. We generalize the definition of
`sibling' for right daughters to include the parent's or grandparent's
(or great-grandparent's etc) sibling, and we use a fictitious node as
the root node's sibling. By extension, the fictitious root sibling is
therefore also the sibling of all nodes on the extreme right branch of
the tree. Fortran pseudo-code illustrating a traversal using each of
the three alternatives is shown here: 

\begin{verbatim} 
do while( not finished )

   if( defer? )then

      YES: store node on special treatment 
      list and advance to sibling

   elseif( accept? )then

      YES: store node on interaction list 
      and advance to sibling

   else

      NO: advance to left child 

   endif

enddo
\end{verbatim}

Figure \ref{fig:tree-trav} shows an example traversal through the tree
structure under the assumption that the `special' characteristic of a
node is that it is a particle and the termination condition that the
next node to be tested is the fictitious node, so that we traverse the
entire tree. For the purposes of this illustration, we do not need to
specify the acceptance criterion. Although any property of a node may
be used to determine its special character, we choose the fact that it
is a particle for simplicity of illustration here. In practical use,
this property could be used to differentiate between particles which
require treatment as neighbors, and those which do not. A decision
about this treatment might be deferred to permit either the traversal
code or the neighbor determination to be optimized (or both) by the
compiler.

Starting from the root node, we descend one level to its left daughter
node. The node does not need to be deferred (i.e. it is not a particle)
and we will assume that it is also not acceptable, so control descends
again to the left daughter node, labeled `A'. Here we will assume that
node A is acceptable, and we place it on an interaction list. Rather
than descending further (the dashed arrow), control passes instead to
its sibling, and upon failing the acceptability criterion descends two
levels until we reach node `B', which is a particle and therefore a
decision regarding its interaction can be deferred. It is placed on the
deferred interaction list, as is its sibling, and control passes
onwards to node C, which we will assume is acceptable and place on the
normal interaction list. Note here that control would have passed to
the same node, labeled D, had any of the previous examinations of C's
parent nodes produced an acceptable result, bypassing all examinations
of their daughter nodes. Continuing on, we examine D and upon finding
it unacceptable, pass control to its daughter and eventually to node
E, which we place on the interaction list. Since the next node
examination is the fictitious termination sibling, the traversal is
complete.

In principle, one or more tests may be required in any particular
traversal to determine the node's special properties, its
acceptability or the `not finished' condition. The flexibility to
choose different criteria but keep the same basic traversal strategy
allows us to tailor tree traversals individually for different
requirements. Two examples of such flexibility are important to point
out for VINE users. 

First, by changing the starting and termination nodes, we may design a
search to traverse all or any part of the tree. For example, in the
prototype above, we traversed the entire tree starting from the root
using the condition that the last node to be examined is a fictitious
termination node. If we had desired instead to search only some
portion of the tree, say all of the nodes and atoms beneath node C,
then we might choose arrival at node D as our termination node and
start the tree traversal at node C. These partial traversals will be
important for the close portion of the traversal for gravitational
forces and for neighbor searches in SPH calculations. Second, by
changing the special property, we can detect atoms very inexpensively
(i.e. as nodes with no children), or we can detect `clumps' as defined
by the condition that the (integer valued) pointer to the next node
and its absolute value are unequal. Later traversals of the deferred
nodes may use the same or entirely different criteria from the full
traversal used to obtain them originally.

\subsection{Finding acceptable nodes and atoms for the gravity
calculation}\label{sec:grav-trav}

In order to determine lists of acceptable nodes and atoms for gravity
calculations, we must adapt the prototype tree traversal above to the
problem at hand and specify acceptance criteria for a node to be used.
For the calculation to proceed quickly, we would like to perform as
few traversals as possible and accept the minimum number of nodes
necessary to produce accurate forces.

We address the first constraint by combining our tree traversal with a
variant of the procedure of \citet{barnes90}, who showed that it was
efficient to perform a single tree traversal and obtain a single list
of acceptable nodes and atoms for groups of nearby particles at the
same time, rather than for individual particles separately. Following
this observation, we group nearby particles into `clumps', as defined
in section \ref{sec:tree-post}, and divide each tree traversal into
two distinct steps corresponding, respectively, to interactions with
distant particles or to close particles.  

In the first step, VINE employs a single tree traversal designed to
obtain a list of nodes that are certainly acceptable for all particles
in the sink clump using the user specified MAC, and a list of neighbor
clumps, defined as clumps which pass the criterion in equation
\ref{eq:neigh-mac}. In the second, VINE performs individual traversals
for each particle in the sink clump, over the list of neighbor clumps.
Identical acceptability criteria are used in both steps, but the
deferral and termination criteria are specific to each. We find that
this division of work allows the total number of traversals to be
substantially reduced, so that the fraction of time spent in
traversals is only $\sim10-20$\% of the total required for the entire
gravity calculation. 

The far traversal uses each clump in the tree in turn as the sink for
the gravitational calculation. It extends from the root to the
termination node and uses the deferral criterion that any clump nodes
found during the traversal are set aside for later processing during
the second step. The node acceptability criterion results in a list of
all non-clump nodes encountered that are acceptable for the gravity
calculation for all particles in the sink clump. It is impossible for
a list of atoms to be generated during this traversal because every
particle is a member of a clump and so is set aside, so no other
criterion is required to handle them.

The close traversal step divides the sink clump into individual sink
particles in turn, and traversals over every neighbor clump complete
the interaction lists for each particle. The traversals proceed from
each neighbor clump and terminate on its sibling, with the special
property set to determine whether the node under examination is
actually a particle. They use the same acceptability criterion as the
far traversal, with an individual particle now replacing a clump as
the sink. Acceptable particles and nodes are placed on a list of atoms
or nodes as appropriate. 

\subsection{Finding neighbor particles for SPH}\label{sec:find-nay}

As for the gravity calculation, VINE's search for the neighbor
particles required for SPH calculations uses a two step far/close
traversal strategy. Here again, dividing the traversal into near and
far components permits a substantial reduction in the total work
required to collect neighbors for the SPH particles. In the far
traversal, we obtain a list of all clumps which are neighbors of our
sink clump, and in which all possible neighbor particles will be
found. Since this traversal is done for an entire clump of particles
at a time, its costs per particle are small because they are amortized
over the whole clump. Close traversals require searches over only the
comparatively small set of neighbor clumps to obtain lists of actual
neighbors, for each active particle.

For both near and far traversals, we desire an outcome exactly
opposite to that in the gravity calculation. Instead of storing
distant nodes on a list to be processed later, we discard them,
keeping only nodes which are close. Because of this difference, we
have found that, in the case of far traversals, searches based on the
template in section \ref{sec:traversal} are actually less efficient
than the level by level (`breadth first') strategy of \citet{hern90},
mainly because the latter permits VINE to retain information from one
traversal to the next. We have therefore implemented it as follows. 

For a given sink clump, we first create a list of all its ancestor
nodes at every level in the tree. Then, for the ancestor on a given
level, we test the acceptability of a list of prospective nodes also
on that level (i.e. that the node and the sink clump's ancestor are
neighbors according to equation \ref{eq:neigh-mac}. If any node is too
distant, we discard it, otherwise we store its children on a stack, or
defer further examination if the node is a clump. Once all nodes are
tested on a level, we descend to the next, where the process repeats
until no nodes remain to be examined. We thus define a progressively
more restricted list of acceptable nodes for the ancestors on each
level in which all neighbors are to be found. Upon completion, we test
all clumps for neighbor status, returning a list to be used in the
close traversal step. This strategy permits later far traversals to
begin at some lower tree level, defined by an ancestor node shared by
the current sink clump and the previously completed sink clump
traversal, for which a more restricted set of prospective neighbor
nodes has already been determined.

Given a list of neighbor clumps, we proceed to the close traversal
phase. For each active particle in the sink clump, each neighbor clump
is examined for neighbor particles, using a variant of our prototype.
The opening criterion is defined again by status as a neighbor
according to equation \ref{eq:neigh-mac}. The deferral criterion by
the condition that the node is a particle of the same species as the
sink particle, i.e. both source and sink particle must be either both
SPH or both $N$-body particles. When the traversals are complete, we
examine all deferred particles for neighbor status at the same time,
thereby permitting code optimizations to be made which improve
performance.

VINE recalculates neighbor lists each time they are needed even
though, in principle, such redundant neighbor recalculation is
expensive and therefore to be avoided. On-the-fly recalculation does
however provide a substantial benefit to the memory footprint used by
the code, because we do not require a large amount of additional
memory be allocated to storing lists of every neighbor for every
particle. Further, we estimate that 30-40\% of each neighbor
determination is actually spent calculating quantities required for
the SPH calculations anyway. To make most efficient use of these
calculations, the neighbor search therefore returns not only the
neighbor identities but also the additional information used in the
SPH calculation. Specifically VINE returns the mass, squared mutual
smoothing length and distance, as well as the components of the
distance and the identity of each neighbor particle are stored and
returned to the driver SPH routine for use in further calculations. 

\subsection{Parallelizing the traversals and their associated
calculations}\label{sec:parallel}

Both tree traversals and the later calculations using data derived
from them are independent of all similar traversals, and produce
results that are stored unique locations as well. Moreover, tree
traversals are `read only' in the sense that no data contained in the
tree are altered. The most natural parallelization strategy will
therefore be to define a loop over all clumps in which one traversal
and all associated calculations are performed in each iteration. Then
the work can be parallelized by distributing different loop iterations
to different processors.

We have parallelized both the gravity and SPH density and force
calculations in this manner. Each iteration of the loop first calls a
routine responsible for the far traversal, then for the close
traversal and finally for the actual summation of terms. Lists of
acceptable nodes and atoms are stored in data structures private to
each processor. Data describing the neighbors for each SPH particle
are similarly defined privately for each processor.

In general, the time required to complete an iteration of the loop
will neither be fixed nor can it be determined easily from a count of
the number of particles in each clump. This is important because load
imbalance between different processors will develop if clumps handled
by one processor have require a systematically larger amount of work
than those handled by another. We might expect that such effects would
tend to average out in large simulations and, indeed, to some extent
they do if a large enough fraction of particles are active. We use the
OpenMP `dynamic' scheduling option, in which loop iterations are
parceled out among processors as they become idle, to improve upon
this average and retain load balance even when relatively few
particles are active.


\section{Gravity}\label{sec:gravity}

In this section, we describe the options available in VINE to
calculate the gravitational forces, using either information obtained
from the tree traversals described in sections \ref{sec:grav-trav}, a
direct summation by either the processor or based on special purpose
`GRAPE' hardware, or a combination of both.

\subsection{Forces obtained from the tree based
calculation}\label{sec:treegrav-opts}

VINE computes accelerations for all active particles in a clump at one
time, using identical interaction lists for the far component of the
traversal but distinct close interaction lists. For sufficiently large
simulations ($\gtrsim$ a few $\times 10^5$ particles), the interaction
lists will typically contain a few hundred to as many as several
thousand entries. Although the node data corresponding to these
entries may total only a few tens or hundreds of kilobytes (quite
small given the memory sizes of today's computers), it will certainly
be much larger than can be accommodated in the fastest level of the
computer's cache hierarchy. More importantly, it will sparsely sample
a set of memory locations spread out over many gigabytes of system
memory, with only one or at most a few entries resident on any single
page. As the summation for each particle proceeds, data for each node
must be retrieved either from a page of main memory or from
secondary/tertiary caches, stored in primary cache and operated upon
by the processor, only to be evicted by later data in the interaction
list and retrieved again for the next particle, an effect known to
computer scientists as cache or TLB `thrashing'. We present in this
section two optimizations that substantially mitigate both sorts of
thrashing.

First, we note that our tree traversal is split into two steps
corresponding to far and near interactions and that it produces
identical node lists for the far step. This means that an additional
optimization, referred to by computer scientists as cache blocking,
can be made that largely eliminates both types of thrashing.
Specifically, we allocate a small array whose dimensions have been
chosen to be exactly the size of the L1 cache (or more precisely, to
the nearest integer multiple of the required single node data volume
which is smaller than the L1 cache size) and load a subset of the node
data from our list into this array. We then cycle through these data
for all of the particles in the clumps that require accelerations.
When calculations for all particles have been completed, we discard
the subset and reload the array with another subset of the nodes,
repeating until the list has been fully exhausted. This strategy
avoids TLB and cache thrashing effects during the computations,
because the required data will almost always be accessed from an array
resident in the fastest level cache available. A small number of cache
misses may still occur, because in addition to these node data, the
components of position, and the so far accumulated components of
acceleration and gravitational potential of the particles for which
calculations are being performed will also be accessed frequently in
the calculation loop. Typically, however, these data will be accessed
so often as to be permanently resident in the registers of the
processor, thereby limiting thrashing from this source.

Although we will find that cache blocking is highly effective in the
force calculation itself, it can do nothing for the tree traversals,
from which the node data are obtained because, by their nature, they
sparsely sample very large volumes of memory only once per traversal.
Nearly every node examination will then require a new TLB entry to be
calculated and one or more new cache lines to be loaded. While few
remedies may improve the performance of the cache behavior, the effect
of TLB thrashing can be mitigated substantially on hardware
architectures for which large pages are available and can be accessed
easily by user programs. For a traversal of a tree of a given size,
the probability that two or more nodes examined during the traversal
will be found on a single page will be higher if the page size itself
is larger, so that fewer are required to span the entire data set
defining the tree, and fewer TLB recalculations must be made. The
magnitude of the mitigation will be dependent on the cost for a TLB
recalculation, the simulation size and on the accuracy required for
the forces, since higher accuracy will translate directly into more
node examinations and longer interaction lists. 

\subsection{Forces obtained via direct summation}\label{sec:dirsum}

VINE includes a run-time switch to calculate gravitational forces via
direct summation, using either the general purpose processor on which
all other calculations are made, or GRAPE hardware if it exists.

\subsubsection{Using the host processor}\label{sec:dirsum-host}

VINE includes a run time option to calculate gravitational forces via
direct summation, using the general purpose processor on which all
other calculations are made. Although it is indeed simple in practice
to compute gravitational forces by direct summation, it is much more
difficult to make that computation fast. At its simplest, the
summation is characterized by repeatedly cycling through the entire
list of particles, incrementing a partial sum of the acceleration on
one particle at a time until the list is exhausted, then repeating the
process until accelerations on all particles have been calculated. On
modern microprocessors, this method suffers from the fact that it
makes absolutely no use of the available cache memory to store data
that are used more than once.

When accelerations on many particles are required, it is possible to
speed up the calculations substantially by using the same cache
blocking techniques that were employed to speed up the approximate
calculation as described in section \ref{sec:treegrav-opts} above.
Instead of cycling through the entire list of particles one after the
other, we load a small number, $n_p$, of positions and masses into the
cache then calculate partial sums of the accelerations on these
particles due to the full list of all $N_p$ other particles in the
simulation, one after the other. Loads of data for a new particle on
the long list occur rarely compared to loads the cached particles, and
so are comparatively cheap compared to the number of computations
performed using that data. When the list of $N_p$ particles is
exhausted, the accelerations for the first $n_p$ particles are
complete. We then load a new set of $n_p$ particle data into the cache
and repeat the process until the list of all particles is complete. 

\subsubsection{Using GRAPE hardware}\label{sec:dirsum-grape}

VINE includes compile time options to calculate gravitational forces
using versions 3, 5 and 6 of the GRAPE hardware, through calls to a
library of communication and calculation routines distributed with the
hardware itself. The simplest approach in using GRAPE boards is to sum
the contributions of all particles on each other directly. While it
does not change the overall $\mathcal{O}(N^2)$ scaling of the
algorithm, it provides a much faster calculation due to the much lower
coefficient in front of the $\mathcal{O}(N^2)$ term. 

The mechanics of the method used to perform the calculations is quite
similar to that used for direct summation on the host. Data for fixed
number of particles, in this case totaling a few megabytes in size,
are transferred to the GRAPE and used as source particles for the
calculation. Small groups of sink particles are then transferred and
partial sums of the accelerations are calculated on them using the
sink nodes in the GRAPE's memory, then returned to the host and added
to the particle's accelerations there. When all sink particles have
been processed for all source particles, the calculation is complete.

\subsection{Forces from the combined tree based and GRAPE based
approaches}\label{sec:grav-grapetree}

VINE includes a run time selectable option to use GRAPE hardware to
calculate gravitational forces from lists of nodes and atoms obtained
from a tree traversal, rather than from a direct summation. This
combination was first discussed by \citet{makino_grapetree91} and
later \citet{athanassoula98} and \citet{kawai2000} reported its 
performance on GRAPE-3 and GRAPE-5, respectively. Efficient use of
the GRAPE/tree option in VINE requires that the code and its run time
parameters be tuned to somewhat different settings than with the
tree/host combination because, with the standard settings, the costs
of communication with the GRAPE far outweigh any speedups due to the
optimized hardware. 

In order to reduce the extreme costs of communication between the
GRAPE and host, calculations must be performed for a much larger group
of particles than will be optimal in the standard form, so that an
interaction list of a given size can be reused many more times before
being discarded. Communication costs are high because the number of
particles on which forces can be calculated with the same interaction
list is limited to the population of a clump, which will be $\sim50$
particles (section \ref{sec:best-clumpsize}). Costs would be reduced
if the maximum clump population were set to very large values, but at
the expense of substantially increasing the time required for SPH
calculations. To retain both fast gravity and SPH calculation, we
define a set of nodes, which we refer to as `bunches', that play the
same role as clumps do in the tree based calculation but contain many
more particles.

There are two competing effects that change the effective calculation
rate. Reducing the bunch population reduces the length of the
interaction list used to calculate forces on its particles. Reducing
the size of this list is important because it must be sent to the
GRAPE, and shorter lists minimize the per-transfer communication cost.
At the same time, smaller bunches mean more transfers, because each
tree traversal produces a different interaction list so that total
data volume sent actually increases. On the other hand, increasing the
bunch population reduces the number of transfers, but increases their
size because the interaction lists are longer, again increasing the
total data volume that must be sent. Minimizing neither the number of
transfers nor their size, independent of the other, will produce
optimal performance. The total time for one force computation on all
particles is instead a complicated function of the size and number of
communications with the GRAPE combined with the number of floating
point operations needed and the desired force accuracy. The
calculation rate also depends on latency and bandwidth of the
connection to GRAPE, but because these are mostly determined by
hardware (there is a slight dependence on the chip set on the
motherboard of the host computer), they are usually beyond the user's
control.

When GRAPE is used in VINE, the far traversals in the gravity
calculation proceed as in section \ref{sec:grav-trav} with the list of
bunches used in place of the list of clumps, creating a single
interaction list for the entire bunch. Additionally, rather than
proceeding to a set of close traversals over neighboring bunches, the
far traversal continues until all nodes and atoms outside the sink
bunch are determined. Finally, all particles in the bunch are
automatically added to the interaction list. We discuss the optimal
choice for the population of bunches in section
\ref{sec:best-clumpsize}. 

Due to the differently tuned traversal, and because GRAPE processors
can compute force interactions to monopole order only, the accuracy of
the forces that result from the calculations will be different from
those obtained from a tree based solution alone, given the same
setting of the MAC. To obtain forces of comparable accuracy, the MAC
setting must also be modified, to a more restrictive value. In section
\ref{sec:grapetree-acc} we define the MAC settings that produce
comparable accuracy for the tree/host and tree/GRAPE modes. 


\section{Overview of the code}\label{sec:overview}

VINE is written in the Fortran~95 language, but in a largely
Fortran~77 style which will be familiar to and comfortable for most
astrophysicists. For this reason, relatively little use has been made
of the newer language revision's features, such as derived types
(somewhat similar to C `structures') or pointers. On the other hand,
extensive use has been made of the `module' format of grouping
procedures together both to improve modularity of the code and in
order to enable the improved debugging features such as argument
checking available when they are used. The code includes a number of
pre-built sub-makefiles for common commercial and open source
compilers, which can be selected by the user and each of which contain
compiler options grouped by the designations `FAST' and `DEBUG'.
Sub-makefiles for other compilers we have not included may easily be
created by users, using one of the existing files as a template.
Although we have made no specific tests to determine an ideal set of
options for VINE for any single compiler (other than to note a factor
of several difference in speed between the two categories just
mentioned), we believe the options selected in each sub-makefile will
provide near optimal performance in terms of either VINE's overall
performance or the ease with which it may be debugged or otherwise
analyzed for coding errors. 

Although some care has been taken to limit the memory footprint of the
code, we have not made it a primary focus in our development. Instead,
we have concentrated on performance improvements and on reducing the
vulnerability of the calculations to errors associated with loss of
precision when, e.g., when two quite similar values are differenced,
or two quite different values are summed. Errors of this sort will
inevitably develop in large simulations because particles are closely
spaced and the precision of numerical representation of their
positions is finite. In this context, and with an expectation of
simulations of ever increasing size in the future, we have made the
default size for all real variables in VINE `double precision' (8
bytes on most systems), for which about 15 digits of precision are
retained at the cost of doubling the required memory relative to
single precision. While even this level of precision does not
eliminate finite precision errors, they will at least be of much lower
magnitude.  In keeping with our philosophy of keeping VINE as modular
and adaptable as possible however, VINE's build system recognizes an
option to change the default to single precision values, to permit
VINE to be used in constrained memory environments or for extremely
large simulations, at or near the capacity of the machine available to
its users.

Table \ref{tab:arrays} contains a listing of all real and integer
arrays, normalized to the total number of particles of all types,
$N_p$, and the number of SPH particles, $N_{\rm SPH}$, and broken down
for each component of the code. The number of arrays allocated to the
tree actually defines an equivalent number of $N_p$ length arrays
because the arrays in the tree hold $2N_p-1$ elements corresponding to
both particles and nodes. The number of arrays is also dependent on
the number of dimensions, $d$, defined at compile time to be either
two or three. In typical operation, we expect that the code 
will include one integrator, use individual time steps, and
include gravity using the multipole summation method and the tree. In
this case, the total amount of memory required for a simulation
containing $N_p$ particles, of which $N_{\rm SPH}$ are also SPH
particles will be
\begin{eqnarray}\label{eq:memory-lf}
M & = & \left[\left(d + 20 + 5N_{proc}\right)I + (13d+6)D\right]N_p \nonumber \\
      & &      + \left[ 2I + 17D\right]N_{\rm SPH}
\end{eqnarray}
when the leapfrog integrator is used or 
\begin{eqnarray}\label{eq:memory-rk}
M & = & \left[\left(d + 20 + 5N_{proc}\right)I + (15d+6)D\right]N_p \nonumber \\
      & &      + \left[ 2I + 22D\right]N_{\rm SPH}
\end{eqnarray}
when the Runge-Kutta integrator is used. If double precision values
require eight bytes and integers require four, then a 3 dimensional
simulation on one processor, with either entirely $N$-body or entirely
SPH particles will require 474 or 584 bytes of memory per particle
when the leapfrog integrator is used (292 or 368 bytes in single
precision), and 522 or 706 bytes per particle when the Runge-Kutta
integrator is used (318 or 410 bytes in single precision). In
comparison, VINE requires 30-50\% more total memory per particle than
the Gadget and GOTPM codes \citep{springel_g2,dubinski03}, when they
are used in double precision mode. Approximately half of the extra
cost is due to fact that VINE includes the quadrupole in the multipole
summation of gravity, which requires 96 bytes/particle to store the
quadrupole moments for all nodes. Other important components of memory
usage are accounted for by additional arrays used to implement
optimizations, allow for modularity or to implement features not
present in other codes. 

For simulations in which machine memory is a significant constraint,
we believe that an optimized version of the code could be developed
with a much reduced memory footprint, at the cost of some increased
computation. For example, gravitational force calculations might
instead be truncated at monopole order rather than quadrupole order, or a
number of temporary arrays throughout the code could be deleted in
favor of recalculation. Particularly in very high resolution
simulations however, we are reluctant to recommend reducing floating
point numbers to single precision, even though it provides a factor of
two footprint reduction for real values, because of the prevalence of
difference calculations in both the gravity and SPH calculations. At
very high resolution, or when particles lie close together in space,
errors in position difference calculations can grow large simply
because the particles share all but the last few digits of their
coordinate values. In such cases, and over many calculations,
catastrophic cancellation can grow to levels important for correctly
realizing the evolution in a given problem \citet{goldberg91}. Though
ultimately impossible to eliminate entirely, we have attempted to
design VINE so as to reduce such problems for its users. 

\singlespace
\begin{deluxetable*}{l|r|r|c|c}
\tablewidth{0pt}
\tablecaption{\label{tab:arrays} Number of Arrays Required by VINE modules}
\tablehead{
\colhead{Module}  & \colhead{Integer } & \colhead{Double }  
                    & \colhead{Integer (sph)} & \colhead{Double (sph)}  }
\startdata
Particle data                   &  2                                &  $4d+5$ & 1 &  8 \\
Integrator data                 &                                   &         &   &    \\
\phantom{\ \ \ \ \ }Leapfrog    &  0                                &  $d$    & 0 &  3 \\
\phantom{\ \ \ \ \ }Runge-Kutta &  0                                &  $3d$   & 0 &  8 \\
Ind. TS data                    &  6                                &  2      & 0 &  0 \\
SPH data                        &  0                                &  0      & 1 &  6 \\
Force law data                  &  0                                &  $6d-6$ & 0 &  0 \\
Tree data                       &  11                               &  $2d+5$ & 0 &  0 \\
Tree build runtime              & 1.0$N_{proc}+d$                   &  0      & 0 &  0 \\
Tree runtime                    & 4.5$N_{proc}$\phantom{\ \ \ \ \ } &  0      & 0 &  0 \\
Pnt. Mass runtime               &  1                                &  0      & 0 &  0 \\
Grape 6 data                    &  0                                &  10     & 0 &  0 \\
\enddata

\end{deluxetable*}

\doublespace


\section{Performance}\label{sec:perf}

After describing the code, its features and optimizations, we move now
to a discussion of its performance. Two separate qualities of the code
affect the performance. First, the code itself has a number of compile
time and run time settings that affect its speed and accuracy, such as
the maximum number of particles contained in a clump or the specific
MAC used or, at a deeper level, the different optimizations made to
the layout of variables in memory as discussed above. Second, the
hardware architecture on which the code is run may be faster or slower
than some other competing architecture, or may have settings which can
be tuned to provide better performance for the code.

In this section, we investigate the sensitivity of calculations with
VINE to the settings of a number of parameters, both in terms of the
speed and of the accuracy of the calculations. We first describe the
initial conditions of three test problems on which we run our tests.
Second, we describe the results of tests designed to determine optimal
settings for the gravity calculations performed using each of the
three multipole acceptance criteria defined in VINE, with and without
the use of GRAPE hardware. Next, we explore the sensitivity of the
gravity and SPH calculations rates to the size of the clumps used to
accelerate the tree traversals. Finally, we describe the serial and
parallel performance of the major components of the code and their
sensitivity to the various optimizations discussed above.

Our preliminary testing showed that some parameter settings (e.g. high
accuracy settings for the MAC used in the gravity calculation) could
result in significant impacts on performance as a function of
processor count due to limitations such as memory latencies due to
NUMA architectures. Although in the end, we found that the most
desirable parameter settings do not appear to be highly sensitive to
such architectural features, we chose to sample the range of processor
counts relatively densely in order to explore sensitivity to such
issues. Dense sampling also enables more direct comparisons on several
different architectures which may have both different maximum
processor counts and different scaling as a function of processor
count.

In each of the comparisons that follow, we show the effect of varying
one of the parameters to which the code's speed is sensitive, while
keeping all others fixed at their optimal values. For reference, and
unless otherwise indicated, we run the tests using 8 processors of the
Origin 3000 listed in table \ref{tab:arch} and we use the Gadget MAC
with an opening criterion set to $\theta=5\times10^{-3}$. This setting
yields gravitational forces for 99\% of particles that are accurate to
a few $\times10^{-3}$, as we show in section \ref{sec:tuningII}. We
use a maximum population of 70 particles in a single clump and we
trigger a complete tree rebuild when the size of any clump reaches
twice its value immediately following the rebuild. We use cache
blocking, tuned to the size of the L1 cache of each
architecture\footnote{Note however that Itanium~2 processors permit
only integers to be loaded into the L1 cache, so for that
architecture, we use cache blocking tuned to the size of the L2
cache}, and the largest page size available on each machine. Finally,
we reorder the particles before performing any of the speed tests. In
these tests, one `calculation of SPH quantities' is dominated
primarily by the calculations of the mass density and the hydrodynamic
forces. It also includes the calculation of the equation of state, the
internal energy derivative, and smoothing length variation. 

\subsection{Description of the test problems}\label{sec:test-probs}

Because the code can be used to follow the evolution of $N$-body
particles alone, SPH particles alone, or both together, we will define
a test problem for each option, in order to demonstrate its
performance in each kind of simulation. These three test problems will
be referred to as the $N$-body test problem, the SPH test problem and
the mixed test problem. In order to understand the performance of the
code as a function of resolution, the mixed problem will have
realizations at several resolutions.

\singlespace
\begin{deluxetable*}{lrr}
\tablewidth{0pt}
\tablecaption{\label{tab:testsims} Test simulations}
\tablehead{
\colhead{Name}  & \colhead{Number of}     & \colhead{Number of    }   \\
\colhead{    }  & \colhead{SPH Particles} & \colhead{$N-$body Particles } } 
\startdata
$N$-body 1      &   0                     &  160000         \\
$N$-body 2      &   0                     &  320000         \\
$N$-body 3      &   0                     &  640000         \\
$N$-body 4      &   0                     &  2000000        \\
$N$-body 5      &   0                     &  7000000        \\
SPH (Initial)   &   3500000               &  0              \\
SPH (Evolved)   &   3500000               &  0              \\
Merger1         &   37500                 &  232500         \\
Merger2         &   75000                 &  465000         \\
Merger3         &   150000                &  930000         \\
Merger4         &   300000                &  1860000        \\
Merger5         &   600000                &  3720000        \\
Merger6         &   1200000               &  7440000        \\
\enddata

\end{deluxetable*} 

\doublespace

\subsubsection{The $N$-body test problem}\label{sec:test-probs-nbody}

The $N$-body test problem is a sphere with density profile of $\rho
\sim R^{-1/4}$ which is set up according to the method described by
\citep{hernquist90} and which is realized with $7 \times 10^6$
particles. A lower resolution version with $2 \times 10^6$ has been
used for the force accuracy tests in section
\ref{sec:tree-acc}. Our system has a total mass of $5.6 \times 10^{12}
M_\odot$, a cut off radius of $175$ kpc and a scale radius of $35$
kpc. We also consider variants of this test with particle counts
ranging to as low as 160000 (see table \ref{tab:testsims}), in order
to study the influence of resolution on bunch size (section
\ref{sec:best-clumpsize}) and force accuracy when GRAPE co-processors
are used. For all tests involving the GRAPE processors, or comparing
the speed of VINE to them, we employ Plummer softening, while all
other measurements involving these tests we use the fixed spline
softening option. In order to maintain systematically consistent
timings at different resolutions, we scale the gravitational softening
length according to the cube root of the number of particles in the
simulation.

\subsubsection{The SPH test problem}\label{sec:test-probs-sph}

The SPH test problem is a set of two conditions defining the initial
and evolved state of a turbulent molecular cloud which undergoes
fragmentation and forms stars. It is identical in setup to the
simulations of \citet{bate2003}, however it was set up using a
different random seed. The molecular cloud in this model is initially
spherical, with uniform density and is realized with $3.5 \times 10^6$
particles. It has mass and diameter of $50M_\odot$ and $0.375$~pc
respectively, and it has an initial temperature of $10$~K. The
free-fall time of the cloud is $t_{ff}=1.9 \times 10^5$ yr and the
late time condition has undergone slightly more than one free fall
time of evolution and is characterized by a a very inhomogeneous
density distribution. For more details, we refer the reader to
\citet{bate2003}. 

\subsubsection{The mixed test problem}\label{sec:test-probs-mixed}

In order to test the code's abilities on a problem which incorporates
both $N$-body and SPH particles, we use a set of initial conditions
for spiral galaxy mergers, similar to those used by \citet{naab2006b}
and \citet{wetzstein2007}, in which two spiral galaxies are initiated
on a parabolic orbit at a distance of $105$ kpc. Each galaxy consists
of a stellar disk and bulge, a gaseous disk and a dark matter halo,
set up according to the method of \citet{hern1993}. The components
have masses of $M_d=3.92 \times 10^{10}M_\odot$ and $M_g=1.68 \times
10^{10}M_\odot$ for the stellar and gas disks respectively, and $M_b=
0.2M_d$ for the stellar bulge and an additional mass of $M_h=3.248
\times 10^{10}M_\odot$ defining the dark matter halo. Both stellar and
gas disc have an exponential surface density profile with scale
lengths of $3.5$ kpc and $10.5$ kpc, respectively. The dark matter
halo is modeled as a pseudo-isothermal sphere with a core radius of
$r_{c,h}=3.5$ kpc and a cut off radius of $r_{cut,h}=35$ kpc. For more
details about the initial conditions, we refer the reader to
\citet{wetzstein2007}. 
 
We have created models of this merger setup at six different
resolutions (see table \ref{tab:testsims}). In the lowest resolution
model, Merger1, each galaxy contained $37500$ particles for the
stellar disk, $18750$ for the gas disk, $3750$ for the stellar bulge
and $75000$ for the dark matter halo. For each higher resolution model
in the sequence, the particle numbers of each species have been
doubled. The fraction of gas particles, for which SPH calculations are
required, thus remains constant for each resolution at $\sim14$\% of
the total. Due both to this comparatively small fraction, and the
inhomogeneous spatial distribution, this problem will therefore
represent a challenging test of the load balancing and parallel
efficiency achievable by the calculations of the hydrodynamic
quantities with SPH. 

\subsection{Tuning the code I: accuracy and speed of various
alternatives for the gravity calculations}\label{sec:tuningI}

The ability to compute forces accurately enough to integrate particle
trajectories correctly using a tree based force calculation is
governed by the choice of the MAC and its setting. In this section we
present quantitative tests of the accuracies for all three opening
criteria implemented in VINE: the geometric MAC (equation
\ref{eq:geometric-mac}), the SW absolute error MAC (equation
\ref{eq:SW-mac}) and the Gadget MAC (equation \ref{eq:gadget-mac}).

We perform our tests on low and high resolution versions of our
$N$-body test problem, as well as the initial and evolved states of
our SPH test problem. The low resolution $N$-body test was realized
with $2 \times 10^6$ particles. For all tests with the GRAPE/tree
option, a GRAPE-6A board ('MicroGRAPE') has been used.

We determine exact accelerations in each of the test problems using
either a calculation of the forces using the direct summation option
in VINE, or the geometric MAC with an opening criterion set to
$\theta=10^{-10}$, effectively equivalent to a direct summation of the
contributions from each particle on the others. These exact values
were used as references to calculate errors in the magnitudes of the
accelerations, defined as
\begin{equation}\label{eq:accel-error-eq}
|\delta \mbox{\boldmath$ a \,$} |
           = {{  | \mbox{\boldmath$ a   \,$} -
                   \mbox{\boldmath$ a_0 \,$}  | }\over
                   { | \mbox{\boldmath$ a_0 \,$}  | } }
\end{equation}
where $\mbox{\boldmath$ a$}$ and $\mbox{\boldmath$ a_0$}$ are the
approximated and exact accelerations, respectively. 

Determination of a specific limit on the acceptable errors will in
general depend on the system simulated, the time span over which the
system must be evolved, and the goals of the simulation. Several
general principles hold for most systems however. First, it will be
desirable to have a narrow error distribution, since the work required
to compute highly accurate accelerations on some particles will be
wasted if the simulation evolves incorrectly due to much larger errors
of others. In general, a reasonable requirement for force accuracy is
that the vast majority, say $99\%$, of the particles have force errors
below $0.1-0.5\%$. In addition, we consider it a reasonable
requirement that none of the remaining $1\%$ of particles have errors
larger than some other, higher threshold, say $1\%$. This second
condition avoids the possibility that a substantial `tail'
distribution of particles exists and which have very large errors
that are large enough to alter a simulation's evolutionary trajectory
even though the population is small. Therefore, we will adopt both of
these limits in our analysis. Again however, we note that the question
of what force error distribution is required for accurate simulation
of a given system, depends both on the system morphology, and on the
goals of the researcher. The MAC settings required for the code in
such cases may therefore be higher or lower than quoted here, though
the values we choose are known to work reasonably well on a wide
variety of problems in the literature.

\begin{figure*}[!t]
   \epsscale{0.9}
    \plotone{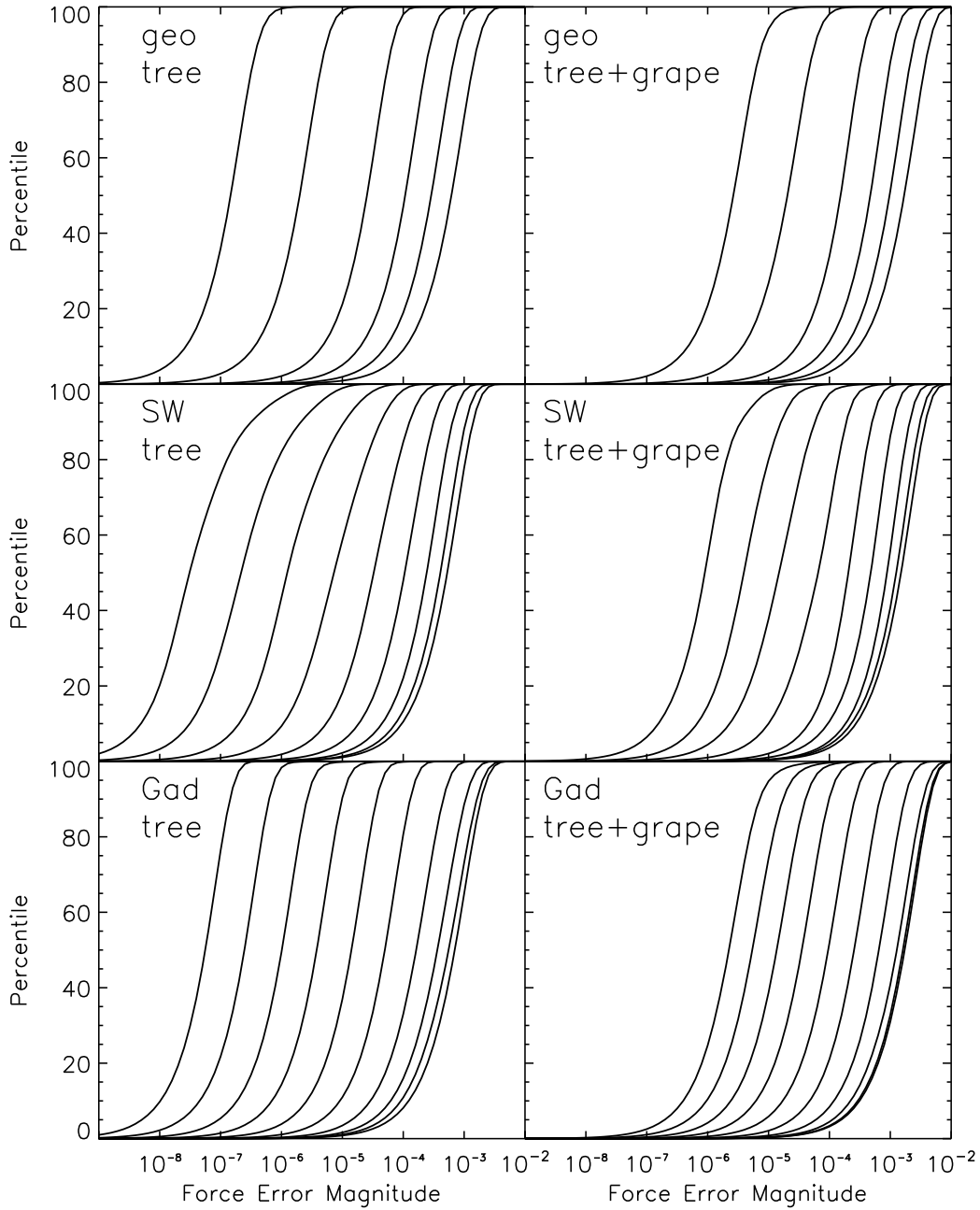}
    \caption{Cumulative distribution of the fraction of particles with
    relative force errors lower than a given value. The left column
    shows the force error distributions for the tree on host, the
    right column for the GRAPE-tree combination. Results for the
    geometric, SW and Gadget MACs are shown in the top,
    middle and bottom rows, respectively. For the geometric MAC, the
    curves represent $\theta=(0.1,0.2,0.4,0.6,0.8,1.0)$, for the
    SW MAC curves correspond to settings of $10^{-6} \le \theta \le
    10^2$ in decade increments and for the Gadget MAC curves
    corresponding to $10^{-9} \le \theta \le 10^0$ in decade 
    increments are shown.}
    \label{fig_force_acc} 
\end{figure*}

Figure \ref{fig_force_acc} shows distributions for each of the three
MACs implemented in VINE, for both the tree based calculation and the
GRAPE/tree combination, performed on a single Itanium~2 processor, or
in combination with a GRAPE-6A. In sections \ref{sec:tree-acc} and
\ref{sec:grapetree-acc}, we discuss these distributions in detail.

\subsubsection{Accuracy using the tree/host based
calculation}\label{sec:tree-acc}

The left panels of figure \ref{fig_force_acc} show the cumulative
error distributions for the tree based force computation performed on
the host. The most relevant parts of the distributions are the overall
widths of the distributions and the upper right part of each curve,
which show the population of particles with the largest errors in the
calculation. These are the most likely to cause visible numerical
errors to the evolution of the system, due to inaccurate integration
of particle trajectories.

For the geometric MAC, the distribution rises steeply towards larger
errors, through an `S' shaped curve, with a small `tail' population at
the high end limit of the distribution and a somewhat larger tail
population at the low end limit. Though we have made no explicit study
of particles in either the high or low error tails, we presume that
most near the high error limit are those for which the absolute
magnitude of the force is small, so that the relative error is
correspondingly larger. For $\theta=1.0$, the high error tail extends
to errors as great as $\sim0.8\%$ and about $25\%$ of the particles
have errors greater than $0.1\%$, though none have errors as large as
1\%. The case of $\theta=0.8$ produces more accurate results, but
still $\approx 4\%$ of the particles have force errors $>0.1\%$. When
$\theta=0.6$, no particle has errors above $0.1\%$. This setting
produces an error distribution somewhat more restrictive than the
requirement we define above, and we conclude that adopting a value of
$\theta=0.6-0.7$ in VINE can be regarded as acceptable for most
scientific applications.

For the SW MAC, the error curve again follows an `S' shape, with
similar maxima, of $\sim0.5-0.8$\%, for the most permissive settings.
Because VINE converts all quantities to dimensionless units for use
inside the code, stating a specific value for the SW MAC setting loses
meaning, since without details of the unit system in place, limits on
the contribution of any given node to the absolute force error become
difficult to interpret. Nevertheless, and with the understanding that
the unit conversions are done with the intent of converting the
relevant physical variables to unit-less quantities near unity, we
provide such a statement. For the particle distribution here, a values
of $\theta < 0.05-0.1$ (in internal code units) appears to be
acceptable to use if our limit described above is applied. For other
unit systems, perhaps implemented by VINE's users, a different setting
may be required.

For the Gadget MAC, the error distributions again rise steeply as a
function of error magnitude and, although the largest errors with the
most permissive settings extend to slightly larger values than for
either of the other two choices, no particles with errors greater than
1\% are found for any setting. Settings with $\theta \gtrsim 10^{-2}$
produce unacceptably large errors in the forces according to the
criteria above, while values $\theta\lesssim 10^{-3}$ provide more
restrictive error limits. We suggest that a reasonable choice is
$\theta\sim1-5\times10^{-3}$, to approximate our chosen error
criteria.

An important characteristic of the geometric and Gadget MACs is that
the shape of the error distribution is both narrow and changes little
as the tolerance parameter, $\theta$, decreases, so that the curve
simply shifts further and further to the right. This is a very
desirable feature not only because errors for the majority of
particles decreases, but also because the largest force errors are
effectively controlled by reducing $\theta$. If instead a significant
population of particles remain with large errors, while the rest
decrease, a simulation may become computationally costly while
remaining insufficiently accurate. This is because particles whose
forces errors are large can affect the evolutionary trajectory of a
simulation disproportionately to their number \citep[for a
particularly dramatic example of such effects, see e.g., the exploding
galaxy problem in][]{salmon_warren94}, so controlling their behavior
is of particular importance for accurate simulations.

In contrast, while the error curves for the SW MAC for the most
permissive tolerance settings appear very similar to those of the
other two, curves representing more restrictive settings do not. The
distribution shifts not only towards lower errors, but also changes
shape. While the low end of the distribution decreases steadily, the
high error tail does not, leading to a large spread in the calculated
error magnitudes for different particles. We believe the widened
distribution is a consequence of the fact that the SW MAC, as
implemented in VINE, constrains the absolute error magnitude of the
force on each particle, while figure \ref{fig_force_acc} shows
relative error magnitudes. In a system where particles have a wide
distribution of force magnitudes, errors for particles with small
magnitudes will decrease less quickly than those with larger until the
opening criterion falls to some critical value for a given particle
and causes additional nodes to be opened in the tree. 

Finally, we note that for the most permissive settings of both the SW
and Gadget MACs, and for the same relative change in the MAC setting,
much smaller differences are seen in the error distributions compared
to changes at more restrictive settings. This is an important feature
of both MACs because they imply an effective upper limit on the force
errors, a clearly desirable feature for numerical realization of any
physical system, all the more so since the error limits remain
comparatively small. 

We believe the reason for the limit is that both MACs enforce a
requirement that tree nodes are only accepted for use in the force
summation if the particle (or clump) on which forces are calculated
lies exterior to the node itself (see \vineI, section 4.2) for any
setting of the MAC. This physical constraint corresponds to the
numerical condition that a given tree traversal will produce the
shortest possible list of acceptable nodes for use in the force
summation, for that MAC. The most permissive settings shown in figure
\ref{fig_force_acc} have clearly entered the regime where convergence
of the generated interaction list towards this minimal list has begun,
and implying that the error limits shown are near those of the minimal
list as well. The geometric MAC does not explicitly contain a similar
condition, however the same effect is still realized in practice
because source and sink nodes may overlap in space for settings with
$\theta>1$, a condition which violates the assumptions underlying the
multipole approximation used to approximate the forces in the first
place. For a more detailed discussion of this effect, we refer the
reader to section \ref{sec:MAC-speed}.

\subsubsection{Accuracy of the GRAPE/tree
combination}\label{sec:grapetree-acc}

Error distributions for the GRAPE-tree based force calculations are
shown in the right panels of figure \ref{fig_force_acc}. As expected
for calculations involving GRAPE hardware, errors for all three MACs
are larger than in the corresponding tree/host based calculation for
identical MAC settings. Except for the most permissive tolerance
settings, error limits an order of magnitude larger are typically
realized with the GRAPE/tree option in comparison to those with the
tree/host option. While maximum errors remain below $\sim1$\%, more
restrictive settings do not shift to smaller errors to nearly the
extent that the tree/host based calculations do, for the same
tolerance parameter.

We interpret each of the effects as consequences of the differences
between the code implementation in either case. The GRAPE/tree
combination makes use of a modified tree traversal (section
\ref{sec:grav-grapetree}), with large clump sizes (`bunches') for the
far traversal and does not terminate with a list of neighbor bunches
for use in a close traversal, but instead calculates many more
pairwise interactions between individual particles. This feature
implies a more accurate calculation of the forces from nearby
particles on each other. At the same time however, the GRAPE/tree
option accounts only for the monopole moment of a node, and so implies
a less accurate calculation of forces from more distant particle
groupings. Therefore, for the same value of the tolerance parameter
$\theta$, the computed accelerations are less accurate than those of
the tree based calculation, which includes quadrupole moments too.

For this problem, error limits at or below the constraints defined
above remain within reach of the GRAPE-6A/tree combination, but
require much more restrictive settings of the opening criterion.
Values of the opening criterion of $\theta\lesssim 0.6$, $\theta
\lesssim 10^{-2}-10^{-3}$ and $\theta\lesssim 10^{-4}-10^{-5}$ are
required for the geometric, SW absolute and Gadget MACs respectively.

\subsubsection{The relative speed of using different
MACs for a given accuracy}\label{sec:MAC-speed}

In this section, we examine the relative efficiency of using each MAC
for producing accelerations of a given accuracy. For these tests, we
use the full $N$-body test problem and both variants of the SPH test
problem. Each test uses the same executable code, running on 8
processors of the Origin 3000. For each run, we specify the MAC and
MAC setting via an input file, run the executable and output
accelerations for all the particles for later analysis. We sort the
errors according to their magnitude to determine the value of the
error magnitudes below which which 50\% and 99\% of particles lie.
Combining information from both error limits, it will be possible to
quantify both an approximate upper limit on the errors and a crude
measure of the overall width of the distribution, from the difference
between the two quantities. The overall shapes of the error
distributions are similar to those shown in figure
\ref{fig_force_acc}. 

\begin{figure*}[!t]
\rotatebox{-90}{\includegraphics[width=120mm]{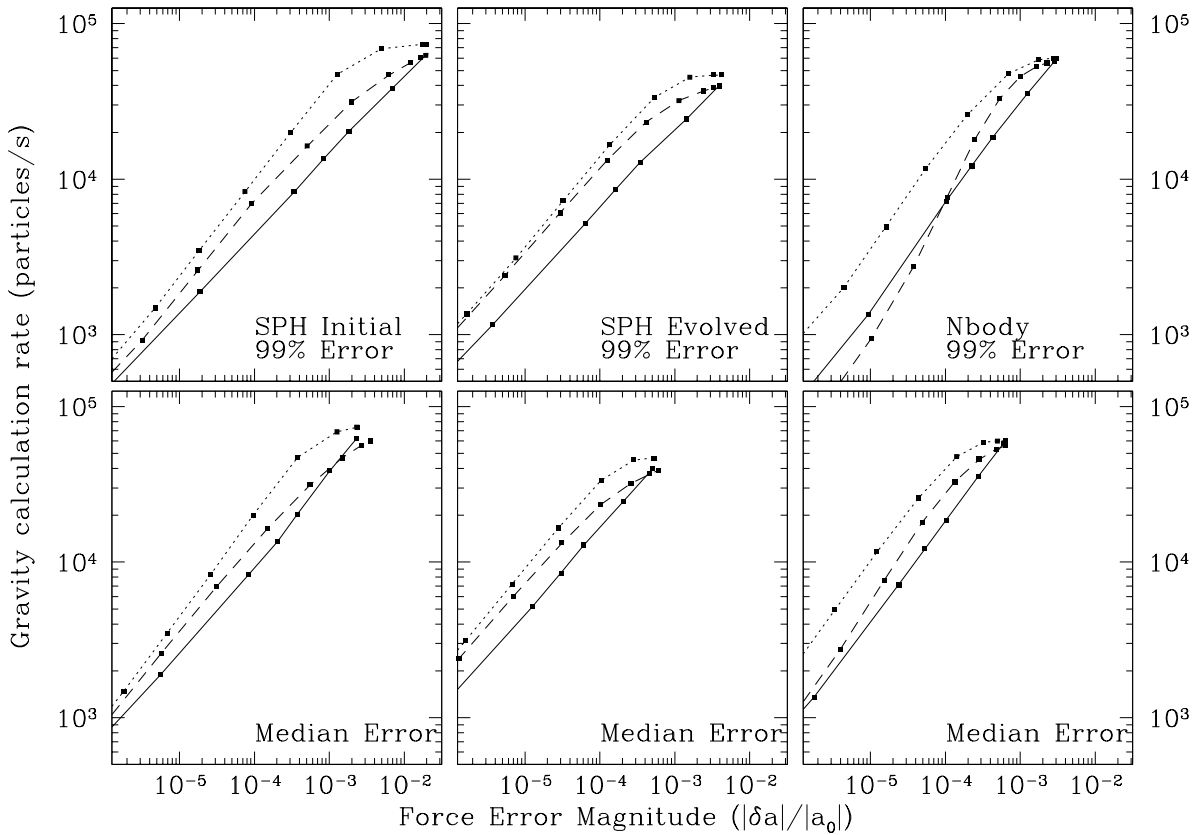}}
\caption{\label{fig:macrate}
The rate of gravitational force calculations for the code using the
three MACs implemented in VINE, as a function of the accuracy. The top
panel shows the rates for which 99\% of the particles have error
magnitudes less than that shown, while the bottom panel shows the
rates for the median error rate, at which 50\% of particles have error
magnitudes less than that shown. The solid curve shows the rates
obtained from the geometric MAC, the dotted curve shows the rates from
the Gadget MAC and the dashed curve shows rates for the SW 
MAC. Points correspond to settings of
$\theta=0.1, 0.2, 0.4, 0.5, 0.6, 0.8$ and $1.0$ for the geometric MAC,
to settings of $\theta$ in decade increments from $10^{-8}$ to $10^0$
for the Gadget MAC, and to settings $\theta$ in decade increments from
$10^{-6}$ to $10^2$ for the SW absolute error MAC, using internal code
units for the forces.}
\end{figure*}

Figure \ref{fig:macrate} shows the rate of gravity calculations per
second as a function of the 50\% (i.e. median) and 99\% error
magnitudes. In their asymptotic, high accuracy limits, the rates for
each MAC are each approximately proportional to the square of the
accuracy. This proportionality is expected from the truncation of the
multipole series at quadrupole order \citep{salmon_warren94}. In that
limit however, the Gadget MAC yields rates that are a factor $\sim2$
higher than those of the geometric MAC, with the SW MAC lying between
and yielding a slightly different proportionality. The difference
appears to be due to the fact that the SW MAC is used in its form as a
limiter on the absolute error magnitude that any given node can
contribute rather than the relative error, as is the case for the
other two MACs. We attribute the overall speed differences between the
MACs to the fact that while the geometric MAC accounts for only the
physical size of a source node, each of the other two also account, in
different approximations, for its internal structure as well. The
additional information proves valuable in allowing nodes to be
selected, which otherwise would need to be opened and examined in more
detail.

The error magnitudes for the most liberal settings of each of the
three MACs in the same test vary only slightly from one another. The
similarities are presumably due to the fact that the three different
opening criteria each reduce to the same physical condition that the
sink point at which the force is calculated must be exterior to the
source node that exerts the force. Although the error magnitudes from
the same test are similar, they differ from one particle distribution
to the next, as do the calculation rates realized for those errors.
The rates are highest in the case of the SPH initial condition and
lowest for the evolved condition, with the Hernquist sphere lying in
between. The differences are significant because we expect that the
speed of a gravitational force calculation will be proportional to
$N_p\log N_p$, and would therefore expect similar rates in the two SPH
conditions and lower rates in the $N$-body test with twice as many
particles.

Some clarity emerges if we also correlate the rates and relative force
errors in each problem and their overall morphology. The relative
force errors are largest in the SPH initial condition, smallest in the
evolved condition and intermediate in the Hernquist sphere--an
opposite trend from that seen for the rates. At the same time, while
the SPH initial condition is quite smooth, the Hernquist sphere is
centrally condensed and the evolved SPH condition is near the onset of
fragmentation and is even more inhomogeneous. We therefore attribute
the rate and accuracy differences to the differences in particle
distributions in each test and conclude that, at least on scales of
factors of two difference in particle count, the morphology of the
particle distribution plays a role much larger than the overall
scaling.

Even for the most liberal MAC settings, the median and 99\% error
magnitudes are well below 1\% and $\sim$2\% respectively for all three
MACs for smooth conditions, and nearer to $\sim0.5$\% for the same
settings in non-smooth conditions. This level of precision differs
from the behavior of many other codes, given the same MAC and setting
\citep[see, eg., corresponding figures
in][]{springel2001,springel_g2,wadsley2004,dubinski03}, first, because
the calculated size of each node (equation \ref{eq:treenode-size}) as
incorporated into the MAC for that node is always an overestimate of
its true size, so that a search will tend to open more nodes than are
strictly required. Second, dividing the traversals into far and near
components for groups of particles means that, for a given setting,
more nodes and atoms are required than would be the case for a single
particle. While the MAC must be satisfied for every node on the
interaction list at the point in the clump closest to that node, most
particles in the clump will be found further away, where the node's
parent might otherwise be acceptable. The trade off is beneficial in
the sense that fewer traversals are required, even though more
interactions result.

Of particular interest for most numerical simulations is the range of
relative errors above a few $\times 10^{-4}$. In this range, the
additional information describing the internal node structure and
incorporated into the Gadget and SW MACs provides its greatest
benefit. The rates obtained from the Gadget MAC decrease from their
most liberal settings by only $\sim$20\% while the calculation becomes
an order of magnitude more accurate, and the setting itself changes
from $10^0$ to $10^{-3}$. The SW MAC exhibits a similar, but less
pronounced characteristic.

Given the results of these tests and those of the previous sections,
we recommend the use of the Gadget MAC in VINE, with a setting of
$\sim5\times10^{-3}$, for simulations to run at both high accuracy and
high efficiency. This setting ensures maximum relative errors of
$\sim1$\% in very smooth conditions, and $\sim0.1-0.2$\% in more
evolved systems, without increasing the cost of the calculation. This
accuracy is also comparable to the expected numerical noise present in
the hydrodynamic quantities of SPH simulations \citep{HW94}, below
which higher accuracy will yield little additional benefit to a
simulation.

\subsection{Tuning the code II: optimal software settings and memory
layout optimizations}\label{sec:tuningII}

In this section, we will discuss the performance of the most costly
components of simulations using VINE that interact with the tree. In
order, these are the gravity calculation, the SPH calculations and the
tree build/revision. We explore the sensitivity of the performance to
various optimizations, in order to quantify the influence that each
makes on the total absolute performance and the scaled performance.
For our purposes, absolute (`raw') performance will be defined simply
by higher calculation rates, while scaled performance will be defined
by efficiency of parallel scaling, relative to the performance on one
processor. 

\subsubsection{The optimum number of particles in a
clump or bunch}\label{sec:best-clumpsize}

The tree traversal strategies described in sections
\ref{sec:grav-trav} and \ref{sec:grav-grapetree} each contain one free
parameter that must be tuned in order to provide the best performance,
namely the number of particles contained in a clump or bunch. In the
case of clump population, both the gravity and the SPH calculations
will be affected because tree traversals are done in both cases, to
search for acceptable nodes and atoms in the gravity calculation, or
for neighbors in the SPH calculations.  

Clumps and bunches may by populated by any number of particles up to
and including a preset maximum. In this section, we explore the
sensitivity of the calculation rate to that maximum population value.
We choose the maximum as our metric because it is easily available for
modification by the user, while the more directly intuitive values of
average or exact populations must be derived from the particle
distribution and the exact tree structure. Over a wide variety of
morphologies, we have found that the population distribution fills
essentially all values from `singleton' clumps up to the maximum, and
the average fall typically at $\sim50-60$\% of the maximum.

\begin{figure}[!t]
\rotatebox{-90}{\includegraphics[width=60mm]{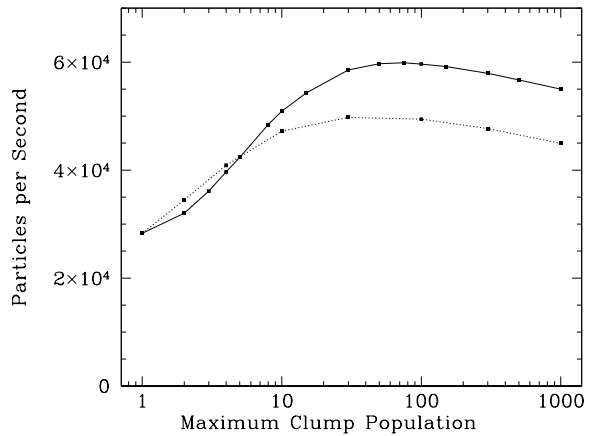}}
\caption{\label{fig:grav-clumppop}
The calculation rate for gravitational forces on particles in the
$N$-body test problem, as a function of the maximum particle
population in a clump. The solid curve shows the rates with cache
blocking, while the dotted curve shows the rates without. Both use the
Gadget MAC, with opening criterion $\theta=5\times10^{-3}$.}
\end{figure}

Figure \ref{fig:grav-clumppop} shows the calculation rates for a
single gravity calculation with the $N$-body test problem as a
function of the maximum number of particles contained in a clump. As
the maximum is increased from a single particle to $\sim30$, the
calculation rate increases by a factor of $\sim2$. Increasing the
maximum further results in much smaller improvements, with a broad
maximum in the rates near $\sim70$ particles per clump. Relative to
the performance with a clump population of one (i.e. individual
traversals for each particle), we obtain a speedup of $\sim2.1$. The
two curves represent the performance obtained with and without cache
blocking (see section \ref{sec:treegrav-opts}), and while both
versions show improvements over the rate for individual traversals,
the unblocked version does not increase nearly as much, saturating at
a factor $\sim1.75$. The differences are important to quantify because
when individual time steps are used in a simulation, only a relatively
small fraction of the particles in a clump will be active. In this
case, the benefits from cache blocking will be reduced by a factor
related to the fraction of active particles in each clump, and will
more closely reflect the rates without cache blocking. Even in the
case when only two particles are active in a given clump, some benefit
of the cache blocking is retained however, so we enable it by default,
except when only a single particle is active in a clump.

\begin{figure}[!t]
\rotatebox{-90}{\includegraphics[width=60mm]{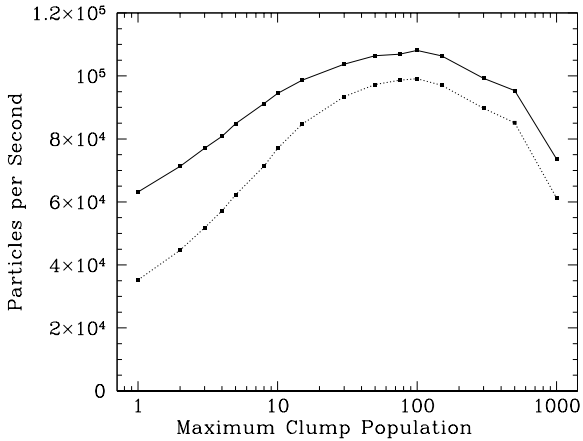}}
\caption{\label{fig:sphspeed-clumpnum}
The rate that all SPH calculations required in the two SPH test
problems can be calculated, as a function of the maximum number of
particles in a clump. The solid curve represents the times for the
initial condition, and the dotted line represents the times for the
evolved state.} 
\end{figure}

Figure \ref{fig:sphspeed-clumpnum} shows calculation rates for a
single calculation of all SPH quantities, including both densities and
the hydrodynamic accelerations, using the initial and evolved particle
distributions in the SPH test problems described above. Here again we
see a substantial improvement as the maximum clump population
increases. In the case of the initial condition, the rate increases by
a factor $\sim1.5$ over that required for clumps defined as single
particles. The evolved state speeds up by a much greater factor of
$\sim3$, presumably due to an increased level of data reuse when
calculations for nearby clumps occur in close succession. Even for the
fastest setting, the time required for the evolved particle
distribution does not match that of the initial state however, where
the optimal rate is about 10\% higher. We have not attempted to trace
the difference to a specific cause, however it is comforting to
realize that in spite of the quite different morphologies, the rate of
calculation changes by only a relatively small amount.

For both the gravity and the SPH calculations, the maximum rate occurs
when the maximum number of particles per clump is $\sim50-70$. We have
therefore chosen to set its default value in VINE to 70 in 3D
simulations, which is also the setting for the maximum number of
neighbors allowed for any SPH particle, before limiters are activated
which push the number downwards again (see additional discussion of
these limiters in \vineI). No detailed tests have been performed to
determine an optimal setting for 2D simulations, however small scale
tests (not shown here) indicate that the optimal value is smaller. As
for 3D, we set the maximum to the maximum number of SPH neighbors,
which is $\sim30$ for 2D simulations.

The speed of the gravity calculation is also sensitive to the clump
population when the tree is used in combination with the GRAPE
hardware. So much so that it even lead us to define a different,
larger grouping (`bunches') for use in combination with GRAPEs, as
discussed section \ref{sec:grav-grapetree}. The reason for the
sensitivity is more complicated when GRAPE is used however because the
time to solution includes contributions from not only tree traversal
and multipole summation computation, but also from communication to
and from the GRAPE processor. 

\begin{figure}[!t]
\epsscale{1.}
\plotone{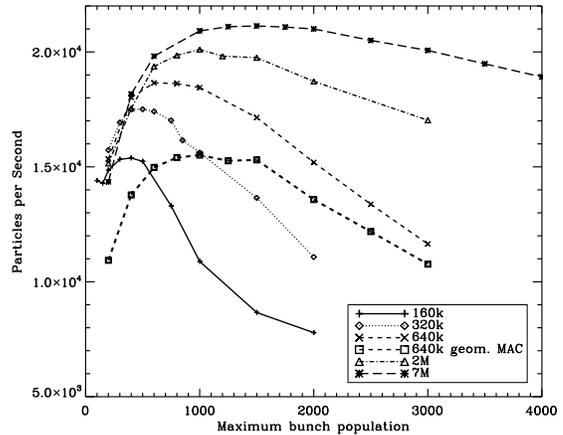}
\caption{\label{fig:grav-bunchpop} 
The calculation rate for gravitational forces on particles in the
$N$-body test problem at several resolutions, as a function of the
maximum particle population in a bunch. We use the Gadget MAC, with
opening criterion $\theta=5\times10^{-3}$. For comparison, one test
with the geometric MAC and $\theta=0.75$ is shown as well. The host
was a single Itanium2 processor and used a GRAPE-6A ('MicroGRAPE')
board for the gravity calculations themselves.} 
\end{figure}

Figure \ref{fig:grav-bunchpop} shows calculation rates for a single
gravity calculation of the $N$-body test problem as a function of
maximum bunch population and using VINE in GRAPE$+$tree mode, and for
a number of resolutions. We use the Gadget MAC with
$\theta=5\times10^{-3}$ for all tests and include one comparison to
the geometric MAC with $\theta=0.75$ for the case with 640000
particles. As was the case for clump populations in the the host-alone
gravity calculation, the calculation rate is sensitive to the maximum
bunch population. The peak rates increase by a factor of $\sim1.5$ or
more over those for smaller or larger populations. Of interest is the
fact that the optimal bunch population changes as a function of the
resolution of the simulation itself, even though the simulations are
of identical configurations. While the case with 160000 particles has
an optimal maximum bunch population of $\sim400-600$, the optimal
population for the $7\times10^6$ particle realization is near $\sim
1200-1800$. Because the calculation speeds change little over a fairly
broad plateau of maximum bunch populations, it is easy to achieve a
near-optimal choice even without extensive testing. The calculation
rate decreases much more steeply for bunch sizes lower than the
optimum than for bunch sizes greater than the optimum, so it is safer
to adopt a higher value if an estimate is required.

We believe that the sensitivity is a consequence of the fact that the
physical size of a bunch containing a given number of particles
decreases as a function of increasing resolution, but the MAC remains
the same. This is important because the preliminary interaction list
for a bunch (i.e. the interaction list that does not include the bunch
members themselves), will contain a volume just outside its boundary
in which most interactions are with particles, rather than with nodes.
If that volume is determined by the MAC setting, through appropriate
choices as to which nodes to open or accept, then much more time must
be spent in portions of the tree traversal that open all nodes to
their fullest extent--an inefficient use of resources. If instead, we
assume that bunches contain more particles, a large fraction of such
tree nodes will be resolved at the level of single particles by
assumption, so that all costs of tree traversal for this volume will
be removed. Taking the argument to the extreme however--allowing only
a single bunch with all of the particles so that the calculation
returns to a pure O($N_p^2$) proportionality--illustrates that there
is clearly a point of diminishing returns where this argument fails.

\subsubsection{Performance of the gravity calculation and the effects
of memory layout optimizations}\label{sec:perform-grav}

In this section, we turn to a discussion of the serial and parallel
performance of the gravitational force calculation, which will
typically be the most costly part of any particle simulation of a self
gravitating system. We also discuss the benefits of each of the memory
layout optimizations discussed in sections \ref{sec:impl-design} and
\ref{sec:treegrav-opts}. For these tests, we use the $N$-body test
problem.

\begin{figure}[!t]
\begin{center}
\rotatebox{-90}{\includegraphics[width=60mm]{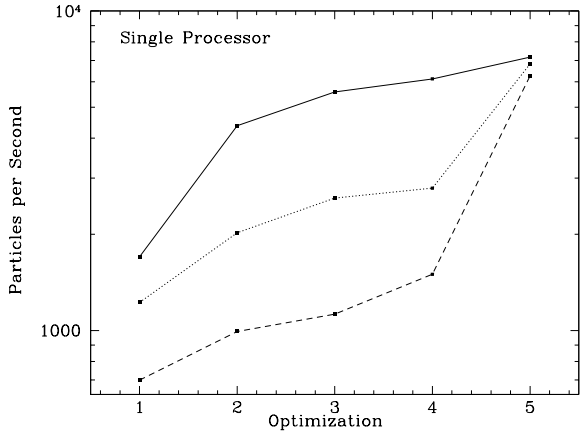}}
\end{center}
\caption{ 
\label{fig:grav-speedups}
The rate that gravitational forces on particles in the $N$-body test
problem can be calculated on one processor under cumulatively more
memory layout optimizations, as defined in the text. The three curves
represent the rates obtained using hardware page sizes of 16MB
(solid), 1MB (dotted) and 64kB (dashed) that are available on the
Origin 3000. }
\end{figure}

In figure \ref{fig:grav-speedups}, we show the speed of a single
gravity calculation performed several times with a different set of
optimizations: each realization contains all optimizations of the
previous levels, and one addition. Defined as in the figure,
optimization level one means that no optimizations are employed, other
than to use the Gadget MAC. Tree traversals are done for individual
particles, all node data are stored in separate arrays and are not
reordered as discussed in section \ref{sec:tree-post}. For level two,
traversals are done on clumps rather than single particles and for
level three, the tree data are reordered to reflect the order of
access in the traversal. Level four arranges node data in two
dimensional arrays so that all data for a single node are located
together in memory. Finally, level five implements the cache blocking
optimization. Although not dependent on any specific code change, the
setting of the hardware page size defines an additional `level' of
optimization which influences calculation speed. To determine the
effects that it may have on the code, we repeat each test for each of
three page sizes: 64kB, 1MB and 16MB.

For the completely unoptimized case, serial performance is 700, 1225
and 1700 particles per second for the small, medium and large page
sizes respectively. With all optimizations included, the calculation
rate increases to 6270, 6830 and 7170 particles/s, for the same page
sizes, respectively. Total performance enhancement from the slowest
(level one with 64kB page size) to fastest (level 5 with 16MB page
size) calculation is a factor $\sim10$, with an additional factor of
$\sim2$ coming from selection of the Gadget MAC in favor of the
geometric MAC as we saw in section \ref{sec:MAC-speed}. Although each
addition results in performance enhancement relative to the previous
level, we do not believe that the specific ordering as defined in the
figure is required for improvements in the code. Though we have not
quantified the contribution of each optimization in isolation, we
expect that each will provide benefits, independent of all others, to
any code that implements them.

For all but the highest optimization level, variation between rates of
identical optimization but different page sizes is much larger than
variation between one optimization level and another with the same
page size. Moreover, identical optimizations provide different
performance enhancements when different page sizes are used. For large
pages, moving to grouped tree traversals provides more than a factor
three speedup, far more benefit than any of the other optimizations.
For small pages, the total speedup provided by the first four
optimizations together provides only a factor two benefit, with a
further benefit of a factor of four from cache blocking. We believe
that while the exact source of the variations is unimportant, they are
most likely due to the relative costs and frequencies of simply
loading a value from cache or memory compared to that of computing a
new TLB entry.

In terms of the `FLOP' rate (floating point operations per second)
achieved by the code, we have profiled tests similar to these with the
{\tt `perfex'} utility available on Origin~3000 systems. We found that
the code achieves approximately 400~MFlops per second, or about half
of the theoretical maximum of 2 flops per clock cycle for 400MHz
R12000 chips. We believe this rate is near `perfect' in the sense that
the mix of operations (i.e. the proportion of adds and multiplies) in
most computationally intensive loop (summing the multiple
contributions from nodes), is able to utilize only some $60$\% of the
possible operations and the tree traversal portion of the calculation
contributes almost no floating operations to the total.

\rotatebox{-90}{\includegraphics[width=60mm]{ms2f7.eps}}

\begin{figure*}[!t]
\begin{center}
\rotatebox{0}{\includegraphics[height=180mm,width=175mm]{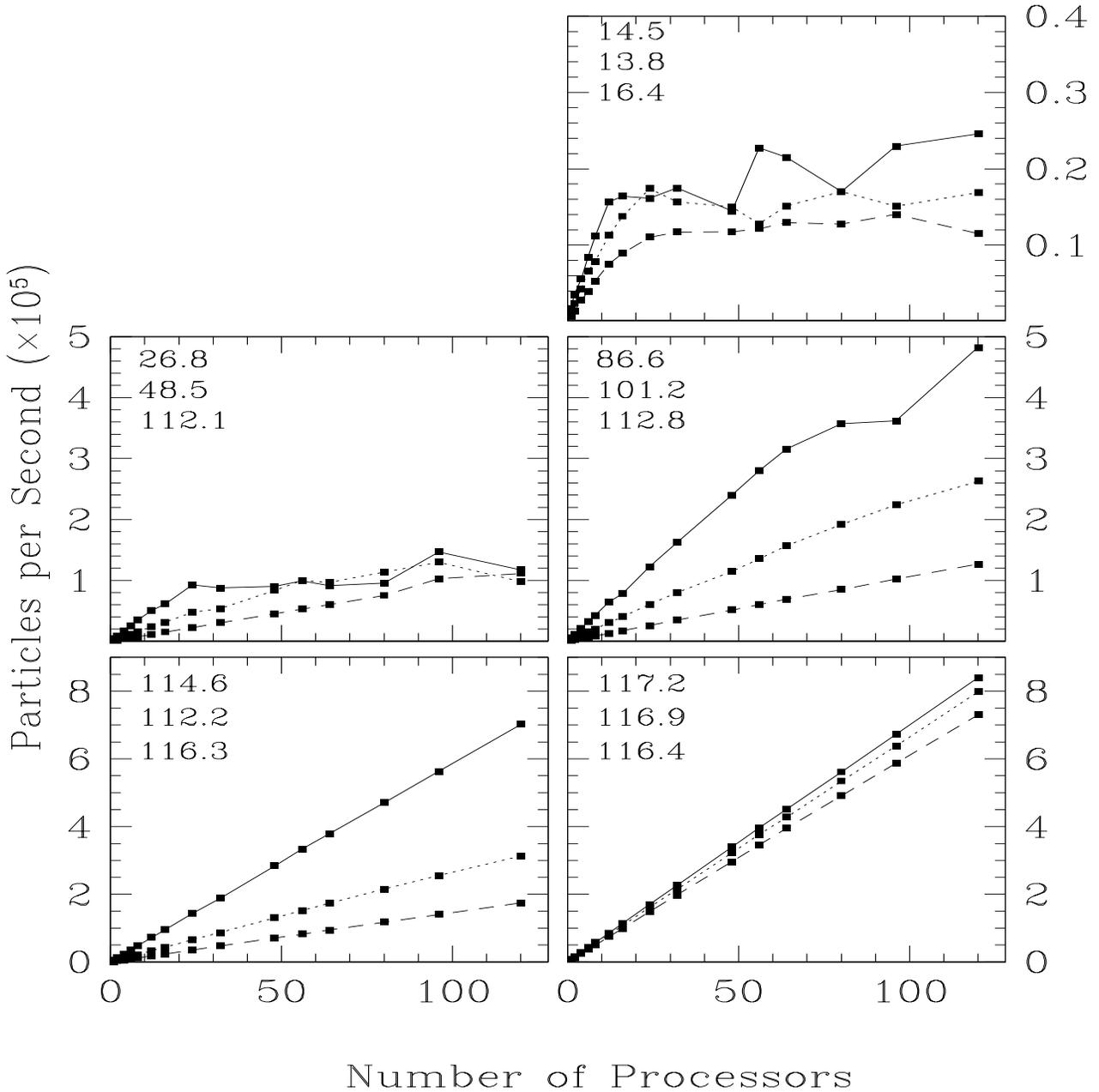}}
\end{center}
\caption{ \label{fig:grav-scaling}
The rate that gravitational forces on particles in the $N$-body test
problem can be calculated under cumulatively more memory layout
optimizations, as discussed in the text. Each panel contains three
curves representing the rates obtained using hardware page sizes of
16MB (solid), 1MB (dotted) and 64kB (dashed) that are available on the
Origin 3000, as a function of number of processors. Top to bottom, the
numbers in the upper left corner of each panel are the scaled parallel
speedup of the code for the 16MB, 1MB and 64kB page sizes,
respectively, out of a maximum of 120 for linear speedup. Note that
the vertical scale at each of the three panel levels are larger in the
lower panels.}
\end{figure*}

In figure \ref{fig:grav-scaling}, we show the speed of a single
gravity calculation as a function of the number of processors for the
same series of optimizations described above. For the unoptimized
version, parallel performance increases near linearly up to a maximum
of only 10 processors and saturates entirely at a speedup of $\sim15$.
As for single processor performance, absolute performance is better
with larger pages and, for successively higher optimization levels,
raw performance continues to increase faster for large page size runs
than for small. While providing superior absolute performance at all
processor counts, parallel scaling performance with large page sizes
saturates on fewer processors. We believe the reason can be tied to
the fact that the NUMA related latencies for obtaining data from
distant nodes are partly masked by the times required to calculate new
page addresses. When fewer TLB entries are required, only the NUMA
latencies remain. 

For the two highest optimization levels, scaled performance for all
variants is only slightly below the `perfect' scaling obtained by
dividing the single processor time by the number of processors. Raw
performance is dramatically different, with only the cache blocking
optimization providing performance that is largely independent of page
size. The discrepancy between scaled and raw performance is important
to note because while better scaled performance is desirable, it is
actually better raw performance that increases scientific
productivity. A code that scales linearly to a large processor count
may in fact perform less well than another that does not scale as
well, but whose speed in an absolute sense is faster. The
optimizations described here increase both the raw and scaled
performance of the code.

It is also important to note that for the fully optimized code, only a
comparatively small 14\% performance deficit for small pages relative
to large pages remains. This is important because large hardware pages
on most systems are difficult or impossible to use effectively in
production environments. Without cache blocking and on systems without
adequate support for large pages, rates would otherwise suffer
dramatically and simulations will proceed far more slowly. The cache
blocking optimization thus ensures a better level of tolerance to
limitations of the hardware itself. We believe the differences that
remain are unavoidable and a consequence of the fact that, while the
force summation can be cache blocked, the tree traversal itself cannot
be. Because the tree traversals sample a very large memory space very
sparsely, TLB entries must be discarded and recalculated much more
frequently when smaller page sizes are used, resulting in slower
execution times. The remaining rate difference is small because the
traversals require only 10-20\% of the total time required for the
gravity calculation itself. 

The results from our analysis of the maximum number of particles per
clump in section \ref{sec:best-clumpsize}, demonstrate that
calculation rates remain significantly higher even when calculations
for only a few particles are performed. The cost of reading the node
data into the cache blocked array and then re-reading it multiple
times is both small and can be amortized effectively over the
calculations of only a few particles. Cache blocking will therefore
remain effective even when only a small fraction of the particles in a
clump require updates, as will often be the case when an individual
time scheme is used. Speed differences due to page size effects will
become somewhat more pronounced however because the ratio of time
spent in the traversal itself, compared to the multipole summation,
becomes less strongly weighted in favor of the summation. In the
extreme case, when only a single particle requires a calculation, we
expect the scaling to be similar to the curves shown for the level
four optimization in figure \ref{fig:grav-scaling}. Since most of the
time in any calculation is spent evolving a much larger fraction of
the particles than this, the benefits of the cache blocking
optimization will remain substantial.

\subsubsection{Performance of the SPH calculation and additional
benefits of particle reordering}\label{sec:perform-sph-reo}

\begin{figure*}[!t]
\rotatebox{-90}{\includegraphics[width=130mm]{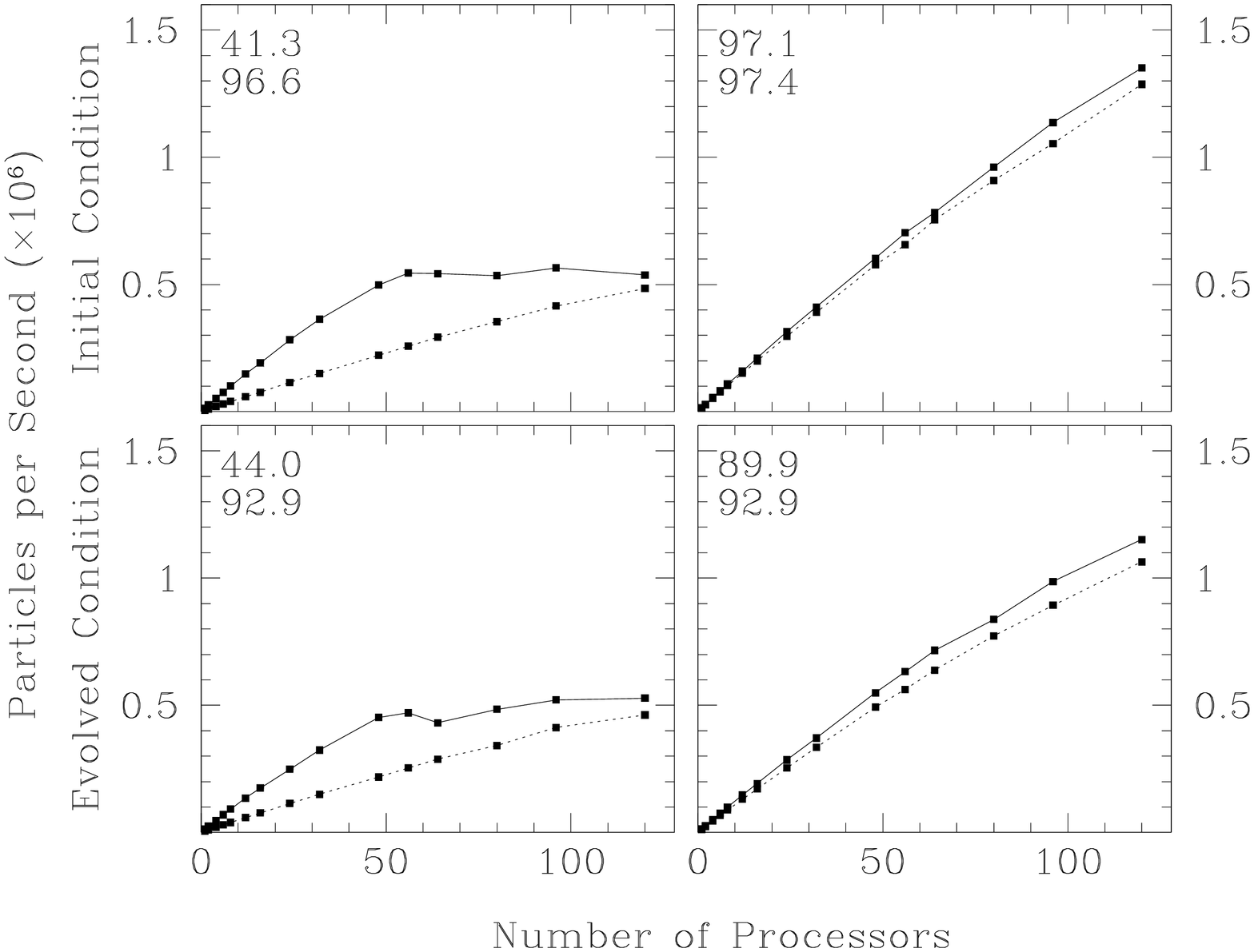}}
\caption{ \label{fig:sph-scaling}
Calculation rate for a single SPH calculation for the SPH test problem
initial condition (top panels) and the evolved condition (bottom
panels), each before (left panels) and after (right panels) reordering
the particles locations in memory. The solid lines indicate results
obtained with 16MB page sizes, and the dotted lines results obtained
with 64kB page sizes. Numbers in the upper left corner of each panel
indicate the scaled parallel speedup for the realizations with 16MB (top)
and 64k (bottom) page sizes plotted in the panel, respectively, out of
a maximum of 120 for linear speedup.}
\end{figure*}

In SPH simulations, the cost of the hydrodynamic calculations is
second only to the cost of the gravity calculation and therefore also
benefits from parallelization and optimization. Here we examine the
speed of a single calculation of all hydrodynamic quantities using the
initial and evolved versions of the SPH test problem, and the
improvements that can be obtained by the particle reordering
optimization discussed in section \ref{sec:tree-post}. While we expect
that, overall, similar speedups in the SPH calculations will be
derived from the various memory layout optimizations described in
section \ref{sec:perform-grav}, we have not tested these specifically.
Instead, we examine only one additional optimization not tested
before--reordering the particle data themselves. We also limit our
tests to only the 64kB and 16MB page size variants.

In figure \ref{fig:sph-scaling}, we show the speed of a single
calculation of the SPH quantities for both the initial and evolved SPH
test problem, before and after reordering the particle data according
to the second reordering variant in section \ref{sec:tree-post}.
Overall scaling of the fully optimized versions are excellent, though
slightly below those of the gravity calculation, with speedups of
nearly a factor between 90 and 100 out of a theoretical maximum of
120. The raw speeds of the SPH calculations, at $\sim13900$ particles
per second per processor for the initial condition and $\sim12900$ for
the evolved condition, are just under twice those of gravity (figure
\ref{fig:grav-scaling}). The raw and scaled performance of the evolved
condition falls below that of the initial condition by $\sim10$\%, but
we are encouraged that performance does not degrade more noticeably in
either case, a possibility not otherwise neglectable if the particle
distribution had played an important role in the scaling. 

The left hand panels of figure \ref{fig:sph-scaling} show performance
prior to the reordering. As for the gravity calculation, the rates are
highly sensitive to page size, with a performance enhancement of a
factor $\sim2.5$ between 64kB and 16MB pages for both initial and
evolved conditions. Further performance enhancements of $10$\% occur
for the large page variants when the particle data are reordered, over
and above that obtained from the reordering already done in the tree
itself, and enhancements of 20\% or more were common in tests
performed on other architectures (not shown here). While scaled
performance using small pages is good, raw performance falls
significantly below the large page variants, even at high processor
counts where performance of the large page variants has saturated.
Reordering the particle data removes all signs of saturation and both
large and small page variants run with nearly equal performance,
demonstrating the effectiveness of the reordering at insulating the
simulation from limitations of the hardware. 

\subsubsection{Performance of the tree build and
revision}\label{sec:perform-build}

\begin{figure}[!t]
\includegraphics[width=90mm]{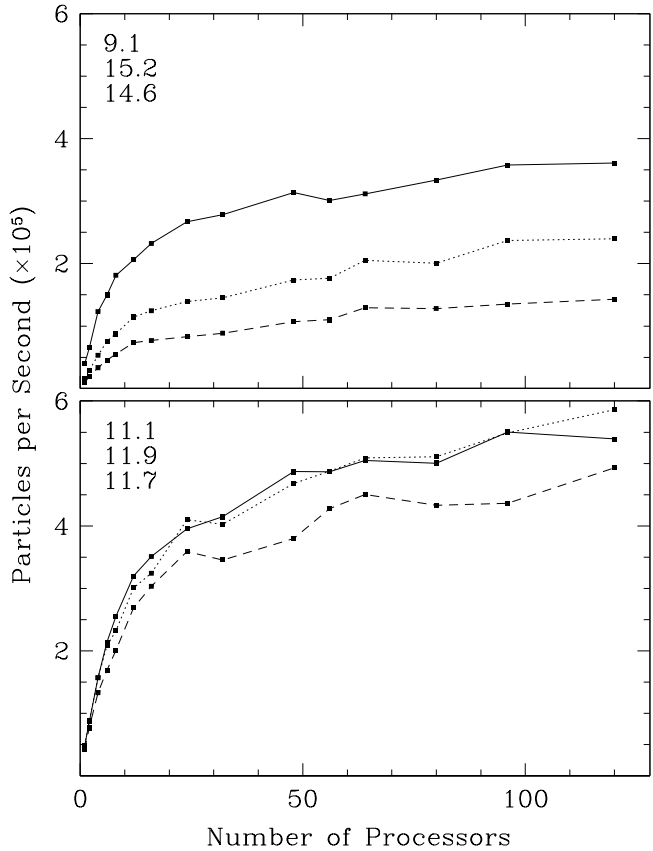}
\caption{ \label{fig:build-scaling}
The speed of a single tree rebuild, before (top panel) and after
(bottom panel) reordering the particle data in memory, as a function
of number of processors. The three curves correspond to the build done
with 16MB pages (solid), 1MB pages (dotted) and 64kB pages (dashed).
Top to bottom, the numbers in the upper left refer to the scaled
parallel speedup for the  16MB, 1MB and 64kB page sizes, respectively,
out of a maximum of 120 for linear speedup.}
\end{figure}

\begin{figure}[!t]
\rotatebox{-90}{\includegraphics[width=65mm]{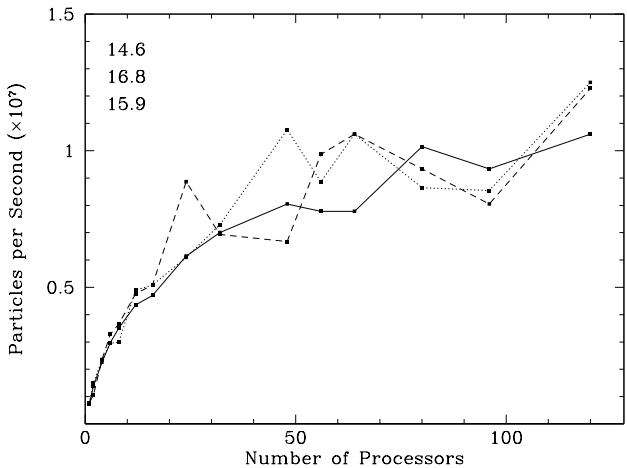}}
\caption{ \label{fig:revise-scaling}
The speed of a single tree revision.}
\end{figure}

Constructing and revising the tree together make up the third most
costly component of particle simulations with VINE. In figure
\ref{fig:build-scaling}, we show the speed of a complete tree rebuild
for the $N$-body test problem, before and after reordering the
particle data in memory, according to the discussion in section
\ref{sec:tree-post}. In figure \ref{fig:revise-scaling} we show the
speed of a tree revision.

On a single processor, the tree build proceeds at rate equivalent to
$5\times10^4$ particles per second with all optimizations, while the
revision proceeds at a rate equivalent to $7.3\times10^5$ particles
per second. These rates correspond to $\sim7$ and $\sim100$ times
faster than one gravitational force calculation for the same particle
distribution. As for the gravity and SPH calculations, parallel
scaling is better with small hardware pages, and raw speed is better
with large pages. Before data reordering, the large page (16MB)
version runs a factor $\sim3$ faster than with small (64kB) pages.
Also as with gravity and SPH, reordering the particle data improves
the raw performance of the tree build, in this case by a factor of
two, due both to better cache and TLB reuse. After reordering, the
speed is substantially less sensitive to page size, indicating that
most of the benefits of large pages can be obtained even on
architectures where they are unavailable, simply by reordering the
particle data in memory.

While both build and revision clearly benefit from parallel operation,
neither yields the linear scaling with processor count seen for the
gravity and SPH calculations. Instead, their scaled parallel speedups
saturate at factor of $\sim11$ and $\sim16$, respectively, on 120
processors. Also, though not shown here, we note that highly clustered
particle configurations can degrade both the absolute performance of
the build and its scalability to some extent. In the case of the SPH
evolved test problem, in which mass density varied by several orders
of magnitude in comparatively `smooth' regions and up to $\sim10^{12}$
in regions where stars had begun to form, we measured performance loss
to be some 20-40\% (depending strongly on the architecture tested)
over the initial condition. Although we cannot be satisfied by either
the saturation or the sensitivity to particle configuration, we are at
least comforted by the fact that performance does not actually
decrease, a very real possibility for parallel codes when
communication costs become significant. 

Despite similar appearances, the factors limiting parallelism in each
case are not identical. For large processor counts, the ultimate
limiting factor affecting both routines is that latency between a
given processor and the data on which it operates is a function of
data placement. It will be both different for each processor and
intrinsic to the NUMA fabric of the system itself, so that even when
ostensibly equal amounts of work are given to each processor, load
imbalance develops due to differences between the times required to
load and store data to and from memory. We believe that data placement
issues such as these are also the source of the scatter in scaling
seen especially in figure \ref{fig:revise-scaling} for processor
counts $>24$, as data distribution among the processors and memory
becomes more dispersed and placements with some processor counts are
more beneficial than others. Although we have performed no systematic
study to verify this hypothesis, in a few, exploratory tests we have
noted significant timing differences between identical runs performed
on different sets of processors spread across different nodes in the
NUMA fabric. Such differences lend support to our conclusion.

Tree construction suffers from two additional limitations which,
together with NUMA, contribute roughly equal proportions to the
overall performance degradation. First, we estimate that $\sim1-2$\%
of the tree build cost remains serialized due to unavoidable
dependencies between different units of work. Second, as tree
construction proceeds to higher levels, fewer and fewer nodes remain
to be shared among processors so that they become progressively more
starved for work, generating relatively large load imbalances,
synchronization and communication costs. These costs become especially
significant for highly inhomogeneous particle distributions, as in the
case of the evolved SPH test problem, because fewer nodes are created
per level, especially near the root. We believe that although a few
improvements could be made to mitigate the effects of these
limitations somewhat, they will never be entirely removable. Section
\ref{sec:perform-size} demonstrates however, that their effects will
become less and less important for larger simulations because a
comparatively smaller fraction of the build time is spent at the
highest levels of the tree, where work starvation is most severe. 

\subsubsection{Frequency of Tree Rebuilds}\label{sec:freq-rebuild}

Combining the information in figures \ref{fig:grav-speedups},
\ref{fig:build-scaling} and \ref{fig:revise-scaling}, we calculate
that a tree build or revision will be, respectively, factors of seven
and 100 faster than a single gravity calculation on one processor, and
so will contribute little to the total simulation time in serial and
small scale parallel operation. Their contributions will become more
significant in highly parallel simulations and when only a fraction of
the particles require force updates, as when individual time steps are
used.

As noted in section \ref{sec:tree-update}, the tree does not need to
be rebuilt after every time step but can instead be used for several in
succession with only an update of the node data to the current time,
saving computing resources better spent directly evolving the
particles. Of course, rebuilds must still occur occasionally because
particles move with respect to one another, causing their parent nodes
to become larger and larger. Larger nodes affect the speed of both the
SPH and gravity calculations because more and more nodes must be
examined for acceptability, and more and more nodes must be included
in the multipole summations. In this section, we attempt to determine
the optimal frequency of tree rebuilds.

\begin{figure}[!t]
\rotatebox{-90}{\includegraphics[width=65mm]{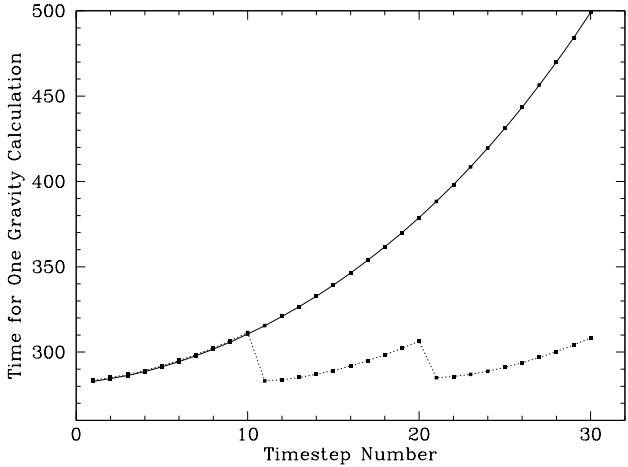}}
\caption{ \label{fig:norebuild-speed}
The time required for a series of time steps of the $N$-body test
problem, for which the gravitational force calculation entirely
dominates the total. The solid curve indicates the times required for
each of 30 time steps with no tree rebuilds, while the dotted curve
indicates the times required with a complete tree build spaced every
ten time steps.}
\end{figure}

When the code is used in global time step mode, the frequency for
rebuilds can be determined easily based on the time required to update
the system immediately after a tree build as compared to the time
required after some number of time steps have elapsed. We illustrate
this slowdown in figure \ref{fig:norebuild-speed}, where the time for
successive force calculations is plotted as a function of the number
of time steps since a complete rebuild, during the evolution of the
$N$-body test problem with global time steps. Over the course of 30
time steps, the time required increases ever more rapidly as particles
and nodes become more and more separated from their initial nearest
neighbor status. Enforcing a tree rebuild every ten time steps returns
the calculation times again to their original high efficiency. Since a
full reconstruction of the tree requires $\sim30$s on 8 processors of
the Origin~3000 on which this test was performed, and the time for one
update increases by an approximately equal amount over 10 time steps,
a reasonable spacing between rebuilds is once per $\sim5-10$ time
steps.

The question of how often to rebuild the tree is not trivial when the
code is used with individual time steps because it is difficult to
determine--on a particle by particle basis--when a calculation
requires sufficiently more time that the cost of a tree rebuild would
be less than that of an update. It would also be beneficial to have
some process by which the user would not be required to tune the
rebuild frequency for a given problem in a global time step
calculation, but rather to have a single parameter by which the
rebuild/revise decision could be made automatically. Since force
calculations using the tree will be sensitive to the physical size of
the nodes in the tree through the node opening criteria, one such
parameter can be defined using the increasing physical size of clumps,
as defined in equation \ref{eq:treenode-size-exact}, as a proxy for
the increasing size of all of the tree nodes. This value is required
for other calculations and will therefore allow the code to make a
decision to rebuild the tree at essentially no additional cost.

\begin{figure}[!t]
\rotatebox{-90}
    {\includegraphics[width=65mm]{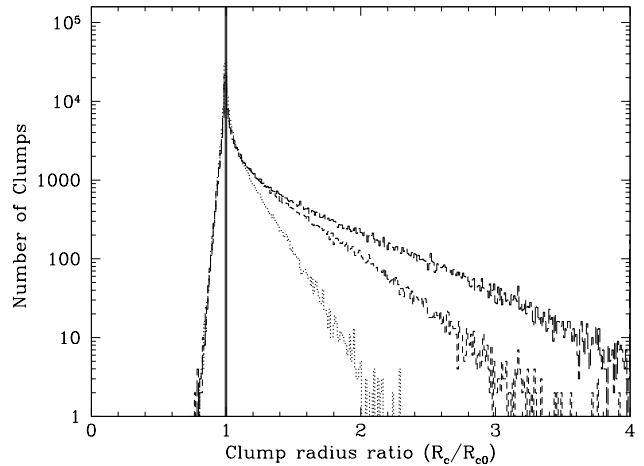}}
\caption{\label{fig:rebuild-clumpsize}
A histogram of the ratios of the physical radius of a clump several
time steps after the tree build, to the value immediately after the build.
The solid curve shows the ratios at time step zero (i.e. all ratios
unity), while the dotted, dashed and long dashed curves show
the distribution after 10, 20 and 30 time steps, respectively.} 
\end{figure}

In figure \ref{fig:rebuild-clumpsize}, we show the clump size
distribution, relative to their original sizes, after a number of time
steps have elapsed since the rebuild, for the same calculation shown
in figure \ref{fig:norebuild-speed}. After 10 time steps have elapsed,
some clumps have expanded to as much as twice their original radius,
after 20 steps, well over three times their original radius, and after
30 steps, over four times their original radius. Decreases in size
also develop, in this problem to about 80\% of the original size. The
ratios are near log-linear in distribution, so that specifying a given
maximum ratio will be an effective proxy for specifying the full
distribution. Noting from figure \ref{fig:norebuild-speed} that an
increase in computation time per time step of 30 seconds (about that of
one tree build), corresponds to a maximum increase in clump radius of
a factor of two, we can conclude that triggering a rebuild when the
size of any clump changes its radius by a factor of two, we will be
able to automatically retain most efficient calculations of the
forces. Experiments on several particle morphologies indicate that
this factor is comparatively insensitive to particle distribution,
though still other distributions may prove more so, and may require
another factor be used. Users may change its value through a setting
in a text input file, read in by the code at run time.

\subsubsection{Performance on problems of different
size}\label{sec:perform-size}

So far, we have demonstrated the parallel scalability of the code on
problems of moderate size, at times when all particles require force
calculations. In practice of course, the code will be used differently
in two important respects. Simulations will be run with different
sizes than we consider and they will be run using an individual time
step scheme, so that only a small fraction of the total number of
particles require updates at any given time. Due to the
unpredictability of load balance and work distribution among time step
bins, it will not be possible to address the latter concern with any
generality. We can still gain some insight into the scalability by
looking at identical problems of different size however. In this
section, we examine the sensitivity of the overall performance and the
parallel scaling of the code using six realizations of the galaxy
merger test problem at different resolution, and which include both
SPH and $N$-body particles, as defined in table \ref{tab:testsims}.

\begin{figure}[!t]
\rotatebox{-90}
    {\includegraphics[width=65mm]{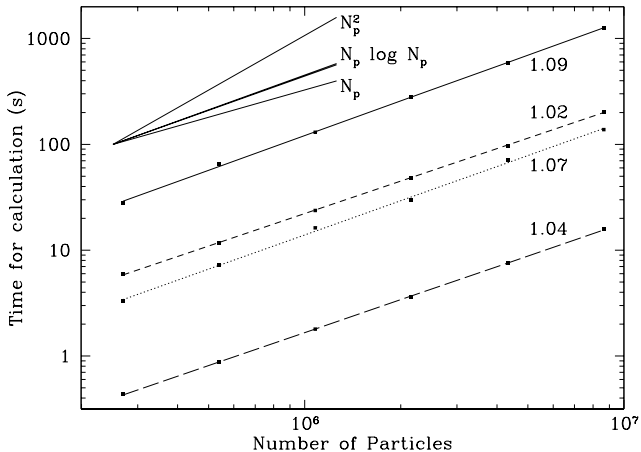}}
\caption{\label{fig:size-single}
Time required to perform one gravity calculation (solid), one SPH
calculation (dotted), one tree build (dashed) or tree revision (long
dashed), for an identical initial condition realized with six
different particle numbers. Points represent actual timings, while the
lines represent a linear fits to the logarithms of each coordinate.
The scaling (i.e. the linear term in each fit) is indicated adjacent
to each. Lines in the upper left corner represent scalings
proportional to $N_p^2$, $N_p \log N_p$, $N_p^{1.09}$ and $N_p$, as
indicated. Only three lines are visible because the curves for the
$N_p \log N_p$ and $N_p^{1.09}$ proportionalities over lie each other.}
\end{figure}

Figure \ref{fig:size-single} shows the time required to perform one
complete calculation of each of the major components of the code on
one processor of the Origin~3000. As expected from theoretical
considerations of the tree algorithm itself, proportionalities for the
gravity and SPH calculations are steepest, with the slope for gravity
corresponding exactly to the value expected over this range for a
curve with $t_{\rm grav} \propto N_p\log N_p$. The proportionality for
the SPH calculations is lower because, while the tree traversals
required for the neighbor searches scale with $N_p\log N_p$,
calculations with the resulting lists of neighbors scale only
proportional to $N_p$.   Both the tree build and revision follow still
shallower proportionalities, so that as problem size increases, their
costs decrease relative to gravity and SPH. As seen for the $N$-body
and SPH tests above, the absolute cost of the gravity calculation is
higher than any other part of the code, but because only $\sim14$\% of
the total particles in the six Merger tests are SPH particles, the
tree build replaces SPH as the second most costly calculation, with
the tree revision an additional order of magnitude cheaper than
either. The excellent scaling of each component over a factor of 30 in
size indicate that simulations of any size with VINE will be possible
with both predictable and affordable cost. 

\begin{figure*}[!t]
\rotatebox{-90}
    {\includegraphics[width=130mm]{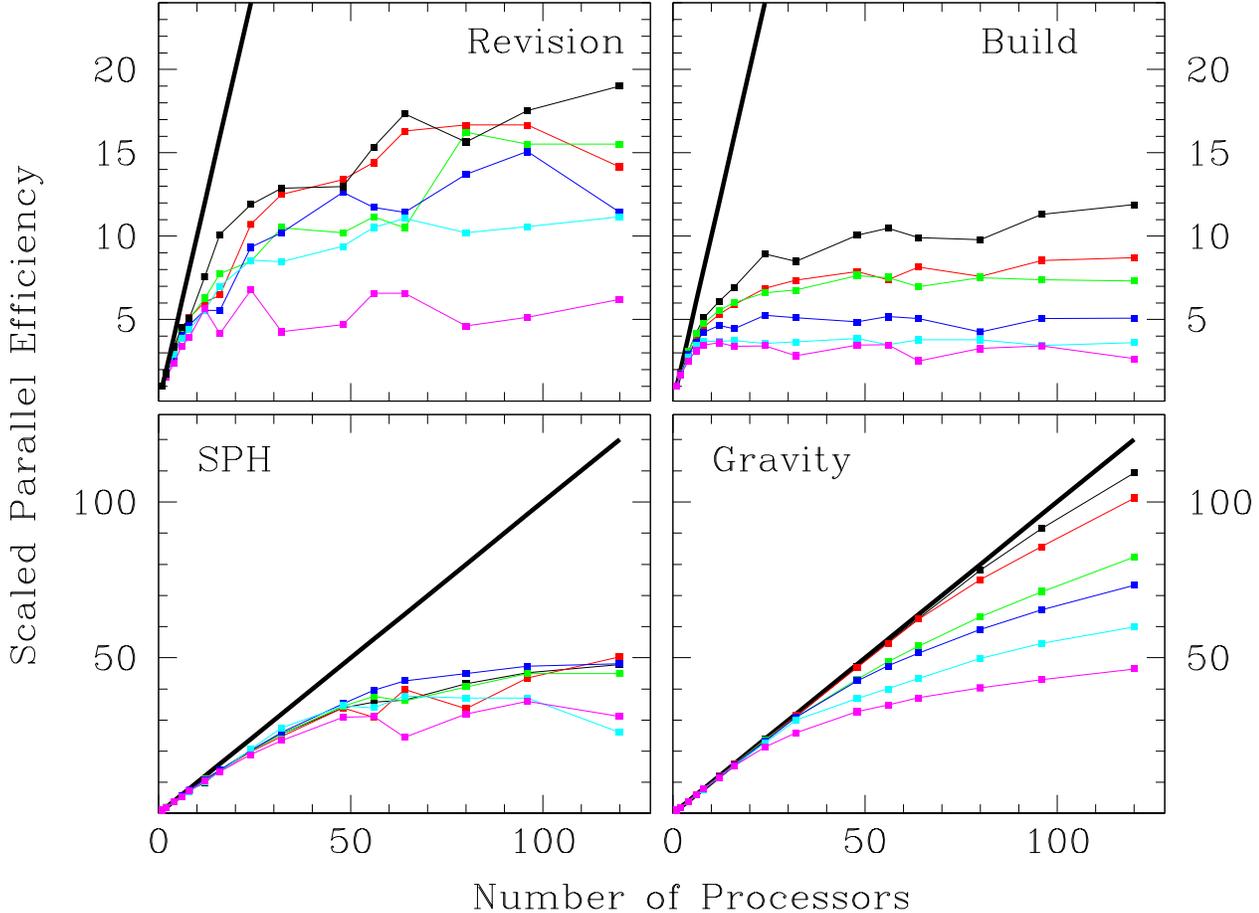}}
\caption{\label{fig:size-parallel}
Parallel scaling of the same calculations shown in figure
\ref{fig:size-single} for realizations at each resolution. The heavy
solid line in each panel delimits perfect linear scaling. In order
from high to low resolution, each curve corresponds to the color
black, red, green dark blue, light blue and magenta. } 
\end{figure*}

Figure \ref{fig:size-parallel} shows the scaled parallel efficiencies
of each code component for each of the realizations at different
resolution. Even at comparatively low resolution, the efficiency of
the gravity calculation is excellent. Only small deviations from
linear scaling are visible for the low resolution (270000 particles)
run up to $\sim32$ processors. Above that count the speedup continues
to increase to the limits of the machine, but no longer follows a
perfect linear relationship, ending in a $\times50$ improvement on 120
processors. Each of the higher resolution variants follows the same
pattern, with the performance knee shifted to progressively higher
processor counts. For the highest resolution version, a small
deviation from linear performance becomes visible only above $\sim80$
processors, dropping to 110 at the 120 processor maximum for the
machine.

The curves for the SPH calculations also show very good scaling up to
$\sim32$ processors though, as for the SPH test problems above, below
that achieved in the gravity calculation. Above 32 processors however,
rather than continuing to improve up to the limits of the machine, all
of the curves instead turn over and saturate near a factor $\sim50$.
Also in contrast to the gravity calculation, relatively little
differentiation between the scalings at different resolutions is
visible. The plateau appears to be due in part to the fact that only
$\sim15$\% of the total number of particles are SPH particles and to
the small total number of SPH particles in these simulations. Both
characteristics increase the likelihood for load imbalance at high
processor counts. The scaling limits are important to quantify
because, in some respects, the fact that SPH particles are distributed
irregularly throughout the tree mimics the behavior of the code when
simulations using individual time steps for each particle are run, and
active particles are distributed irregularly throughout the tree.

For both construction and revision scaling is good up to $\sim 5-8$
processors at all resolutions, but saturates at higher levels of
parallelism, as we found in our tests in section
\ref{sec:perform-build}. Above eight processors, the curves for
different resolutions become differentiated from one another. The
curves for the tree build are relatively smooth functions of processor
count, and are well distinguished from both their higher and lower
resolution cousins. Performance of the lowest resolution variant does
not increase beyond a factor $\sim3-4$ speedup, but the saturation
level increases in progressively higher resolution variants,
eventually to a speedup of $\sim12$ for the highest resolution case.
For still higher resolution simulations, we expect this trend to
continue because a relatively smaller fraction of the total will be
spent associating nodes near the root of the tree, where the total
number of unassociated nodes is small and parallelism ineffective.

The revision scalings display much more irregular patterns. For our
highest resolution variant, scaling increases steeply to $\sim 16$
processors, then more shallowly to a factor $\sim 20$ at the high end
limit of our study. Lower resolution variants fall off at
progressively fewer processors, but the curves sometimes intersect
each other and, in some cases, actually decrease slightly before
increasing again. We attribute the irregularity to the fact that the
revision requires only a very small amount of total time (typically
much less than a second, especially for the lower resolution tests),
so that sensitivities to systematic effects such as data placement in
the NUMA hierarchy play a proportionately larger role. Attempts to
repeat a subset of the timings appear to confirm this hypothesis, as
timings for different runs could differ by as much as several tens of
percent in otherwise identical runs, done at different times on
different distributions of processors.

Overall, these tests demonstrate that the code's efficiency scales
very well both to very large and to very small problem sizes. For
small problems, performance may no longer increase linearly beyond
$\sim30-50$ processors, but in no case do we find the overall scaling
to decrease with additional processors, even for very low resolution
simulations where even small load imbalance and interprocessor
communication can become significant. This is important for
simulations utilizing individual time steps for each particle because
only a fraction of the total complement of particles will require
updated force calculations at any given time. 

\subsection{Performance on other architectures}\label{sec:perform-arch}

The Origin 3000 architecture on which the tests in sections
\ref{sec:tuningI} and \ref{sec:tuningII} were performed is only one of
a number of shared memory systems on which VINE can be run. It is
also, at this writing, a retired architecture for which no succeeding
system exists using the same processor, though the Altix machine plays
a similar role using Intel Itanium~2 processors and a somewhat similar
interconnect fabric to connect processors and memory. Other large
scale, shared memory parallel systems are commercially available from
IBM, Sun and HP. Desktop machines with multiple processors have also
become common and we can only expect this trend to continue to larger
scales in the future as machines with more processors, and processors
with more cores per chip, become commonplace.

In this section we compare the performance of the code on a several
hardware architectures and compilers, referenced in table
\ref{tab:arch}. Because many of these machines are from different
computer `generations', a direct comparison of one processor type
against another will not be particularly meaningful as an indicator of
some overall best architecture or processor, and readers are cautioned
to be mindful of this fact in making such comparisons. On the other
hand, correlated with user experience of speedups expected between one
generation and the next within a single processor family, some
inferences may be justified. In any case, the comparisons will be
useful as a relative indicator of how the code performs on exactly the
same problem across a wide variety of machines, possibly assisting
heavy users of the code in the selection of one machine over another
to which to commit resources.

\singlespace
\begin{deluxetable*}{lllr}
\tablewidth{0pt}
\tablecaption{\label{tab:arch} Computer Architectures Tested}
\tablehead{
\colhead{Machine}  & \colhead{CPU type} & \colhead{Compiler } &
\colhead{Identifier} }

\startdata
SGI Origin 3800 &            128$\times$ R12000 (400MHz)    & Mipspro 7.4.2   & 1 \\
IBM p690        &  \phantom{0}32$\times$ Power~4 (1.3GHz)   & xlf 9.1         & 2 \\
AMD Opteron     &  \phantom{00}2$\times$ Opteron (2.4GHz)   & Pathscale 2.3.1 & 3 \\
IBM p575        &  \phantom{0}16$\times$ Power~5 (1.65GHz)  & xlf 9.1         & 4 \\
SGI Altix 350   &  \phantom{0}56$\times$ Itanium~2 (1.5GHz) & Intel 9.1.33    & 5 \\
SGI Altix 350   &   (GRAPE-6A related tests)                & Intel 10.0.25   &   \\
\enddata

\end{deluxetable*} 

\doublespace

\subsubsection{Serial performance}\label{sec:arch-compar-ser}

\begin{figure}[!t]
\includegraphics[width=85mm]{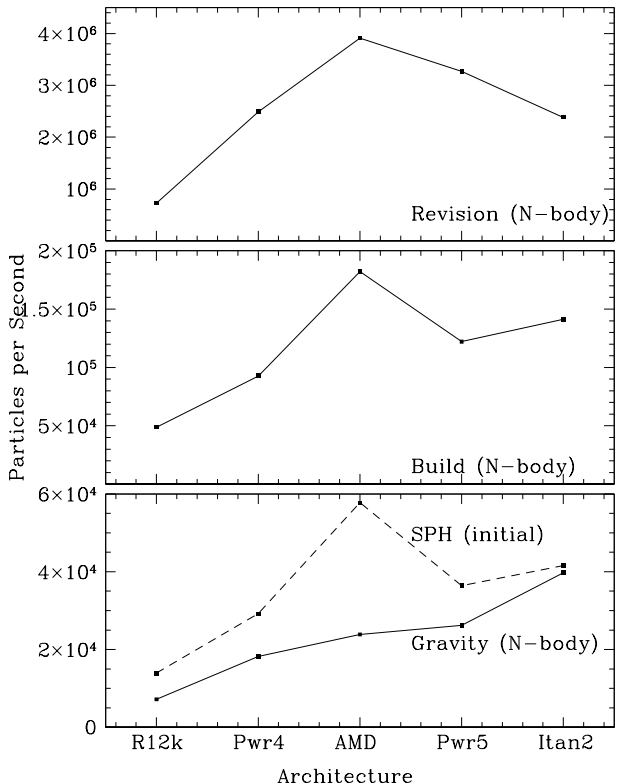}
\caption{\label{fig:arch-single}
Calculation rates for the gravity, tree build and revision
calculations on the $N$-body test problem, and the SPH initial
condition, running on a single processor of the specified type. Each
plot is ordered from left to right according to increasing performance
of the gravity calculation. The SPH evolved condition (not shown)
follows the same pattern as does the initial condition, at an overall
speed $\sim10$\% slower than the initial condition. }
\end{figure}

In figure \ref{fig:arch-single}, we show calculations rates for single
processor tests, on five different processors for each major component
of the code. For the gravity calculation, the clear `winner' is the
Intel Itanium~2 processor, with calculation rates of a factor 50\%
faster than its nearest competitor, the Power~5. Its performance is
also a factor $\sim5.5$ faster than the R12000 processor on which the
detailed performance analyses above were made. For all other
components, performance of the AMD~Opteron is highest, at a factor
$\sim$1.4 faster than its nearest competitor (Itanium) for SPH and
$\sim4.1$ times faster than the R12000. Comparable figures apply to
the tree build and revision as well.

A number of features of figure \ref{fig:arch-single} are of interest.
First, the superior performance of the Opteron is, to some extent, to
be expected since it is more recently made than any other processor in
our sample. The fact that its performance on the gravity calculation
falls between that of the two Power processors is a significant
drawback to its performance overall because of its large total cost
for most simulations of interest in astrophysical contexts. Also,
although its serial performance is very good, no machines are
currently commercially available which exploit the high performance in
large scale shared memory configurations, so the higher serial
performance may only be useful for comparatively small simulations,
where highly parallel operation is not required. 

The raw performance differences between the Power~4 and Power~5
processors are remarkably small, corresponding roughly to differences
in clock speeds between the two processors. In comparison, differences
between their performance on the well known
SPEC~CPU2000\footnote{http://www.spec.org} benchmark suite would lead
one to believe that improvements nearer a factor two or more were to
be expected. Also, their performance is disappointing on the SPH and
tree calculations relative to the much older R12000 processor, at
factors of only $\times2$ and $\times2.5$ faster, respectively, though
the gravity calculations do somewhat better at factors of $\times2.5$
and $\times3.6$.

When we compare the rates derived from the merger test simulations
done in \vineI\  with those shown in figure \ref{fig:arch-single}, we
notice an important discrepancy. The rates for the merger test in
\vineI, of $\sim70$~kparticles/s for 8 Itanium~2 processors, are far
lower than would be expected from a simple extrapolation of the
$\sim40$~kparticles/s for one Itanium~2 processor shown here. The rate
differences appear to be due to two effects. First, the mass
distributions are not the same. While the $N$-body test problem a
spherically symmetric system with $r^{-1/4}$ density profile, the
merger simulation consists of two more or less separated galaxies at
various times during their merger evolution. Second, and more
importantly, the merger simulation consists of several types of
particles, each with their own masses and gravitational softening
lengths. When larger softening lengths are used, the node size of all
parent nodes increases proportionally (see equations
\ref{eq:treenode-size} and \ref{eq:treenode-size-exact}), and will
therefore result in many additional node examinations during the tree
traversal, and the addition of many additional nodes to the
interaction lists themselves. The rate differences between these
different configurations do not affect the conclusions in \vineI, that
VINE is $\sim4\times$ faster than Gadget-2 however, since the speed
comparisons for that test were done using exactly the same particle
configurations for each code.

As is to be expected given its much greater age relative to the
others, the performance of the R12000 processor falls substantially
behind that of all the other processors in our sample. This is
important because the performance characteristics of VINE discussed
above will be better on architectures commonly available today in
proportion to the speedups seen in this plot. It is of some interest
that the magnitudes of the speed increases are somewhat smaller than
might be expected from a naive application of Moore's
Law.\footnote{Although by its actual definition, Moore's Law refers to
a time scale for doubling the component density per processor, we
apply it here in its commonly misused form as a speed doubling every
18-24 months.} Some deviations from the expected speedups may be due
simply to hardware features that VINE exploits more fully on one
processor family rather than another, or on one workload rather than
another. The fact that some of VINE's calculations speed up more than
others across different processor types illustrates clearly that no
single architecture dominates all aspects of the calculations required
for astrophysical simulations.

\subsubsection{Parallel scaling}\label{sec:arch-compar-par}

\begin{figure*}[!t]
\rotatebox{-90}
    {\includegraphics[width=125mm]{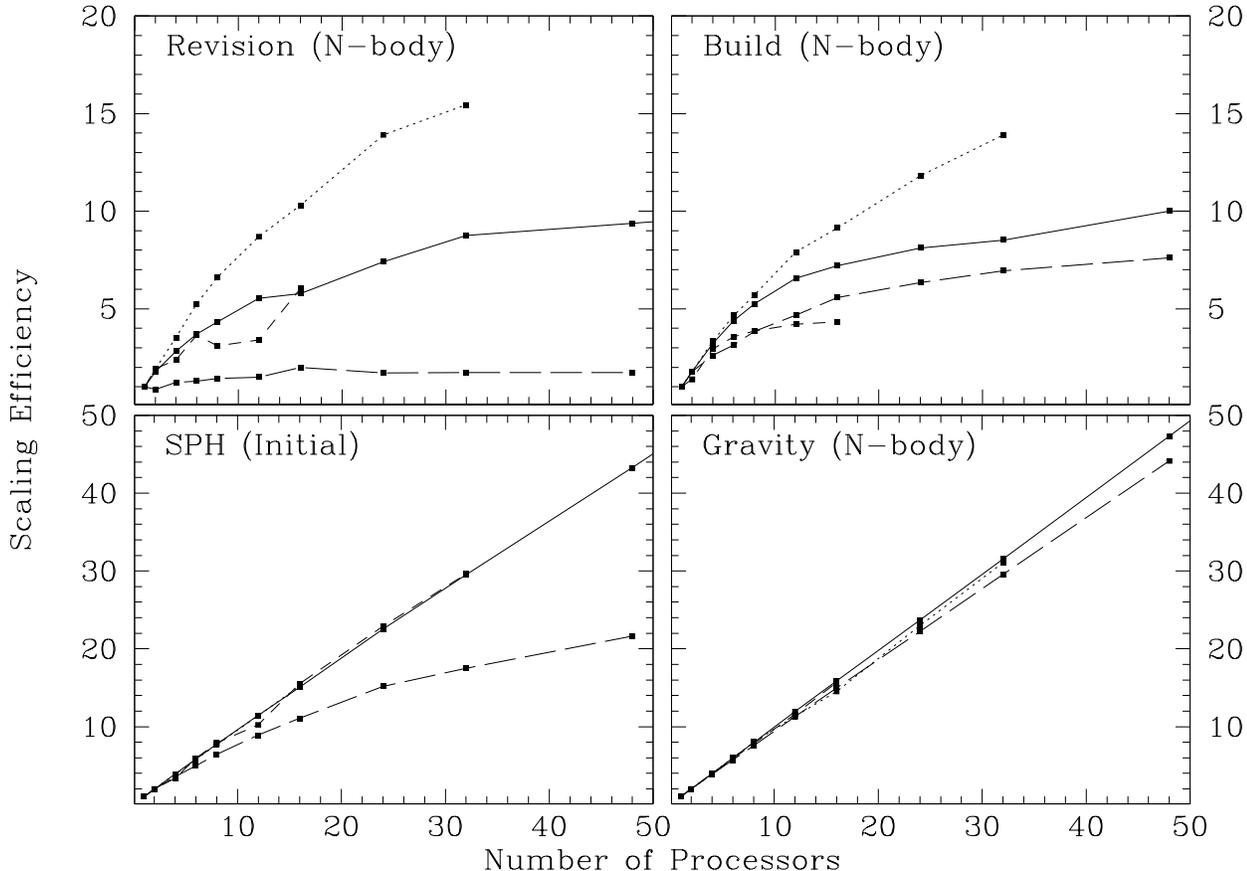}}
\caption{\label{fig:arch-parallel}
Parallel scaling of the major components of the code, for a number of
different architectures. Tests of the SPH evolved condition mirrored
the results of the initial condition and are neglected here. The
curves correspond respectively to the SGI~Origin~3000 (solid),
IBM~p690 (dotted), IBM~p575 (short dashed) and SGI~Altix (long
dashed). Curves for each architecture extend different distances along
the $x$ axis depending on the size of the machine on which the tests
were made.}
\end{figure*}

Figure \ref{fig:arch-parallel} shows scaled parallel efficiencies for
the four major components of the code on several architectures. The
parallel scaling of the gravity calculation is near linear with
processor count for all architectures tested, up to the limit imposed
by the size of the machine itself. Of the four architectures,
performance on the Altix fared comparatively least well, falling to
`only' $\times29.5$ and $\times44$ speedups out of 32 and 48
processors on the two largest tests, respectively. In comparison, the
Origin~3000 and Power~4 based p690 machines gave speedups of 31.5 and
31.1, respectively, when run on 32 processors. 

The scaling of the SPH calculations is also excellent, with near
linear speedup in the SPH calculations on three of the architectures
tested, to the limits of each machine. Performance on the Altix is an
exception however, deviating from linear speedup near $\sim8-16$
processors and eventually dropping to a comparatively poor factor
$\times22$ speedup on 48 processors. Much larger disparities, among
all of the systems in our test, are present in the tree build and
revision procedures. The Power~4 based p690 realizes the best
performance on both, reaching speedups of $\sim9$ and $\sim16$ on 16
or 32 processors, respectively, while the other three realized
speedups between $\times5$ and $\times10$, even when run on more
processors. Parallel scaling of the tree revision on the Altix machine
saturated at an even lower plateau of a factor $\sim$2 speedup, no
matter how many processors were used. 

Of some interest are the differences in scaling between the Power~4
and Power~5 based machines and, to a lesser extent, also between the
Origin and Altix machines, due to the architectural similarities
between each pair. In fact, limited correlation can be seen in the
performance between the two pairs, even though the code and data are
identical. We have not attempted to explore the origin of the
performance disparities, but nevertheless can make some conjectures,
based on the kinds of calculations made in each of the different
procedures. Specifically, that the differences are due to the
complicated interactions between the processors and the memory they
use. Procedures with a small memory footprint, even if they consume
huge amounts of processing time as do the gravity and SPH
calculations, scale consistently well. Procedures that cycle through
large volumes of memory but perform few calculations, as do the tree
build and revision, scale less well and inconsistently across
architectures.

It is of some disappointment that the Altix architecture,
theoretically scalable to higher processor counts than any other, does
not perform as well as the others on the latter workload, since a
non-negligible fraction of VINE's total workload includes such
operations. As an example of the consequences that will result in
practice, we point out the scaling behavior of the SPH calculation.
Examination of its various components on Altix show that the density
and hydrodynamic force calculations, which together account for
$\sim98$\% of the total time on one processor, and which do a large
amount of work using a small volume of memory accessed repeatedly, do
in fact scale linearly with processor count. On the other hand, while
other components, such as the equation of state calculations or the
smoothing length derivative which each cycle through a large volume of
memory performing only a few operations per particle, require only
$\sim2$\% of the total time on one processor. their parallel
performance is quite poor, in some cases actually decreasing with
processor count rather than increasing. Specifically for the
Origin/Altix architecture pair, we may ask the question `can we
attribute the scaling differences between our tests to differences in
memory distribution mechanisms in IRIX and Linux?'

The comparisons in this section neglect several architectures and
processors entirely, largely because we were unable to secure time on
such machines, and because of simple limitations in the amount of
effort required to perform the desired tests. Of the machines in our
comparison, the best overall serial performance for users of VINE will
be obtained on machines using Itanium~2 processors, especially for
simulations of purely $N$-body systems. Due to their excellent
performance in all other components, AMD Opteron processors may also
provide comparable performance in simulations with both self gravity
and hydrodynamics. No tests of the parallel scalability of machines
using the AMD processors were performed because at the time these
tests were performed, highly scalable, shared memory machines using
AMD processors were not yet commonly available\footnote{More recently,
machines incorporating dual and quad core processors in multi-socket
motherboards have become commonly available, though we have not yet
made tests on these machines}. Best overall performance will be
achieved when using machines based on Power processors. 

\subsubsection{Effect of large pages on other architectures, and
difficulties associated with their use}\label{sec:page-arch}

In sections \ref{sec:perform-grav}, we demonstrated the large
sensitivity of the code's speed to the hardware page size on the
Origin 3000 architecture using MIPS R12000 processors. In general, we
cannot expect the same sensitivity across all processor families
because some include circuits to calculate TLB entries directly, while
others, like the R12000, compute new entries in software by the
operating system. In this section, we explore the sensitivity of the
code to page size on the Power~5 processor, in order to demonstrate
that the sensitivity is not specific to a single processor family, and
to demonstrate that the performance optimizations we have made also
provide benefits on other architectures as well. For reasons we
discuss below however, we point out to readers that direct speed
comparisons with results in other sections should not be made, because
the tests were run at a different time and different settings were
used to produce the results. 

\begin{figure*}[!t]
\begin{center}
\rotatebox{-90}
  {\includegraphics[width=125mm]{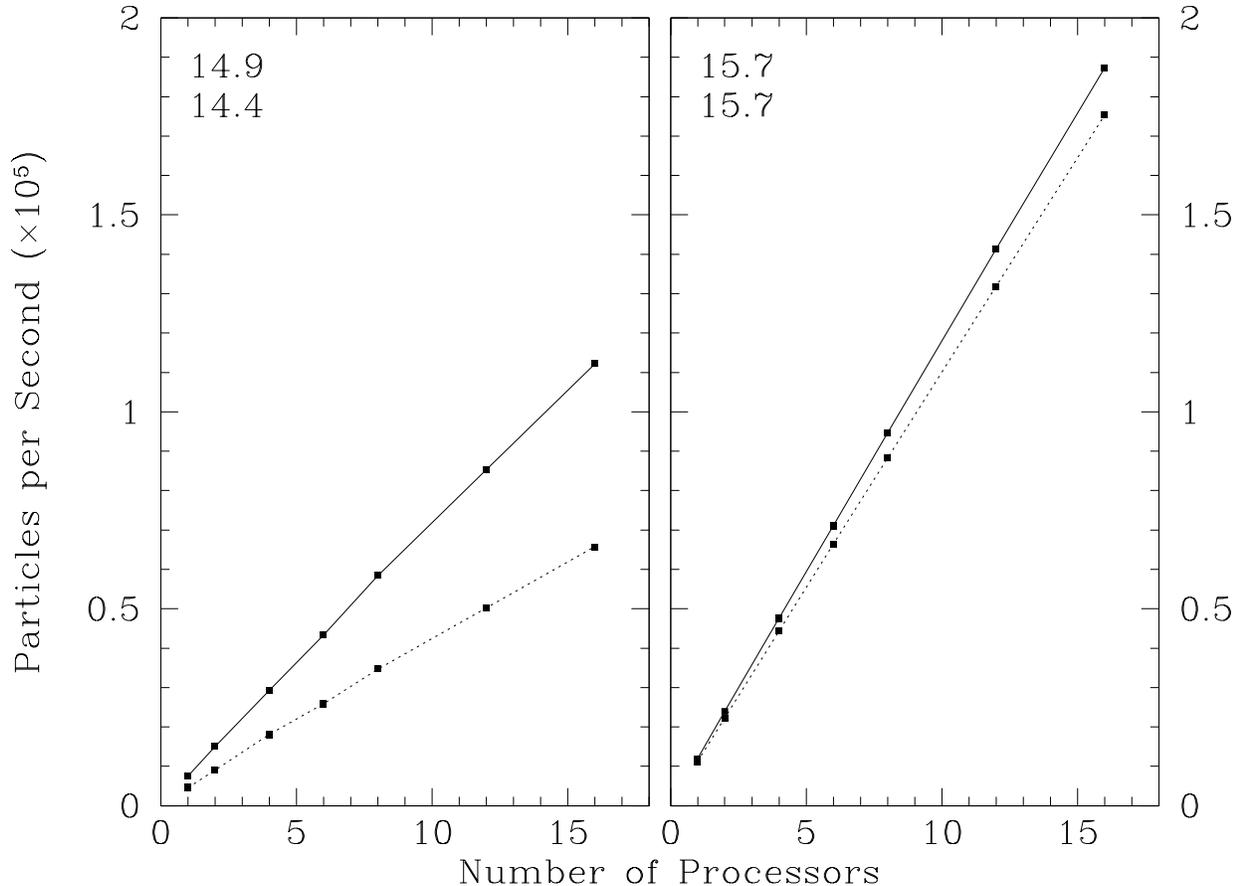}}
\end{center}
\caption{\label{fig:power-pagesize}
Gravitational force calculation rates obtained using VINE on the
Power~5 architecture. As in the lowest two panels of figure
\ref{fig:grav-scaling}, the left panel shows the performance without
the cache blocking algorithm and the right panel shows performance
with cache blocking. The solid and dotted curves correspond to
runs performed with 16MB pages and with 4kB pages respectively.
Scaled parallel efficiency is shown in the upper left corner of each
panel for the 16MB (top) and 4kB (bottom) results, out of a machine
maximum of 16. } 
\end{figure*}

Figure \ref{fig:power-pagesize} shows gravitational force calculation
rates obtained from simulations run using Power~5 processors, as a
function of processor count. As with MIPS, performance of the
non-cache blocked version of the code is greatly enhanced when large
pages are available and are used, in this case by a factor 1.65. The
differences decrease to only $\sim5$\% in the cache blocked versions,
where we expect the effect should be minimized. We conclude that the
sensitivity of the code speed to hardware page size is not specific to
any one processor family, but reflects a general property of the code
itself. The magnitude of the sensitivity will vary somewhat from
family to family, depending on the implementation of the TLB refill
process itself. The benefits seen on the Origin architecture may be
particularly large due to this fact, because MIPS processors require
software intervention for their TLB refills. 

In contrast to the R12000 processor, the difference between non cache
blocked and cache blocked versions with large pages is also
substantial. Cache blocking improved performance by a factor $\sim1.6$
on the Power~5, compared to only $\sim17$\% on the R12000. As we
expect, both processors improve by more substantial margins when small
pages are used, so that the both the large and small page variants
obtain similar performance. We conclude that other sources of memory
latency are much more significant on the Power architecture than on
the Origin 3000, though we have not attempted to isolate the exact
origin more fully. 

Given our conclusions, it is important to note as well that the ease
and flexibility of actually using large pages can vary widely between
architectures, and even between operating systems using the same
architecture (e.g. Linux or AIX on Power processors). On the Origin
3000 architecture, where the sensitivity to page size was initially
seen as the code was developed, use of large pages was straightforward
and we were regularly able to test the effects of various alternative
optimizations on the code speed. Even on this system however,
consistently obtaining the same page distribution from run to run
(when the desired page size is unavailable, the operating system falls
back to other page sizes) was frequently difficult or impossible,
yielding benchmark data contaminated by one or more data with a mixed
distribution of page sizes. To minimize such disruptions in timings
discussed in this work, we monitored the page sizes used in the run
with system tools designed for the purpose, and reran tests where
contamination was unmistakable. 

On other architectures, such as the IBM Power series for example,
special coordination with system administrators was required to make
any tests at all, because of instabilities introduced into the machine
and operating system by the use of large pages. Although we were able
to perform the series of preliminary tests illustrated in figure
\ref{fig:power-pagesize}, large page use in production was disabled on
the machines due to the resulting unacceptably severe instabilities in
the operating system that resulted from their use. For this reason,
later performance tests using identical code settings as were used in
figures \ref{fig:grav-speedups} and \ref{fig:grav-scaling} were not
possible. On still other systems, large pages can only be utilized if
the code is rewritten to take advantage of special memory allocation
calls, such as those implemented in Linux through the `hugetlbfs'
infrastructure. We did not pursue this option, due to its cost both in
programming time and to portability.

Fortunately for users of VINE both on the Origin architecture and on
others where large page use is more restricted, the optimizations
described in this paper demonstrate that a properly implemented code
may realize many of the benefits of large pages, even when only small
pages are available on the machine where the code is run. Performance
will still be less than optimal when the code is used in its
individual time step mode however, because the effectiveness of the
cache blocking optimization will be reduced.

\subsection{Performance of the GRAPE$+$tree option in
VINE}\label{sec:grape-tree-perf}

As we continued to develop VINE, more advanced versions of GRAPE
hardware became available and modules implementing interfaces to them
were added to VINE's code base. In this section, we discuss the
performance of the combined GRAPE$+$tree option in VINE, in which
we use tree traversals to reduce the total number of nodes needed to
determine the force on a given particle, but send those nodes to a
GRAPE co-processor board for calculation rather than using the host
processor itself.

We use several variants of our $N$-body test problem, each at
different resolution, and each using the optimal maximum bunch
population for that problem size as discussed in section
\ref{sec:best-clumpsize}. In addition, we test the speed with a
2$^{21}$ particle homogeneous sphere configuration, which we have
configured to have identical characteristics to the same configuration
discussed in \citet[][hereafter FMK05]{fukushige2005}. The system of
units is chosen such that $M=G=1=-4E$ where $M$ is the total mass, $G$
the gravitational constant and $E$ the total energy. The sphere has a
cut-off radius of $r_{max}=22.8$ and is modeled with $N=2097152$
particles, using Plummer softening with $\epsilon=7.6 \times 10^{-3}$.
As in FMK05, we use a single GRAPE-6A (`Micro-Grape') board, which in
our case was attached to an SGI Altix computer with 1.5~GHz Itanium~2
processors.

All tests have been done on a single processor. Reference
calculations, against which force accuracy are measured, were
performed using either using VINE's direct summation mode on the host,
or alternately with VINE in tree/host mode using a geometric MAC
setting of $\theta=10^{-16}$. Both cases use double precision floating
point values, as is standard in VINE. Test calculations using the
Gadget MAC use the form defined in equation \ref{eq:gadget-mac}, which
provides an estimate of the multipole truncation error at hexadecapole
order. An alternate form, described in \citet{springel_g2}, implements
an opening criterion based on truncation error at quadrupole order,
which presumably makes it more appropriate for calculations involving
GRAPE. We have chosen not to implement this option however in order to
make the tree traversal process as nearly identical as possible
between tests with and without GRAPE. In any case, the two criteria are
closely related through the formula
\begin{equation}\label{eq:quad-hex}
\theta_H = \left({{h}\over{r}}\right)^2 \theta_Q,
\end{equation}
where $\theta_H$ and $\theta_Q$ are the opening parameter for the
hexadecapole and quadrupole based criteria, respectively, $r$ is the
separation, and $h$ is the size of the node under consideration. For
more liberal settings for which performance is highest, nodes openings
will be driven by equation \ref{eq:neigh-mac}, rather than equation
\ref{eq:gadget-mac}, so that $h/r\lesssim 1$ and the criteria become
identical. For more conservative settings where $h/r < 1$, two
parameters become scaled variants of each other. Because we plot
calculation rates against force accuracy, rather than against opening
parameter, differences in the parameter value with one or the other
criterion do not directly affect the comparisons.

\subsubsection{Performances as a function of accuracy and problem
size}\label{sec:grape-tree-facc}

\begin{figure*}[!t]
\rotatebox{-90}{\includegraphics[width=120mm]{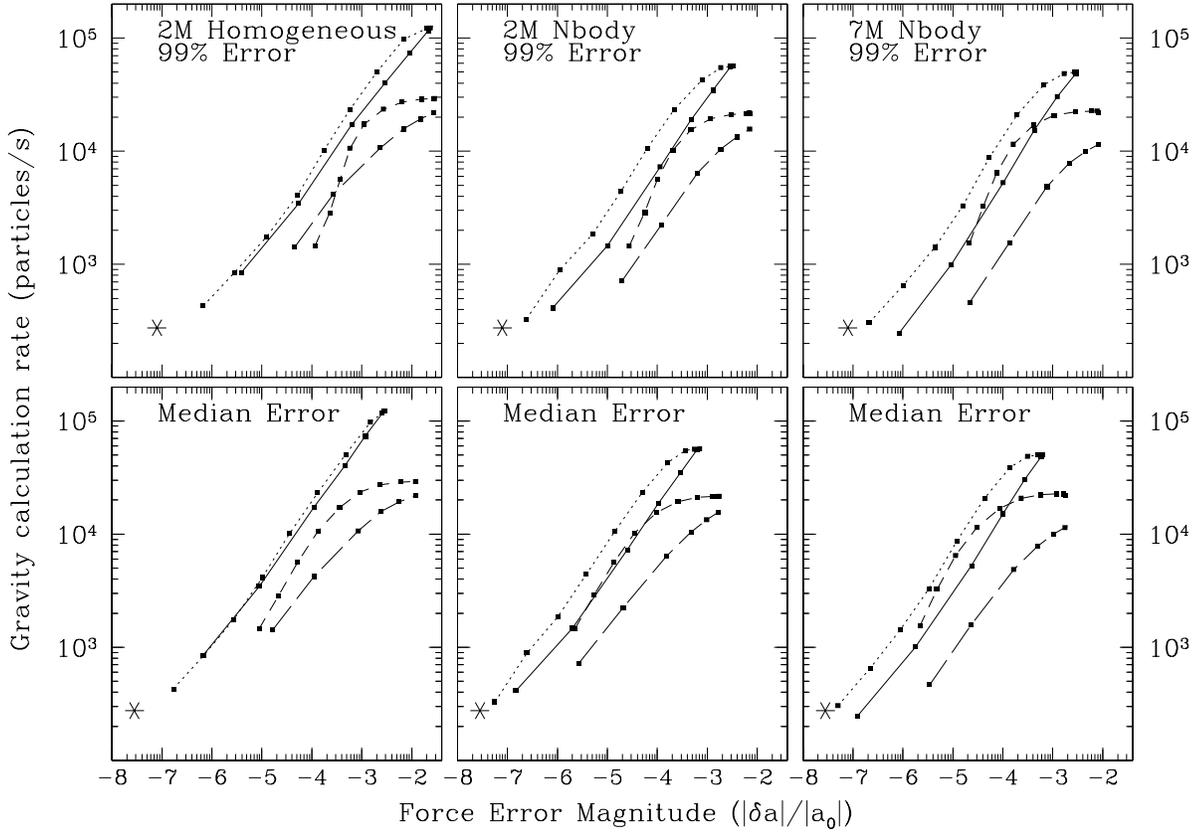}}
\caption{\label{fig:grapetree-perf}
The calculation rate for gravitational forces on particles in a
homogeneous sphere configuration (left panels) and in the 2 and 7
million particle $N$-body test problems (center and right panels,
respectively). Top panels show rates for which 99\% of the particles
have error magnitudes less than that shown, while the bottom panels
show rates at which 50\% (i.e. the median) of particles have errors
less than that shown. The solid and dotted curves represent the rates
using the geometric MAC and Gadget MACs respectively, both using VINE
configured to perform tree$+$host calculations. The long and short
dashed curves represents VINE configured to perform GRAPE-6A$+$tree
calculations using the geometric and Gadget MACs, respectively. Points
on each curve represent settings of $\theta=(0.1,0.2,0.4,0.6,0.8,1.0)$
for the geometric MAC tests and $10^{-9} \le \theta \le 1.0$ in decade
increments for the two calculations with the Gadget MAC. The single
asterisk in each panel shows the speed and accuracy locus of the
GRAPE-6A running in direct summation mode. Labels on the $x$ axis
define the exponent of the force accuracy only, rather than the full
numerical value, in order to avoid confusion.} 
\end{figure*}

Figure \ref{fig:grapetree-perf} shows the rate of gravity calculations
per second as a function of the 50\% (i.e. median) and 99\% error
magnitudes, for the tree/host combination using the Gadget and
geometric MACs and the GRAPE-tree combination using only the Gadget
MAC. As we saw in figure \ref{fig:macrate} for parallel calculations
on the Origin architecture, using the Gadget MAC allows superior
performance relative to the geometric MAC at the same force accuracy,
both for the tree/host combination and the GRAPE/tree combination. In
no case do error limits rise above 1-2\%, even for the most liberal
MAC settings, for which several points lie essentially on top of each
other in the figure. Only for the homogeneous sphere test, do the error
levels rise to this level--for the other two tests, the upper limits
are reached at a factor of 2-4 smaller errors. The overall trends for
the tree/host calculations closely resemble those seen in figure
\ref{fig:macrate}, except for a smaller difference between the
behavior of the two MACs in the homogeneous sphere problem here, and
the SPH initial condition above, presumably because of the relatively
smaller value of the softening length in the present case. Quite
different behavior occurs in the GRAPE-6A$+$tree curves, especially
with the Gadget MAC, in which an essentially flat performance plateau
exists over nearly an order of magnitude in error limits.

Rates as high as 120kparticles/s are obtained for the homogeneous
sphere configuration and $>$50kparticles/s for the two $N$-body test
problem configurations.\footnote{Note however that the rates for the
7M particle $N$-body test problem rise lie some 10\% higher here than
in figure \ref{fig:arch-single} because the present calculations use
Plummer softening, in order to compare more directly to the GRAPE
calculations, rather than fixed softening length spline softening.} In
every case, the tree/host calculation rates rise to values well above
those of the GRAPE-6A$+$tree rates for the same error limits. Also,
the GRAPE-6A$+$tree curves produce much higher error limits for a
given MAC setting than do the tree/host versions, ultimately yielding
error limits a factor 2-3 larger at the most liberal MAC settings. The
differences are the consequences of two effects which may partially,
but incompletely, offset each other. First, while the tree/host
calculations includes quadrupole contributions in each interaction,
the GRAPE-6A$+$tree calculations include only the monopole term, so
that difference of one order in truncation error in the multipole
expansion exists. Larger errors at a given setting are expected as a
result. Secondly, many more particle-particle interactions will
typically be included in the GRAPE-6A$+$tree calculations because
bunch sizes optimal for this method are much larger than the optimal
clump sizes used in the tree/host method. Smaller errors at a given
setting are expected in this case, since the particle-particle
contributions contribute nothing to the total error. Taken together,
clearly the loss of accuracy due to the lower truncation order
dominates.

For error limits similar to those produced with the tree/host
calculation using the Gadget MAC, for which we use the
$\theta=5\times10^{-3}$ setting derived in section \ref{sec:MAC-speed}
as our reference, the GRAPE-6A$+$tree calculation requires a setting
of $\theta\approx 1\times10^{-4}$. Coincidentally, both settings
produce the most restrictive accuracy limits for negligible decreases
in performance below more liberal settings. For the geometric MAC,
equivalent accuracy in the tree/host calculation requires a setting of
$\theta\approx0.75$, while the GRAPE-6A$+$tree tests require a value
$\theta\approx0.5$. When used in VINE with these settings, the
tree/host option in the $N$-body tests are a factor of about two
faster, producing error limits below a few $\times 10^{-3}$ for 99\%
of the particles, while in the homogeneous sphere test, the tree/host
calculations are nearly a factor of four faster.

Though computationally costly, we have extended the high accuracy end
of each error curve in figure \ref{fig:grapetree-perf} to relative
errors as small as $10^{-8}-10^{-7}$ for the tree/host settings.
Although the settings for which these limits would be obtained would
rarely (if ever!) be used in practice, they serve another purpose
here. Namely, they allow us to make comparisons between VINE, used in
tree/host mode, at the same accuracy as GRAPE-6A processors, used in
direct summation mode. Calculations rates for VINE used in tree/host
mode with a setting of the Gadget MAC of $\theta\sim10^{-9}$ typically
fall within a factor of $\sim2-3$ of the calculation rates for the
GRAPE-6A processor used in direct summation mode.\footnote{Note that
because calculations on the GRAPE-6A are done using fixed point
arithmetic, with some intermediate operations performed with fewer
than the full 64-bits expected on a host processor \citep{makino2003},
the error limits determined for the GRAPE-6A typically lie near values
of a few $\times10^{-8}$, rather than $\sim10^{-15}$ expected from
double precision floating point limits standard on the host processor.
Such direct summation calculations on the host, performed in order to
obtain reference values for the force calculations, yield rates of
$\sim20$-$30\times$ slower than the GRAPE-6A.} In each case, GRAPE-6A
in direct summation mode is the faster option and provides superior
absolute performance. It is of interest to note however that the
performance advantage is not overwhelmingly large and, when factors
such as code complexity and hardware cost are considered, becomes even
less so. Assuming host and GRAPE processor are of equal cost,
performance when using GRAPE must be at least twice that of the host
alone in order to provide a more cost effective solution to a given
problem. By this metric, direct summation using GRAPE-6A VINE and
host/tree VINE running in parallel on 2-3 host processors, provide
roughly equivalent performance.

\begin{figure}[!t]
\rotatebox{-90}{\includegraphics[width=65mm]{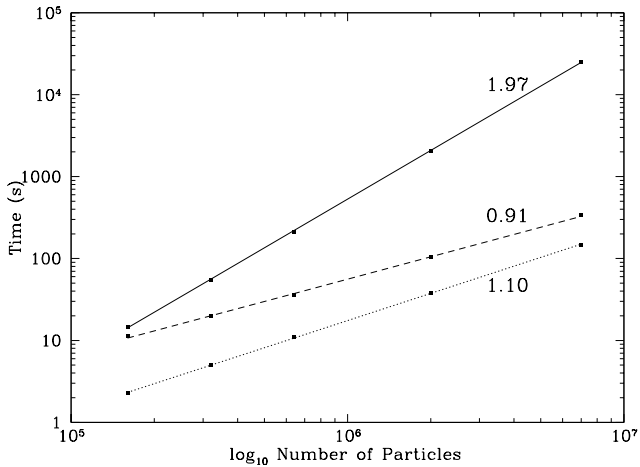}}
\caption{\label{fig:grape-resolution}
Time required to compute gravitational forces of all particles on each
other as a function of problem size, for three alternative calculation
methods employed in VINE. The three lines correspond to the time
required for the GRAPE-6A direct summation option (solid), the
GRAPE-6A$+$tree option (dashed) and the tree/host option (dotted).
Each line is a linear fit to the times defined by the points
immediately adjacent to it. The slope of each fit is included next  
to each line.}
\end{figure}

As a final test of the performance of VINE with GRAPE, we compare the
calculation times obtained by using the code in GRAPE-6A direct
summation mode, in GRAPE-6a$+$tree mode and tree$+$host mode as a
function of the size of the problem, again using variants of the
$N$-body test problem. Figure \ref{fig:grape-resolution} shows the
time required to compute gravitational forces with VINE using GRAPE-6A
direct summation mode, GRAPE-6A$+$tree mode and host$+$tree mode, each
as a function of the number of particles in our $N$-body test problem.
Times for the latter two variants are derived using the settings
defined above which, as nearly as possible, reproduce the error
distributions with similar accuracy constraints. As we expect from
basic theoretical considerations of the costs of directly summing the
gravitational forces of particles on each other, the slope of the
GRAPE-6A direct summation lies very near its predicted value of two.
The small deviation below the expected slope is readily explainable if
we account for the communication costs of sending and receiving
information between the GRAPE-6A and the host. While negligible at
large problem sizes, they account for some 5\% of the total time for
the smallest problem size.

Of greater interest are the two other curves, for GRAPE-6A$+$tree and
for tree/host. While the slope for the tree$+$host calculation falls
quite near the expected $N \log N$ behavior (see also section
\ref{sec:perform-size} and figure \ref{fig:size-single}), the slope
for the GRAPE-6A$+$tree variant lies well below it--below even a
linear proportionality. While it seems paradoxical to combine an
algorithm whose performance is formally $\mathcal{O}(N_p \log N_p)$
(i.e. tree traversal based force calculations) with another whose
performance is formally $\mathcal{O}(N_p^2)$ (i.e. direct summation)
and obtain one whose performance is better than $\mathcal{O}(N_p)$,
the paradox is readily explainable. Due to the calculation speed of
GRAPE hardware, communication costs dominate if the number of
calculations is not too large. The turnover towards the expected
$\mathcal{O}(N^2)$ scaling takes over only for interaction lists
longer than a few $10^4$. As will be shown below in section
\ref{sec:grape-tree-reconcile}, VINE's tree traversals reduce the
number of interactions per particle to well below this limit and are
extremely fast in and of themselves, so we are clearly in a
communication dominated performance regime. This conclusion is also
confirmed more directly by measurements of the portions of the
calculation in which communication occurs. Formally, and as predicted
by arguments in \citet{athanassoula98,kawai2000} and others, the
communication dominated regime should scale linearly with problem
size, in contradiction to our finding. We note that their arguments
neglect the possibility that the optimal bunch population will change
as a function of problem size and, though we have made no specific
investigation of the sensitivity of scaling to this parameter, we make
the tentative conclusion that this factor is responsible for the
effect we see.

At all problem sizes in our study, the performance of the tree/host
calculation exceeds that of both the GRAPE-6A direct summation, and
the GRAPE-6A$+$tree option in VINE. Given the clearly different
scaling behavior of the two methods however, this statement will not
remain true for all problem sizes. The magnitude of the performance
difference, even for the largest problem sizes in our study, indicate
that the crossover point lies at problem sizes of several tens of
millions of particles, a scale at which other factors, such as
parallel operation of the code, will play a much larger role in
determining the optimal computational method. Superior performance of
a tree-based method over that of direct summation is more or less to
be expected for all but the smallest problems of course, since the
algorithms and the accuracies obtained from them are so different. On
the other hand, superior performance of the tree/host calculations
over the GRAPE/tree calculations in VINE may not be as readily
accepted, since GRAPE hardware is specially designed and optimized to
be an efficient gravitational force solver and, as discussed in FMK05,
the GRAPE-6A hardware specifically for operation in conjunction with
tree based methods. The excellent performance of VINE in tree/host
mode demonstrates that careful attention to details of code
optimization can overcome such advantages. We caution however, that
before concluding that VINE in tree/host mode provides a equal or
better alternative to the use of GRAPE hardware, several other factors
must be accounted for, each of which will be addressed in the next
section.

\subsubsection{Reconciling inconsistencies between the results of our
tests and those of
\citet{fukushige2005}}\label{sec:grape-tree-reconcile}

In section \ref{sec:grape-tree-facc} we found that the performance of
VINE's GRAPE-6A$+$tree combination was substantially slower than that
using its tree/host combination at the same accuracy. A closer
examination of the timings reveals that the performances is also
substantially slower than that quoted in FMK05 as well--while they
quote timings of $\sim15$-20~s for force calculations in the
homogeneous sphere configuration, our best times were no better than
$\sim72$~s. Before concluding that force calculations using VINE on a
host processor are indeed faster than those using VINE in combination
with GRAPE-6A or, more generally, any other code in combination with
GRAPE, we must understand the origin of the differences between their
timings and our own. 

Since communication costs dominate the computation rate, the most
obvious possible explanation of the discrepancy is that VINE requires
a far larger data volume per interaction to be transfered to the GRAPE
than does the code used in FMK05. Here, we show that differences in
data transfer volumes are in fact responsible for essentially all of
the measurable performance differences. They originate in two
important differences between VINE and the FMK05 code: the total
number of nodes sent to the GRAPE board for one force calculation, and
the volume of data sent per node.

We discuss the latter difference first. The data required for the
GRAPE-6A to compute the force due to a given source node minimally
include the three components of its position and its mass. In
addition, the GRAPE-6A also requires a node index, to avoid potential
self interactions between source and sink nodes, a destination board
identifier and a memory address to define where the node will be
stored on that board. VINE utilizes the `{\tt g6a}' communication
library included as the software component of the GRAPE-6A
distribution. Communication done using this library also transfers
three components of velocity for each node as well as a number of data
related to the derivatives of acceleration, which were saved from
calculations at previous time steps. In total, 18 words (72 bytes) of
data are transfered per node using this library, though the library
calls actually used in VINE set the values of many quantities to zero. 

In contrast, an alternate library, referred to in the distribution as
the `{\tt g65}' library, provides an emulation layer for codes written
to use calls to a previous generation library distributed with the
GRAPE-5 hardware. Most significantly in the present context, this
library requires that only the minimal set of data be transfered to
the GRAPE-6A for each node, and uses reduced precision for the data
that are sent so that the communication cost is reduced to 6 words (24
bytes) per node. In spite of several independent attempts by two of
the present authors, we were unable to successfully implement code in
VINE to take advantage of the emulation library, although code
successfully interfacing directly with the GRAPE-5 library is known to
function correctly in VINE. Our tests therefore reflect the higher
costs of the full `{\tt g6a}' library, with its comparatively
inefficient communication. In contrast, tests done in FMK05
implemented code to call the emulation library (2007, T. Fukushige:
personal communication), and so reflect a communication cost per node
sent which is three times smaller than our own. Differences in
communication cost translate to a factor somewhat less than three in
speed, because while the costs of tree traversals on the host and
computation on the GRAPE-6A are comparatively small, they are not
entirely negligible. We will discuss additional implications of this
important difference below.

Even accounting for the difference in data volume per node sent,
inconsistencies in timings remain between our tests and FMK05. The
remaining factors for which we have not accounted are differences in
the total number of nodes transfered by each code. Is the total number
of nodes sent to the GRAPE in our tests much larger than in the tests
performed in \citet{fukushige2005}? Unfortunately, this question is
difficult to answer with precision because the tree construction and
traversal algorithms in our code and theirs are quite different. We
cannot expect the interaction list lengths obtained for the force
calculation on a given bunch to be similar, nor can we expect the
average bunch populations to be identical. While one algorithm may
produce shorter interaction lists, it may also produce much different
force accuracies as a consequence, or may require many more total
nodes be sent to the GRAPE because the bunches themselves are not as
large. \citet{fukushige2005} provide timing data only as a function of
the (in their case, geometric) MAC setting itself, rather than the
accuracy limits provided by that setting. Therefore, comparisons of
similar calculations cannot be made. Nevertheless, we will be able to
draw some important conclusions from the analysis, so we will proceed. 

\begin{figure}[!t]
\rotatebox{0}{\includegraphics[width=85mm]{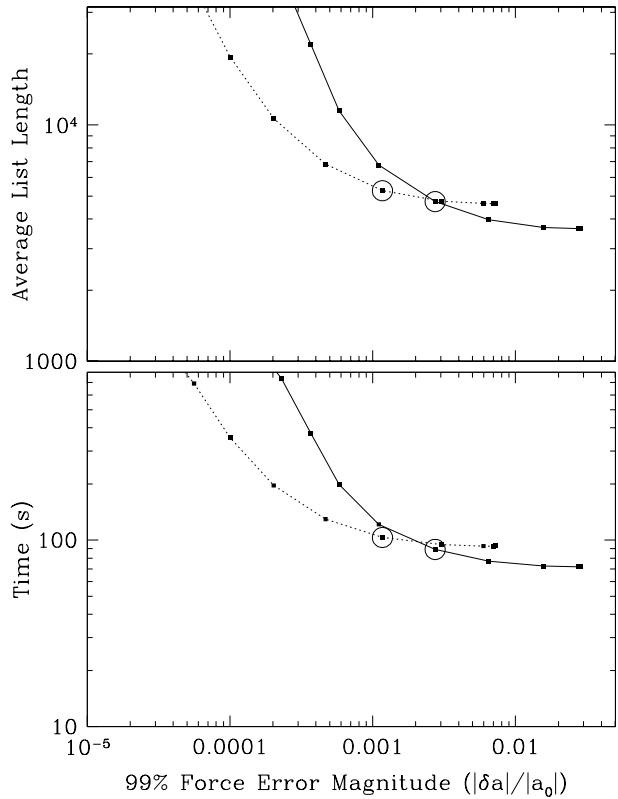}}
\caption{\label{fig:facc-intlength}
Top panel: The average length of the interaction list for each bunch
sent to the GRAPE-6A for the homogeneous sphere and the 2M particle
$N$-body test problem, each using the Gadget MAC with the same
settings as were used in figure \ref{fig:grapetree-perf}. Bottom
panel: the time required for a full force calculation on all particles
for the same configurations. Solid curves characterize the homogeneous
sphere configuration, while dotted curves define the 2M particle
$N$-body test problem configuration. The circled points in each curve
denote the location defined by the MAC setting of
$\theta=1\times10{-4}$, recommended above.}
\end{figure}

Figure \ref{fig:facc-intlength} shows the average length of the
interaction lists and computation time, each as a function of the 99\%
limit on force accuracy, using the Gadget MAC. Values for the
geometric MAC are larger and are not shown. If we assume that the near
flat performance plateau at and above our recommended Gadget MAC
setting of $\theta=1\times10^{-4}$ corresponds to a similar feature in
figure 5 of FMK05, between $\theta\approx0.7$ and
$\theta=1$, using a geometric MAC, then we may gain some insight into
the performance differences by comparing values derived from our
figure and theirs.

For both configurations, the average interaction list lengths at the
most liberal MAC settings begin near 3850 and 4650 nodes per bunch for
the homogeneous sphere and $N$-body test problem, respectively. The
lengths increase some 10-20\%, to near 4750 and 5280 nodes per bunch
at our recommended MAC setting and thereafter increase much more
rapidly, similar to the behavior seen in FMK05. In comparison to the
values of $\sim3200$ nodes per bunch quoted for their recommended
geometric MAC setting of $\theta=0.75$, ours are some $\sim$50\%
larger. Also, in comparison to their average bunch population of 836
(obtained from their Table 2), ours are smaller, near 600 particles
per bunch. In combination, the two quantities determine the total
number of nodes sent to the GRAPE-6A, and a simple calculation leads
to the result that VINE requires slightly more than twice the number
of nodes be sent to the GRAPE-6A than does the code used by FMK05. 

\subsubsection{Final notes on GRAPE$+$tree performance in
VINE}\label{sec:grape-tree-notes}

Naively multiplying together the factor of $\sim2$ difference in node
count sent to the GRAPE-6A by VINE with the near factor of three
difference in data volume per node, we arrive at a potential speed
difference of a factor of as large as $\sim$~5-6 between the
GRAPE-6A$+$tree code used in VINE compared to that in FMK05. This
value is quite consistent with the differences in overall timings at
the fastest rates quoted in their work and ours, though ambiguity
remains regarding timing comparisons for forces computed to greater or
lesser accuracy. This consideration may be of some significance
because of our finding in section \ref{sec:grape-tree-perf} that the
geometric MAC setting of $\theta=0.5$ is required to obtain force
accuracies similar to those found for optimal settings in VINE with
the Gadget MAC. Calculation rates derived from the FMK05 analyses
decrease by a nearly factor of two with this setting as compared to
their optimal rates. In the absence of performance tests compared at
the same accuracy, we can make no more definitive statements comparing
the true relative performance of the two codes.

Nevertheless, we must conclude that as currently implemented in VINE,
the GRAPE$+$tree option is not optimal, due to the inefficient
communication arising from our use of the {\tt g6a} library.  Users
who wish to take full advantage of GRAPE-6A hardware with VINE may be
well advised to incorporate appropriate modifications to either the
VINE source code or the {\tt g65} emulation library, to enable its
successful use. Such modifications have not been a high priority in
our work because of the comparatively good performance of the
tree/host option. Assuming an ideal factor of three performance
enhancement using the reduced communication emulation library, the
relative performance compared to VINE in tree/host mode favors the use
of GRAPE-6A$+$tree by only $\sim$50\%--a factor much smaller than can
be had by simply running the same job on two or more processors in
parallel. Such parallelism is highly desirable for the simulations
most commonly run in our scientific work, because they are typically
much larger than can be completed within the short turn-around time
spans required for maximum scientific productivity. They also
typically include physical processes (e.g. hydrodynamics with SPH)
beyond gravitational effects, which of course cannot utilize the GRAPE
hardware but can be efficiently parallelized on the host.

We are hopeful that VINE will be able to demonstrate speedups when
using the GRAPE$+$tree combination in conjunction with future GRAPE
hardware currently in development, such as GRAPE-7, as these use
faster versions of the PCI bus interface for the data transfer.


\section{Summary}\label{sec_summary}

In this paper we have described the algorithms used in our particle
based numerical code VINE, the optimizations we have made to those
algorithms to improve their performance on microprocessor based
computers and, finally, a number of benchmarks designed to illustrate
the benefits of each optimization in the most costly calculations
required for simulations of astrophysical systems. VINE is written in
standard Fortran~95, and is known to compile and run without
modification on a variety of common hardware platforms, from small
scale desktop workstations to large scale shared memory parallel
supercomputers. It realizes excellent performance on both, in both
serial and parallel operation. It includes options to model a number
of basic physical processes commonly required in models of
astrophysical systems, and has been designed to be extensible, so that
including additional physical processes in models will be
comparatively straightforward.

Although we believe that many members of the computational
astrophysics community would benefit from using VINE as an important
or primary component of their computational tool box, we recognize
that others with substantial investments in other codes might prefer
to continue working with them instead. We believe that such users will
not find great difficulty in either porting their physical models to
VINE and using it as an alternative to verify or otherwise check
results from their own codes, or in modifying their codes to include
many of the optimizations described here, providing benefits of
similar magnitude.

\subsection{Overall performance and
scaling}\label{sec:overall-performance}

We have discussed the performance and scalability of various
components of the code in great detail, but have left questions of the
scalability of the code as a whole to \vineI. Nevertheless, some
comments are relevant here as well. We are very encouraged to see that
the parallel scalability of the gravity and SPH calculations are
excellent. Also, though we have made no specific study of it,
performance appears to remain good even when the code is used in an
environment where only a small fraction of particles require forces on
any given time step update.

Nevertheless, there are features which limit VINE's performance, of
which its users should be aware. Perhaps the most important are the
current costs of tree builds and revisions and their parallel
scalability. Even though comparatively small in absolute cost, these
operations may become expensive when VINE is used in conjunction with
individual time steps because their cost is the same, no matter how
many particles are updated. They may also become comparatively more
costly in massively parallel operation, if VINE were run on hundreds
of processors. In studies not detailed here, we have found that when
significant populations of particles are present in $\gtrsim 12$ time
step bins (i.e. a factor of $\sim4000$ difference in $\delta t$), tree
updates become comparable to force evaluation costs on a moderately
parallel (32-64 processor) $\sim3$ million particle simulation. Costs
and scaling will improve as simulation size increases of course, but
cannot entirely be removed.

\subsection{Additional Optimization}\label{sec:more-opts}

VINE is very well optimized for high performance and includes many
features, but it is a programming truism that no code is ever
complete. VINE is no exception, and there are a number of areas where
it could be made more flexible, or its performance could be improved.
In \vineI, we discussed additional optimizations that may benefit the
code at a comparatively high level, such as changes to the integrators
used. Here, we discuss a number of additional, lower level
optimizations that we believe may be beneficial to VINE users. 

The two most costly components of VINE are the gravity and SPH
calculations. For gravity, the optimizations most likely to be
beneficial will be from including additional terms in the multipole
summation, which is currently truncated at quadrupole order. Including
terms to octupole or hexadecapole order would increase the accuracy of
the calculation for a given MAC setting, but would also increase the
cost of the calculation both in memory and in time. Indirectly, the
additional terms will increase the cost to update the nodes in a tree
revision, already a comparatively inefficient component of the code.
In order to be beneficial, increased accuracy must be both necessary
for the calculation and require less total cost to achieve than with
the present code used with a more restrictive MAC setting. Because
models of many astrophysical systems require force accuracies of only
0.1--1\%, for which the speed of the calculation is already at its
maximum, the actual benefit will be minimal. To date, we have
therefore not implemented such higher order terms.
\citet{salmon_warren94} demonstrate that for higher accuracy
requirements, including higher order multipole moments will be
beneficial and when such accuracy is required, we would recommend
their inclusion.

We saw in section \ref{sec:perform-size} that the parallel performance
of the SPH calculations saturated at a factor $\sim50$ speedup in the
Merger test simulations in which both SPH and $N$-body particles were
present. In part, we attributed the comparative performance loss,
relative to the pure SPH test problems, to the fact that both SPH and
$N$-body particles shared the same tree structure, with only a small
fraction of the total modeling the gas as SPH particles. In this
context, a useful optimization would be to separate each particle
species into a distinct tree structure. Then, tree traversals required
to obtain neighbor information would require far fewer node
examinations, and load balancing between different processors would be
less sensitive to the specific distribution of particles in the tree.
We estimate that building separate trees for particles of different
types would improve the SPH calculation rate by roughly a factor of
two in the Merger simulations. We have not implemented such separation
because benefits will only be substantial when SPH particles are a
small fraction of the total, so that the total computation time is
less significant. 

Of the four major components of the code described in this paper, the
performance of the tree build and revision scale with least efficiency
to large processor counts. They can therefore represent a significant
proportion of the total time required to complete a given simulation,
especially when the option to use individual time steps for each
particle is active. Even though the cost of a single call to either
component is small, that cost is constant no matter how many particles
are updated, so that their cost relative to the average number of
particles updated per call can be high. Further optimizations that
remedy their performance limitations will therefore be of greatest
benefit in speeding up entire simulations.

In the case of the tree revision, essentially all of the cost is
contained within a single traversal of the data arrays holding the
position, mass and multipole moment information for each node. Similar
performance bottlenecks occur for other simple loops like those that
perform particle extrapolations and updates, and are due largely to
the variable memory latencies (`NUMA'--see section
\ref{sec:impl-para}) intrinsic to large scale shared memory
architectures. Unfortunately however, we have already mined the area
of memory layout optimizations and further improvements from these
techniques are not likely to yield additional benefits. Instead, a
more profitable optimization will likely be to update only parts of
the tree corresponding to active regions of the calculation, rather
than its entirety. The benefit will not come without cost however
because partial updates mean increased levels of error in the
calculations that depend on the tree data, and some very fast
technique must be developed to define which parts of the tree are
active enough to require a full update, and which may be left for
later.

Parallel performance of the tree construction also saturates at
factors of $\sim10-12$, but for other reasons than memory latency. In
present form, some portions of the build remain un-parallelized, of
which the operations involved in assigning particles to specific hash
grid zones comprise the largest fraction, at $\sim1-2$\% of the total
build time on one processor and correspondingly more in parallel
operation. While nominally parallelizable, attempts to do so resulted
in particularly poor scaling and actual slow downs relative to serial
performance and have therefore been disabled. Alternative assignment
strategies that improved this scaling would yield immediate benefits
in the overall scaling of the build.

Secondarily, it may be beneficial to dynamically adjust the hashing
factor used to assign particles to the temporary grid according to
whether the particle distribution is particularly smooth or
particularly inhomogeneous. A low average number per zone will benefit
highly inhomogeneous distributions, where a few zones may contain a
very large number of particles while many others remain empty. A high
average number per zone will benefit smooth distributions where the
cost of expanding the search volume increases relative to the cost of
examining a single zone. Finally, due to the modular structure of the
tree build and tree traversals, it is possible to replace the binary
tree currently implemented with another structure to serve the same
purpose, so long as a similar post-processing step can be taken to
establish identical relationships between nodes on any right branch of
the tree and their parent node's sibling, as described in section
\ref{sec:tree-post}. Whether any of these changes will be of net
benefit is unknown.

\subsection{Availability of the code}\label{sec:wrapup}

The code is available to the public under GNU General Public License
version 2, via download from the ApJS website, directly from the 
authors or via download at the USM website:
http://www.usm.lmu.de/people/mwetz and at both
http://arXiv.org/abs/0802.4245 and http://arXiv.org/abs/0802.4253, by
choosing the "Other Formats" sub-link and then "Download Source".


\acknowledgements
We wish to thank Willy Benz for his generous gift to so many, over so
many years, of his SPH wisdom and the original code on which VINE is
based. We also wish to thank the referee, Volker Springel, for his
comments on this manuscript. Some of the computations reported here
were performed using the UK Astrophysical Fluids Facility (UKAFF), on
which this code was largely developed, and to whom AFN owes gratitude
for financial support. AFN wishes to thank UKAFF system administrators
Chris Rudge and Richard West for their helpful and continued
cooperation with changes to system configuration that enabled various
performance tests to be made during the development of this code.
Portions of this work were performed on the SGI-Altix 3700 Bx2
supercomputer at the University Observatory, Munich, which was partly
funded and is supported by the DFG cluster of excellence "Origin and
Structure of the Universe" (www.universe-cluster.de). Portions of this
work were carried out under the auspices of the National Nuclear
Security Administration of the U.S. Department of Energy at Los Alamos
National Laboratory under Contract No. DE-AC52-06NA25396, for which
this is publication LA-UR 08-0430. Some of the computations presented
here used facilities at the Rechenzentrum Garching of the
Max-Planck-Gesellschaft. We wish to thank Matthew Bate for making a
set of initial conditions for our SPH tests runs available to us.  MW
acknowledges financial support by Volks\-wagen Foundation under grant
I/80~040.

\bibliographystyle{apj}
\bibliography{lit}

\end{document}